\documentclass[12pt,a4paper,oneside]{book}
\usepackage{graphics}
\usepackage{amsmath,amssymb}
\begin{document}

\begin{titlepage}
\begin{center}
{\Large
THESIS\\
}
\vspace{0.5in}
{\Huge
Study of Optimization Problems\\ by Quantum Annealing\\
}
\vspace{4.5in}
{\Large
Tadashi Kadowaki\\
}
\vspace{0.5in}
{\large
Department of Physics, Tokyo Institute of Technology\\
}
\vspace{0.5in}
{\large
December, 1998
}
\end{center}
\end{titlepage}

\setlength{\baselineskip}{18pt}

\setcounter{page}{1}
\renewcommand{\thepage}{\roman{page}}

\chapter*{Acknowledgments}

I would like to express my sincerest gratitude to Professor
Hidetoshi Nishimori for his guidance, useful discussions
and above all his encouragements.

I acknowledge useful discussions
with Professor Seiji Miyashita, Dr. Nao-michi Hatano, Dr. Yukiyasu Ozeki,
Professor Kazuyuki Tanaka and Dr. Jun-ichi Inoue.
I also thank all members of the Nishimori group and the condensed-matter
theory group in the Physics Department of Tokyo Institute of Technology
for stimulating discussions.

Numerical calculations were performed on the CRAY C916/12256
in the Computer Center, Tokyo Institute of Technology.

Most of the work on computers was performed by various free or open
source softwares on Debian GNU/Linux system.
I would like to thank people who support free or open source softwares
and the community of Debian GNU/Linux.

Finally, I thank my family for spiritual and financial supports.

\chapter*{Preface}
\addcontentsline{toc}{chapter}{Preface}

On the occasion to submit my thesis to the preprint server,
I would like to refer to previous and recent studies which are
not referred to in the original thesis.

In the early days of quantum annealing study, Amara et al. used
the imaginary time Schr\"odinger equation to search the global
minimum state of the classical system[1]. 
Finnila et al. introduced a controlled and scheduled quantum
effect to the Monte Carlo Simulation[2].

Recently, thermal and quantum annealing in the disordered Ising
magnet, $\mbox{LiHo}_{0.44}\mbox{Y}_{0.56}\mbox{F}_4$, with
transverse magnetic field were compared[3,4].
Numerical studies of quantum annealing for the disordered Ising
model and the protein folding problem were performed by a few
different groups[5,6].

A quantum computer algorithm is also related with quantum annealing.
The algorithm uses the adiabatic evolution of a time dependent
Hamiltonian[7,8,9,10].

\noindent
\newline
1 May 2002
\newline
TK

\vspace{5ex}
\noindent
{\Large \bf Bibliography}

\vspace{2ex}
\begin{enumerate}
\renewcommand{\labelenumi}{[\theenumi]}
  
\item
  P. Amara, D. Hsu and J. E. Straub,
  {\it Global Energy Minimum Searches Using an Approximate Solution
  of the Imaginary Time Schr\"odinger Equation},
  J. Phys. Chem. {\bf 97} (1993) 6715.

\item
  A. B. Finnila, M. A. Gomez, C. Sebenik, C. Stenson and J. D. Doll,
  {\it Quantum Annealing: A New Method for Minimising Multidimensional
  Functions},
  Chem. Phys. Lett. {\bf 219} (1994) 343, chem-ph/9404003 (1994).

\item
  J. Brooke, D. Bitko, T. F. Rosenbaum and G. Aeppli,
  {\it Quantum Annealing of a Disordered Magnet},
  Science {\bf 284} (1999) 779, cond-mat/0105238 (2001).

\item
  J. Brooke, T. F. Rosenbaum and G. Aeppli,
  {\it Tunable Quantum Tunneling of Magnetic Domain Walls},
  Nature {\bf 413} (2001) 610, cond-mat/0202361 (2002).

\item
  Y. H. Lee and B. J. Berne,
  {\it Global Optimization: Quantum Thermal Annealing with Path Integral
  Monte Carlo},
  J. Phys. Chem. A {\bf 104} (2000) 86.

\item
  G. E. Santoro, R. Marto\v{n}\'{a}k, E. Tosatti and R. Car,
  {\it Theory of Quantum Annealing of an Ising Spin Glass},
  Science {\bf 295} (2002) 2427.

\item
  E. Farhi, J. Goldstone, S. Gutmann and M. Sipser,
  {\it Quantum Computation by Adiabatic Evolution},
  quant-ph/0001106 (2000).

\item
  E. Farhi, J. Goldstone, S. Gutmann, J. Lapan, A. Lundgren and D. Preda,
  {\it A Quantum Adiabatic Evolution Algorithm Applied to Random
  Instance of an NP-Complete Problem},
  science {\bf 292} (2001) 472, quant-ph/0104129 (2001).

\item
  A. M. Childs, E. Farhi and J. Preskill,
  {\it Robustness of adiabatic quantum computation},
  Rhys. Rev. A {\bf 65} (2002) 012322, quant-ph/0108048 (2001).

\item
  E. Farhi, J. Goldstone and S. Gutmann,
  {\it Quantum Adiabatic Evolution Algorithms versus Simulated Annealing},
  quant-ph/0201031 (2002).

\end{enumerate}

\chapter*{Abstract}
\addcontentsline{toc}{chapter}{Abstract}

We introduce quantum fluctuations into the simulated annealing process of
optimization problems, aiming at faster convergence to the optimal state.
Quantum fluctuations cause transitions between states and thus play the
same role as thermal fluctuations in the conventional approach.
The idea is tested by the two models, the transverse Ising model and the
traveling salesman problem.

Adding the transverse field to the Ising model is a simple way to
introduce quantum fluctuations.
The strength of the transverse field is controlled as a function of time
similarly to the temperature in the conventional method.
The goal is to find the ground state of the diagonal part of the
Hamiltonian with high accuracy as quickly as possible.
We also consider the traveling salesman problem.
This model can be described by the Ising spin, so that we also apply the
same technique to the transverse Ising model.

We solve the time-dependent Schr\"odinger equation numerically for small
size systems with various types of exchange interactions of the Ising
model.
Comparison with the results of the corresponding classical (thermal)
method reveals that the quantum method leads to the ground state with
much larger  probability in almost all cases if we use the same
annealing schedule of the control parameters.
We check the case of large-size systems by using the
quantum Monte Carlo method.
The simulation supports the results of the small-size systems,
while the dynamics of the Schr\"odinger equation and the quantum
Monte Carlo method are not the same.
We find that the simulated annealing by quantum fluctuations has
a better performance than the conventional method for the ground state
search of the Ising-spin systems.

The calculation of the traveling salesman problem is also performed as
an application to the general optimization problems.
We obtain the same feature of the fast convergence to the optimal state
as the transverse Ising model by using the quantum fluctuations.

We also find that the relaxation time is quite short for quantum systems
by numerical simulations.
We consider this is one of the reasons why the annealing in quantum
systems have a better performance of finding the optimal state in
comparison with classical systems.

\clearpage
{\Large List of Papers\\}

\begin{enumerate}
 \item T. Kadowaki and H. Nishimori,
``Quantum annealing in the transverse Ising model'',
Phys. Rev. E {\bf 58} (1998) 5355.

\item T. Kadowaki and H. Nishimori,
``Monte Carlo study of quantum annealing'',
in preparation.
\end{enumerate}

\vspace{2ex}

{\Large List of Papers Added for Reference\\}

\begin{enumerate}
\item T. Kadowaki, Y. Nonomura and H. Nishimori,
``Exact Ground-State Energy of the Ising Spin Glass on Strips'',
J. Phys. Soc Jpn. {\bf 65} (1996) 1609.

\item M. Yamana, H. Nishimori, T. Kadowaki and D. Sherrington,
``High-Temperature Dynamics of Spin Glasses'',
J. Phys. Soc Jpn. {\bf 66} (1997) 1962.
\end{enumerate}

\tableofcontents

\chapter{Introduction}

\setcounter{page}{1}
\renewcommand{\thepage}{\arabic{page}}

\section{The Combinatorial Optimization Problem}
\label{sec:1.1}

The combinatorial optimization problem is to find a minimum or maximum
value of a function of very many independent variables,
where the variables take discrete values.
This function, usually called the cost function or objective function,
represents a quantitative measure of the ``goodness'' of some complex
systems.
A famous example is the traveling salesman problem \cite{TSP1,TSP2} to
find the shortest or lowest-cost route to visit given cities.
The techniques to find the shortest route have been studied for a long
time, because such a problem actually happens in various situations,
and people have to obtain the solution.
One of the most important application of the optimization problems today
is the circuit design of the computer systems.
The cost function is the wiring length in this case.
Long wiring makes delays in electric circuit,
and the speed of computer slows down.

The algorithm to find the shortest tour of the traveling salesman
problem or the wiring length of the circuit design in polynomial
time of $N$, where $N$ is the size of the problem, is not found yet.
When $N$ increases, the complexity of the calculation increases
exponentially and goes over the limit of the computational power.
This kind of problems are called the ``Nondeterministic
Polynomial-time solvable hard or complete ($NP$-hard or $NP$-complete)''
problems.
(Details of the definitions are explained in the next section.)
Any algorithm of solving an $NP$-hard or $NP$-complete problem in
polynomial time is not found so far and people consider that there does
not exist such an algorithm.
Therefore we consider approximate algorithms by which we can obtain an
approximate solution or an exact solution with some probability in
polynomial time.
In this thesis we focus on how to solve the $NP$-hard or $NP$-complete
problems approximately as fast as possible.

\section{The $NP$-hard and The $NP$-complete Problems}
\label{sec:1.2}

In this section we explain several levels of difficulty of optimization
problems.
The $NP$-hard and $NP$-complete problems belong to the class quite hard
to solve.
The difference between the classes $NP$-hard and $NP$-complete is
not in the degree of difficulty but the way to answer the questions;
``yes or no'' for $NP$-complete and in general statements for $NP$-hard.

\begin{figure}[htb]
\scalebox{0.8}{\includegraphics{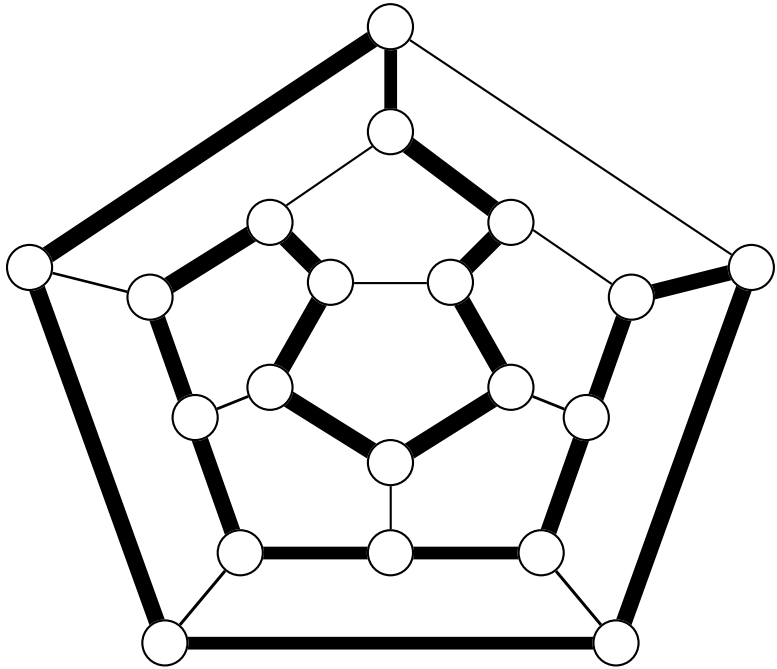}}
\caption{
The Hamilton graph is the graph which has a Hamilton circuit.
The bold line is the Hamilton circuit.
} 
\label{fig_Hamilton}
\end{figure}

First, let us consider the class $NP$ before the definition of the classes
$NP$-hard and $NP$-complete.
The class $NP$ is a subclass of the solvable class of decision
problems whose answer is expressed by either ``yes'' or ``no''.
For instance, the Hamilton circuit problem is the problem to check the
existence of the Hamilton circuit which is the closed loop constructed
by edges and all the vertexes in a given graph.
An example of the Hamilton circuit problem and its Hamilton circuit are
shown in Fig.~\ref{fig_Hamilton}.

Roughly speaking, the $NP$ problem is the problem which can be verified
in polynomial time by the people who know the evidence to prove the
existence of the solution.
For instance, one can answer ``yes'' for the Hamilton circuit problem
shown in Fig.~\ref{fig_Hamilton}, when the Hamilton circuit (the
evidence in this case) is given as a bold line in Fig.~\ref{fig_Hamilton}.
The verification time of the evidence, checking the closed loop which
contains all the vertexes, is proportional to $N$.
To be precise, the $NP$ problem is defined as a problem which can be
solved by the nondeterministic Turing machine in polynomial time of $N$.
However, the definition of the nondeterministic Turing machine is
complicated, and the reader who is not interested in the detail can
skip the following two paragraphs.

Before presenting the definition of the nondeterministic Turing machine,
we explain the definition of the usual (deterministic) Turing machine.
The nondeterministic Turing machine is the extended model of the Turing
machine.
The definition of the Turing machine is constructed with
\begin{itemize}
\item the tape which includes alphabets (as commands and data).
\item the state of the machine.
\item the transition function of the state by using the alphabet read
      from the tape.
\end{itemize}
The Turing machine is a simple model of actual computers.
For the deterministic Turing machine, the transition function is
single-valued.
Turing machine reads an alphabet from the tape and modifies its
configuration (the state and alphabets on the tape) from the previous
configuration according to the given alphabet.
A unique modified state is given by this transition.

\begin{figure}[htb]
\scalebox{0.7}{\includegraphics{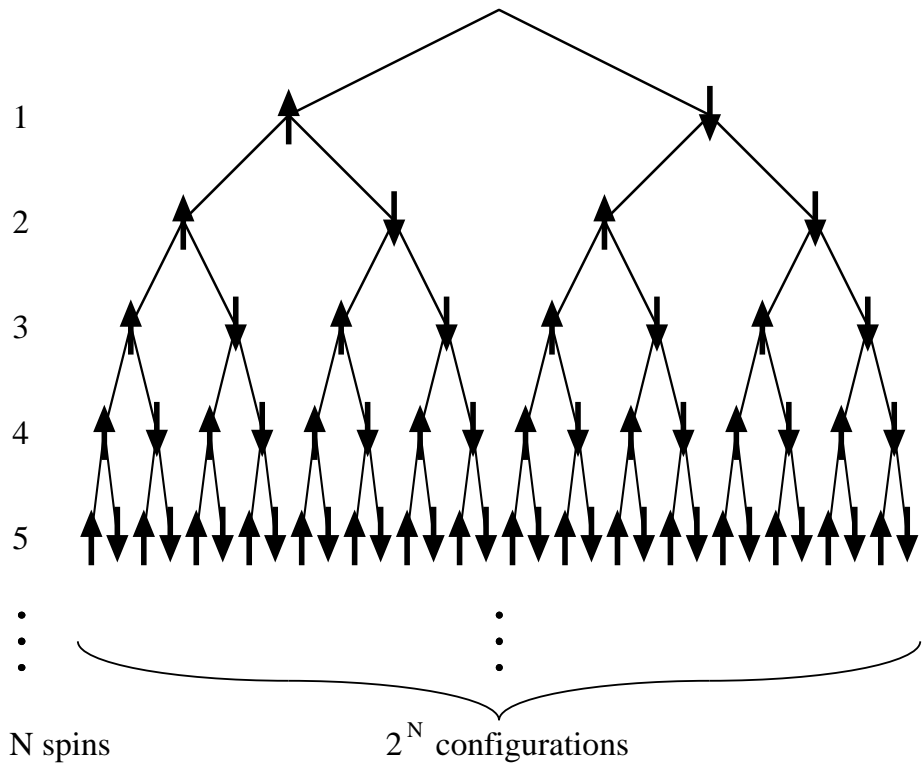}}
\caption{
The path means the spin configuration.
The number of the leaves (the bottom ends of the tree) equals to the
number of configurations as $2^N$.
} 
\label{fig_tree2}
\end{figure}

On the other hand, the transition function of the nondeterministic
Turing machine is multi-valued.
The feature of the nondeterministic Turing machine quite differs
from conventional computers.
The configuration of the nondeterministic Turing machine becomes
multiple configurations.
This can be understood loosely in a way that the nondeterministic Turing
machine makes replicas of possible configurations by a single step and
the number of replicas increases exponentially as a function of the step.
This feature is quite useful to solve complicated problems.
For finding the ground-state energy of Ising-spin systems, for example,
we have to enumerate all possible spin configurations, if the
ground state is nontrivial.
These configurations are illustrated by the tree structure in
Fig.~\ref{fig_tree2}.
The direction of spins is assigned at each branching point and the level
corresponds to the location of the spin.
Each path means the configuration.
The nondeterministic Turing machine can check all paths in parallel.
At the first step, the machine makes a replica and the number of
machines is two.
Each machine makes a replica at each step so that the number of machines
is $2^N$ at the $N$th step.
The calculations in the same level of the tree are performed at a time.
By this procedure, we can enumerate $2^N$ configurations in $N$ steps.

All $NP$ problems can be verified in polynomial time even on a
deterministic Turing machine, but the difficulty of solving the
problems is not the same.
The problems are divided into some subclasses --- the subclasses $P$ for
easy problems and $NP$-complete for difficult problems.
The problem which has the algorithm to be solved in polynomial time by
the deterministic Turing machine belongs to the class of Polynomial-time
solvable ($P$).
Obviously $NP$ contains $P$, because the nondeterministic Turing machine
includes the deterministic Turing machine in its definition.

\begin{figure}[htb]
\begin{center}
\scalebox{1}{\includegraphics{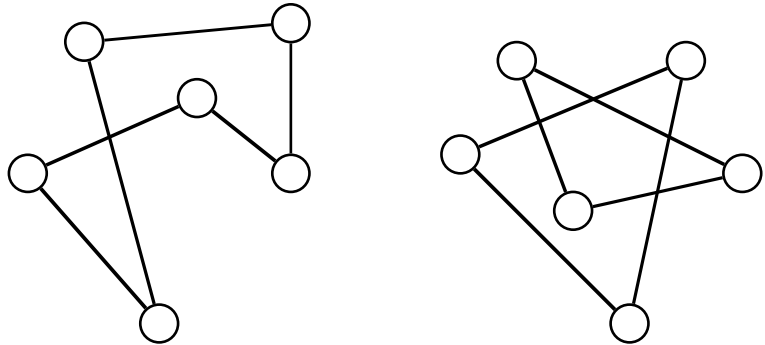}}
\end{center}
\caption{
The Hamilton circuit problems with the answer ``yes'' and ``no''.
The left graph is the single loop (the Hamilton circuit) and right one
is not the single loop.
} 
\label{fig_Hamilton_ex}
\end{figure}

For an example of the $P$ problem, we explain the Hamilton circuit
problem for the graph whose maximum degree of the edge in a vertex is
limited to two belongs to $P$.
The typical graphs are shown in Fig.~\ref{fig_Hamilton_ex}.
The answers of the left and the right problems are ``yes'' and ``no''
respectively.
One can check the existence of the Hamilton circuit by the following
way:
\begin{enumerate}
\item Start from a vertex and jump to the neighbor vertex connected by an
edge. (One can not jump to the neighbor which has already been visited once.)
\item Repeat 1 until no vertex to jump is left.
\item If the last vertex is connected to the starting vertex and all the
vertexes have been visited, the answer is ``yes''.
The answer is ``no'' other wise.
\end{enumerate}
This procedure needs polynomial time.
If the maximum degree of the edge in a vertex is not limited to two,
this procedure does not work.

Let us consider the difficult classes $NP$-hard and $NP$-complete.
If any problems in $NP$ can be solved by using a particular algorithm
of the problem repeatedly polynomial times at most, the problem is
called ``$NP$-hard''.
In other wards, we can solve any $NP$-problems by calling the subroutine
for a particular $NP$-hard problem (maybe once except for other $NP$-hard
problems).
The $NP$-hard problem is more difficult than any $NP$ problems or
as difficult as the other $NP$-hard problems in polynomial order.
The algorithm of the $NP$-hard problem is the most general algorithm
in $NP$ class.
The $NP$-hard problems by definition are not limited to belong to the
class $NP$.
For instance, the traveling salesman problem is not an $NP$ problem
but an $NP$-hard problem because the answer is not ``yes or no'' but the
tour length and route.
The subclass ``$NP$-complete'' is the set of the $NP$-hard problems
which belong to the class $NP$.
The $NP$-complete problems are more difficult than any other $NP$
problems and as difficult as the $NP$-hard problems.
The Hamilton circuit problem is known as an $NP$-complete
problem~\cite{Aho}.

\begin{figure}[htb]
\scalebox{0.8}{\includegraphics{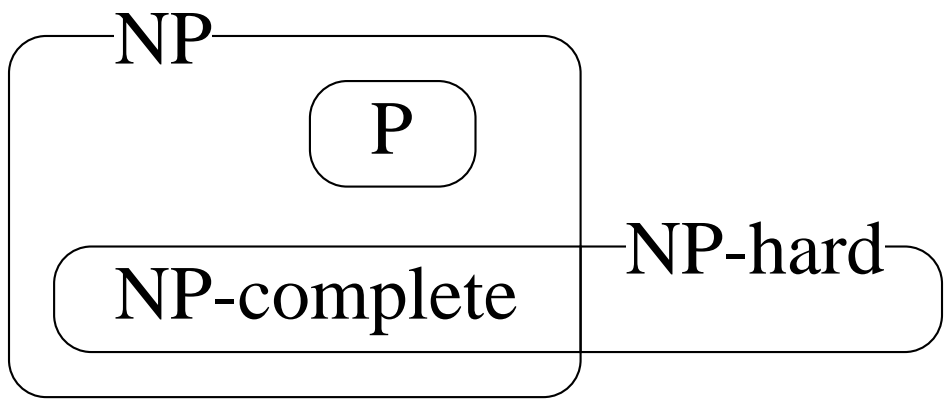}}
\caption{
The class of $NP$ problems assuming that $P \neq NP$.
Under this assumption it can be shown that there are $NP$ problems that
are neither $NP$-complete nor $P$.
} 
\label{fig_prob_class}
\end{figure}

The difficulty of the problems can be expressed symbolically as
$P \leq (NP - P - NP\text{-complete}) \leq NP\text{-complete} \approx
NP\text{-hard}$.
It is a trivial consequence of their definition that if one of the
$NP$-complete problems can be solved in polynomial time, then all the
problems of the $NP$ class are also polynomial.
In this case one would have $P = NP$.
So far no polynomial algorithm has been found for an $NP$-complete
problem, and the question of whether $P$ is equal to $NP$ is still open,
although the general belief is that $NP$-complete problems are not
polynomial.
The most probable situation is sketched in Fig.~\ref{fig_prob_class}.
The $NP$ problems can be divided into three subclasses, $NP$-complete, $P$
and the subclass which does not belong to the previous two subclasses.
The class $NP$-hard is separated into two subclasses, $NP$-complete and
the other part by whether the problem is the decision problem or not.
Empirically the time it takes to solve an $NP$-hard and an $NP$-complete
problems tends to scale exponentially with the size $N$.
All the problems can be solved by the enumeration method by which
all the configurational space of the problem are enumerated and this
calculation needs the time proportional to $e^N$.

\section[Spin Glass Models and Optimization]
{Spin Glass Models and Their Relation to Optimization}
\label{sec:1.3}

Combinatorial optimization problems appear in various disciplines.
In statistical mechanics, the energy is equivalent to the cost function
of the optimization problem.
Until the 1970s, almost all the systems to be studied were homogeneous so
that the energy landscape has a simple structure and the optimal solution
(the ground state) is trivial.
Although such a system has many variables, the degrees of freedom can be reduced
to small numbers.
For instance, spins turn to the same direction in the ground state of the
ferromagnetic model and the spin variables are reduced to single-spin variable.
This reduction comes from the homogeneous interactions.
In the 1970s the study of the spin glass started.
The spin glass system is a non-homogeneous system and the ground state
is nontrivial.
The background of the physics of spin glasses and its relation to the
combinatorial optimization problem is explained in this section.

The study of spin glasses started in 1972 when Cannella and
Mydosh~\cite{Cannella} investigated the AC susceptibility of a dilute
magnetic alloy Au-Fe and found out that the AC susceptibility has a
nontrivial sharp cusp.
In the alloy, the interaction $J_{ij}$ between two spins of Fe atoms
is expressed as the RKKY interaction~\cite{RKKY1,RKKY2,RKKY3}
\begin{equation}
J_{ij} \propto \frac{1}{r^3_{ij}}\cos(2 k_{\text{F}} r_{ij}) \ ,
\end{equation}
where $r_{ij}$ is the distance between two spins
and $k_{\text{F}}$ is the Fermi wave number.
From this interaction and a random distribution of Fe atoms in space,
it is possible that the interaction is a positive or a negative random
variable.

Edwards and Anderson proposed the so-called Edwards-Anderson (EA)
model~\cite{EA} to introduce the effect of the RKKY interaction in a
diluted magnetic alloy.
The EA model has nearest-neighbor interactions $J_{ij}$.
The value of the interaction distributes as
\begin{equation}
P(J_{ij}) = \left ( \frac{z}{2 \pi J^2} \right )^{1/2}
            \exp \left [ - \frac{z}{2 J^2} \left ( J_{ij}
               - \frac{J_0}{z} \right )^2 \right ] ,
\end{equation}
where $z$ is the number of nearest neighbors.
The average of the distribution is $J_0/z$, and the standard deviation
is $J/\sqrt{z}$.
The Hamiltonian is given by
\begin{equation}
{\cal H} = - \sum_{\text{n.n.}} J_{ij} S_i S_j \ .
\end{equation}
The EA solution is not solved exactly.
Sherrington and Kirkpatrick~\cite{SK} investigated an infinite-range
model, the so-called SK model.
The Hamiltonian is given by
\begin{equation}
{\cal H} = - \sum_{<ij>} J_{ij} S_i S_j - h \sum_i S_i \ ,
\end{equation}
where $<ij>$ denotes a summation over all spin pairs.
This is the reason of the name of the infinite-range model.
The summation is different from that in the EA model.
If the average of distribution of interactions is equal to zero and
the external field is vanishing, a second order phase transition
at a finite temperature $T_{\text{c}} (=J/k_{\text{B}})$ occurs,
and the spin glass phase realizes below $T_{\text{c}}$.

\begin{figure}[htb]
\begin{center}
\scalebox{0.5}{\includegraphics{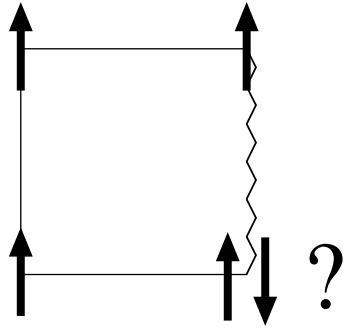}}
\end{center}
\caption{
The frustrated plaquette.
The solid and the zigzag lines are the ferromagnetic and
antiferromagnetic bonds.
}
\label{fig_frust_bond}
\end{figure}

The statistical property as the SK model was almost solved by further
studies~\cite{AT,Parisi1,Parisi2,Parisi3,TAP,Tanaka_Edwards,Mackenzie,
Nemoto,Dominicis},
but it is difficult to analyze the EA model by the same approaches.
The EA model is often studied by computer simulations, the Monte
Carlo simulations or other methods.
The great interest of the EA model or the $\pm J$ model whose
interactions are distributed binary (at $J$ and $-J$) is whether
the same picture of the SK model --- the picture of the
ultrametricity and the structure of the $P(q)$~\cite{Parisi3,Dominicis}
--- exists or not.
One can check this picture from the calculation of the distribution
function $P(q)$ where the overlap $q$ is calculated from the pure states
which are the equilibrium states separated by high energy barriers and
can be obtained by the Monte Carlo simulation.
If the simulation starts from a random configuration which is equivalent
to the configuration at high temperature, the system is stuck in a
metastable state and it takes long time to escape from the metastable
state because the basins of the pure states and the metastable states
are separated by high energy barriers.
To avoid this difficulty of the simulation, one can start the simulation
from the ground state, which is close to the pure state at low temperature.
However, finding the ground state is also difficult.
The reason is the following:
Spins on the some plaquettes in the EA model do not satisfy all the bonds
illustrated in Fig.\ref{fig_frust_bond}.
This situation occurs, when the product of all the bonds in the plaquette
has a minus sign.
The frustrated plaquettes are located at random so that we have to
consider the global structure of the configuration which pays as small
penalty of the energy as possible.
Therefore the ground state is not the trivial configuration.
The difficulties in studying the EA model are the slow relaxation and the
non-trivial configuration of the ground state.
The latter problem is a typical example of the combinatorial
optimization problem.

\begin{figure}[htb]
\begin{center}
\scalebox{1}{\includegraphics{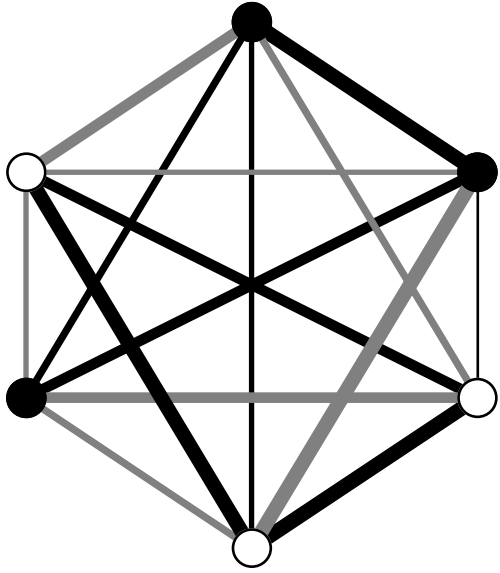}}
\end{center}
\caption{
The bipartition problems with $N=6$.
The open and closed circles are the divided members of the group.
The black and the gray lines mean ``likes'' and ``dislikes'' between the
members.
}
\label{fig_bipartition}
\end{figure}

The SK model is connected with the bipartition problem directly.
We divide the group into two parts.
``Likes and dislikes'' between the members is expressed as the numerical
value.
The task is to find the best partition of the members.
The case of $N=6$ is illustrated in Fig.~\ref{fig_bipartition}.
The members are divided into open and closed circles, and the black and
the gray lines mean ``likes'' and ``dislikes'' between the members.
Assigning $+1$ to the members in one part and $-1$ to the members in the
other part, we can regard that the members are expressed by Ising
spins $\{\sigma_i\}$.
The cost function of the ``likes and dislikes'' between two members $i$
and $j$ can be obtained as $J_{ij}\frac{\sigma_i \sigma_j + 1}{2}$.
If the two members $i$ and $j$ are in the same part, ``likes and
dislikes'' is counted as $J_{ij}$.
If the two members are in the different parts, the value is zero.
We can calculate the total cost of ``likes and dislikes'' by summing up
all the pairs as:
\begin{eqnarray}
\text{Total Cost} & = & \sum_{(ij)} J_{ij}
                        \frac{\sigma_i \sigma_j + 1}{2} \nonumber \\
& = & \frac{1}{2} \sum_{(ij)} J_{ij} \sigma_i \sigma_j + \text{const.}
\end{eqnarray}
This cost function is exactly the same as the Hamiltonian of the SK
model.
The bipartition problem and the finding the ground state of the SK model
are known as the $NP$-hard problems~\cite{neuralcomp,Shinomoto}.

\section{Simulated Annealing}
\label{sec:1.4}

As shown in the previous section, large computational time is needed
to solve an $NP$-complete or an $NP$-hard problem exactly.
The technique of simulated annealing (SA) was first proposed
by Kirkpatrick {\it et al.} \cite{Kirk}
as a general method to solve optimization problems.
This method is an approximate algorithm to find the optimal solution
in finite time, but the solution converges to the optimal state
in the infinite-time limit.
The probability to find the ground state increase with time and converges
to one in the infinite-time limit.
For practical applications, it is important to find a sufficiently good
solution, where ``good'' means that the cost function is close to the
optimal value.

The idea is to use thermal fluctuations to allow the system
to escape from local minima of the cost function so that the system
reaches the global minimum under an appropriate annealing schedule
(the rate of decrease of temperature).
Let us consider a simple example of a system described by Ising
spins $\{\sigma_i\pm 1\}$, binary variables.
Spins interact each other as $-J \sigma_i \sigma_j$ and the total
energy (cost function) of this system is $\mathcal{H}=- \sum_{ij} J
\sigma_i \sigma_j$, where ${ij}$ runs over all possible pairs of
sites which interact with each other.
If the two spins $\sigma_i, \sigma_j$ take the same value, the energy
becomes lower.
This system is the simplest model of a ferromagnet.
In the optimal configuration, all spins align in the same direction,
up or down.

Once we define the energy of the system, we can consider
statistical mechanics of this system.
The spin configuration $\{\sigma_i\}$ is modified by thermal noise.
The probability with which the configuration $\{\sigma_i\}$ appears is
proportional to the Boltzmann factor
$\exp(-E(\{\sigma_i\})/k_{\text{B}}T)$, where $T$ is the temperature and
$k_{\text{B}}$ is the Boltzmann constant.
The probability has to be normalized and the normalization factor is $1/Z$.
The inverse of the factor, the so-called partition function, is
expressed as:
\begin{equation}
Z = \sum_{\{\sigma_i=\pm 1\}} \exp(-E(\{\sigma_i\})/k_{\text{B}}T)
\end{equation}

\begin{figure}[htb]
\scalebox{0.8}{\includegraphics{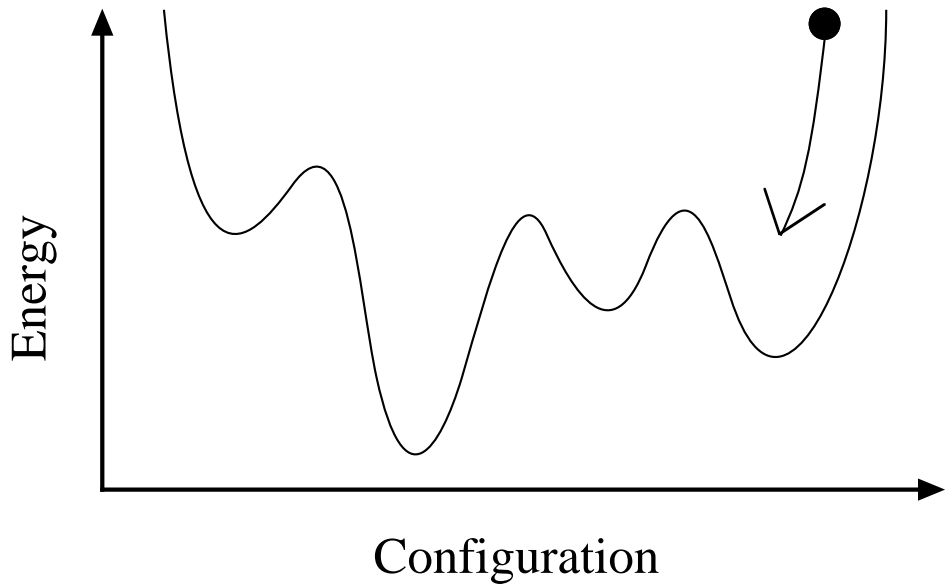}}
\caption{
Cooling from high temperature to low temperature.
If the cooling rate is too fast, the system may be trapped in a local
minimum. 
} 
\label{fig_SA_ZT}
\end{figure}

\begin{figure}[htb]
\scalebox{0.8}{\includegraphics{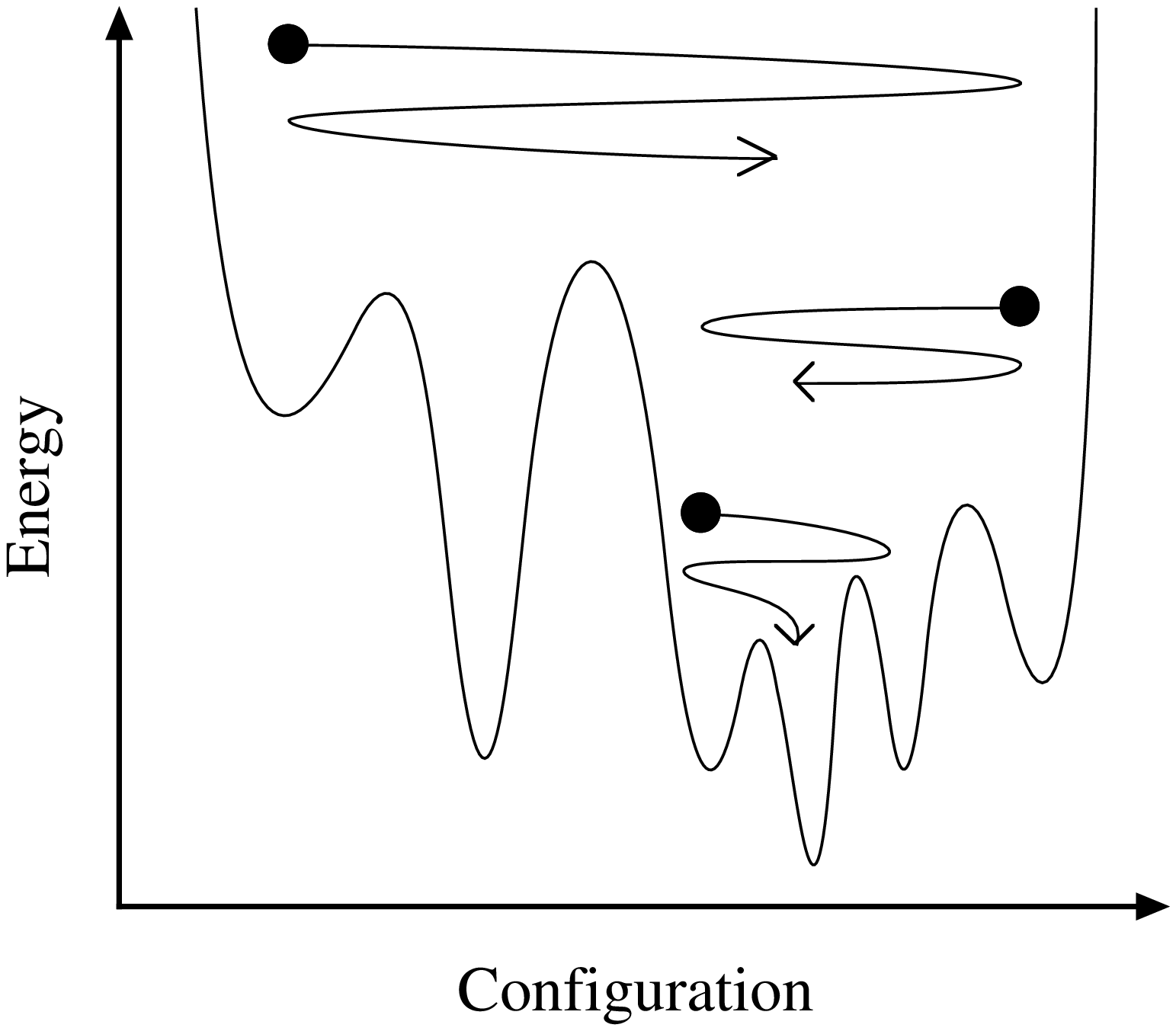}}
\caption{
The system seek the various levels of the energy-landscape structure
corresponding to the temperature.
The system searches the global (local) structure at high (low) temperatures.
} 
\label{fig_SA_img}
\end{figure}

The lower the temperature becomes, the larger the probability of the spin
configuration of the lowest energy becomes.
The system freezes to the ground state as the temperature goes to zero.
However, the system does not always freeze to the ground state in rapid
cooling.
As illustrated in Fig.~\ref{fig_SA_ZT}, the system may be trapped in
a local minimum.
Therefore, the temperature should be cooled attentively.
At high temperature, the system changes the configuration almost
randomly and searches the global structure of the energy landscape.
The system seeks the area in which the valley of the global minimum is
included when the temperature is decreased.
Finally, the system is stuck at the valley of the global minimum at zero
or sufficiently low temperature (lower than the energy gap of the ground
state and the first excited state).
The idea of this cooling is illustrated in Fig.~\ref{fig_SA_img}

By the way, how can we cool the system slowly enough?
Geman and Geman proved a theorem on the annealing schedule
for combinatorial optimization problems~\cite{Geman}.
They showed that any system reaches the
global minimum of the cost function asymptotically
if the temperature is decreased as
$T= c/\log t$ or slower,
where $c$ is a constant determined by the system size and
other structures of the cost function.
When the product of the temperature and the Boltzmann constant
$k_{\text{B}}T$ become less than the energy gap between the ground
state and the first excited state, the system jumps to the excited state
rarely by thermal fluctuation.
Writing such a temperature as $T_0$, we obtain the time to find the
ground state as $t\sim \exp(c/k_{\text{B}}T_0)$.
This bound on the annealing schedule may be the optimal
one under generic conditions
although faster decrease of the temperature often
gives satisfactory results in practical applications
for many systems.

\section{Quantum Annealing}
\label{sec:1.5}

In SA thermal fluctuations are introduced in the optimization problem so that
transitions between states take place in the process of search for the
global minimum among many states.
Thus there seems to be no reasons to avoid use of
other mechanisms for state transitions if
these mechanisms may lead to better convergence properties.
One such possibility is the generalized transition probability
of Tsallis \cite{Tsallis}, which is a generalization of the
conventional Boltzmann-type transition probability
appearing in the master equation and thus used
in Monte Carlo simulations.
In this method, the system converges to the optimal state under
power-low decrease of the
temperature~\cite{Tsallis,Penna,Andricioaei,Nishi}.
This schedule is faster than the log schedule of the conventional SA.
However, the generalized transition probability does not satisfy the
condition of the detailed balance, the sufficient condition for
convergence to the equilibrium state, and the physical meanings of the
system is not trivial.

We seek another possibility of making use of quantum tunneling processes
for state transitions, which we call quantum annealing (QA).
In particular we would like to learn how effectively quantum
tunneling processes possibly lead to the global minimum
in comparison to temperature-driven processes used in the
conventional method of SA.

In the virtual absence of previous studies along such
a line of consideration,
it seems better to focus our attention on
a specific model system, rather than to develop a general
argument, to gain an insight into the role of quantum
fluctuations in the situation of optimization problem.
Quantum effects have been found to play a very similar role
to thermal fluctuations in the Hopfield model and the image
restoration problem in a transverse field in thermal
equilibrium~\cite{NN,Tanaka}.
This observation motivates us to investigate dynamical
properties of the Ising model
under quantum fluctuations
in the form of a transverse field.
We therefore start the study of the QA by the transverse
Ising model with a variety of exchange interactions.
The transverse field controls the rate of transition
between states and thus plays the same role as
the temperature does in SA.
We assume that the system has no thermal fluctuations
in the QA context and the term ``ground state''
refers to the lowest-energy state of the Hamiltonian
without the transverse field term.

Static properties of the transverse Ising model have been
investigated quite extensively for many years \cite{Chak,Lan,Alvarez}.
There have however been very few studies on
the dynamical behavior of the Ising model with a transverse field.
We refer to the work by Sato {\it et al.} who carried out
quantum Monte Carlo simulations of the two dimensional EA model in an
infinitesimal transverse field, showing a reasonably fast approach
to the ground state~\cite{Sato}.
We present a new point of view by comparing the
efficiency of QA directly with that of classical SA in reaching
the ground state.

In the next chapter, we consider the time-dependent Schr\"odinger
equation for QA and the classical master equation for SA numerically for
small-size systems with the same exchange interactions under the same
annealing schedules.
We explain the model and define the measure of closeness of the system
in QA to the desired ground state.
This measure is compared with the corresponding classical
probability that the system is in the ground state.
Calculations of probabilities that the system is in the ground
state at each time for both classical and quantum cases
give important implications on the relative efficiency of the
two approaches.
The numerical results for QA and SA suggest that QA generally gives a
larger probability to find to the ground state than SA under the same
conditions on the annealing schedule and interactions.
We also calculate the analytical solutions for the one-spin case
which turns out to be quite non-trivial.
Explicit solutions yield very useful information to clarify several
subtle aspects of the problem.

In Chapter 3, we extend our method by using the Monte Carlo method
to calculate large-size problems.
We adopt the master equation as the time-evolution equation of SA
in contrast to the Schr\"odinger equation of QA.
The original definition of SA is expressed in terms of the Monte Carlo
method, which is equivalent to the master equation.
It is natural to consider the applicability of Monte Carlo method to QA.
The time-evolution of the wave function is expressed as:
\begin{equation}
| \Psi(t) \rangle = \mathrm{T} e^{-i \int_0^t dt \mathcal{H}(t)}
                               | \Psi(0) \rangle ,
\end{equation}
where the symbol $\mathrm{T}$ is the time ordering operator.
There are two approaches to take into account the quantum effects or
the time evolution gradually.
The first one is to divide the Hamiltonian into some parts which are easy to
diagonalize (for instance, into three terms which include $\sigma^x$,
$\sigma^y$, $\sigma^z$, respectively) as
\begin{equation}
e^{\mathcal{H}} = \lim_{m \rightarrow \infty}
  ( e^{\mathcal{H}_0/m} e^{\mathcal{H}_1/m} \cdots )^m .
\end{equation}
This division is the Suzuki-Trotter decomposition.
By inserting the complete set $\sum_i |i\rangle \langle i| (=1)$,
the system is mapped to a $(d+1)$-dimensional classical system.
Once the system is expressed as a classical system, we can perform the
Monte Carlo method, the so-called quantum Monte Carlo method.
The second one is the division of the short-time evolution as
\begin{equation}
e^{-i \int_0^t dt \mathcal{H}(t)} = \lim_{\Delta t \rightarrow 0}
   e^{-i \Delta t \mathcal{H}(t)} \cdots e^{-i \Delta t \mathcal{H}(0)} ,
\end{equation}
the path integral Monte Carlo method.
The dynamics of the two approaches are not exactly the same as the dynamics
of the Schr\"odinger equation.
The performances of the two approaches are the same or better in comparison
with the dynamics of the Schr\"odinger equation in some sense (a detailed
discussion is shown later).
We finally adopt the quantum Monte Carlo method for the calculation of
QA, because this method is similar to the original SA and we can find
the optimal state precisely by the time evolution.
The results of the simulations show that QA performs in finding the
optimal state better than SA in spite of the disadvantage in the
calculation of the quantum effect (the calculation on $(d+1)$-dimensional
systems is larger than $d$-dimensional system).
We also calculate the difference between the quenched system and
annealed system.
The calculation of the quenched system is almost the same as the study by
Sato {\it et al.}~\cite{Sato}.
In general, quenched systems tend to be stuck in local minima.
We find that the classical quenched system is stuck more often than the
quantum quenched system.
We consider this is one of the reasons why QA works better than SA.

So far, the systems are physical models.
The problems of the ground state finding of physical models occupy the
small area of the combinatorial problems and we have to check
applicability to the general problems.
One of the well-known problems is the TSP.
We apply QA to the traveling salesman problem (TSP) in Chapter 4.
TSP can be mapped to the Ising spin system so that we can use the same
framework of QA.
The modification of the configuration of TSP in SA is not the same as the
modification of the usual one-spin dynamics of SA.
Four spins are changed at a time in TSP.
The quantum-effect term of the Hamiltonian (the cost function) has to be
changed to the product of the four spin operators to fit the dynamics of
SA.
However, we adopt the same quantum-effect term, the transverse field,
because of the difficulty of the calculation.
In spite of this, QA also works and improves in comparison to SA.
This fact suggests that QA can be applied to various problems in which
we can not define the quantum effect exactly.

\chapter[Analysis by Differential Equations]
{Analysis of Small-size Systems by Differential Equations}
\label{chap:2}

In this chapter, we introduce the quantum annealing (QA) for small-size
systems by using the Schr\"odinger equation and compare the results of
QA and the simulated annealing (SA).
The models (problems) we deal with here are the models in statistical
physics, precisely the Ising model.
The transverse field is introduced as the quantum effect and scheduled
as a decreasing function of the time.
This field flips single spin at a time so that the transition is quite
similar to the usual single-spin-flip dynamics of SA.

The analysis is performed numerically and analytically for SA and QA.
The system size of the numerical calculation is up to eight.
The calculation limit of the size is around fourteen by the
middle-class supercomputer today. (for a fourteen-spin system,
calculation needs the storage about 2G Byte!)
The models we use as test-bed are ferromagnetic, frustrated and
random-interaction models.
We also study for the single-spin quantum system with longitudinal and
transverse field.
The exact solutions are obtained for some schedules.
The single-spin system is too small to discuss the property of QA, but
the result provides some information for the results of the numerical
calculations.

\section{The Transverse Ising Model}
\label{sec:2.1}

Let us consider the following Ising model with longitudinal
and transverse fields:
\begin{eqnarray}
\mathcal{H}(t) & = &
 -\sum_{ij} J_{ij}\sigma^z_i\sigma^z_j -h\sum_i\sigma^z_i
 -\Gamma(t)\sum_i\sigma^x_i 
  \label{Hamiltonian}\\
 & \equiv & \mathcal{H}_0 - \Gamma(t)\sum_i\sigma^x_i,
\end{eqnarray}
where the type of interactions will be specified later.
The term of longitudinal field was introduced to remove
the trivial degeneracy in the exchange interaction term
coming from the overall up-down symmetry that effectively
reduces the available phase space by half.
The $\Gamma (t)$-term causes quantum tunneling
between various classical states
(the eigenstates of the classical part $\mathcal{H}_0$).
By decreasing the amplitude $\Gamma(t)$ of the transverse field
from a very large value to zero, we hopefully drive the system
into the optimal state, the ground state of $\mathcal{H}_0$.

The natural dynamics of the present system is provided by
the Schr\"odinger equation
\begin{equation}
i \frac{\partial \left|\psi(t)\right\rangle}{\partial t} =
 \mathcal{H}(t) \left|\psi(t)\right\rangle.
   \label{Schroedinger}
\end{equation}
We solve this time-dependent Schr\"odinger equation numerically
for small-size systems.
The representation to diagonalize $\mathcal{H}_0$ (the $z$-representation)
will be used hereafter.

The corresponding classical SA process is described by
the master equation
\begin{equation}
\frac{d P_i(t)}{d t} = \sum_j \mathcal{L}_{ij} P_j(t),
 \label{master}
\end{equation}
where $P_i(t)$ represents the probability that the system is
in the $i$th state.
We consider single-spin flip processes with 
the transition matrix elements given as
\begin{equation}
\mathcal{L}_{ij} = \begin{cases}
 \dfrac{\exp(-E_i/T(t))}{\exp(-E_i/T(t))+\exp(-E_j/T(t))} &
   (\text{single-spin difference}) \\
 - \sum_{k\neq i} \mathcal{L}_{ki} & (i=j)\\
 0 & (\text{otherwise})
\end{cases} .
 \label{rate}
\end{equation}

In SA, the temperature
$T(t)$ is first set to a very large value and then is gradually
decreased to zero.
The corresponding process in QA should be to change $\Gamma (t)$
from a very large value to zero.
The reason is that the high-temperature state in SA
is a mixture of all possible states with almost equal probabilities,
and the corresponding state in QA is the linear combination of
all states with equal amplitude in the $z$-representation,
which is the lowest eigenstate of the Hamiltonian (\ref{Hamiltonian})
for very large $\Gamma$.
The low-temperature state after a successful SA is the ground
state of $\mathcal{H}_0$, which should also be the eigenstate
of $\mathcal{H}(t)$ as $\Gamma (t)$ is reduced to zero sufficiently
slowly in QA.
Another justification of identification of $\Gamma$ and $T$ comes
from the fact that the $T=0$ phase diagram of the Hopfield model
in a transverse field has almost the same structure as the
equilibrium phase diagram of the conventional Hopfield model
at finite temperature if we identify the temperature axis of the
latter phase diagram with the $\Gamma$ axis in the former~\cite{NN}.
We therefore change $\Gamma (t)$ in QA and $T(t)$ in SA from
infinity to zero with the same functional forms
$\Gamma (t)=T(t)=c/t, c/\sqrt{t}, c/\log (t+1)$
($t: 0 \to \infty$) or $-ct$ ($t: -\infty \to 0$).
The reason for choosing these functional forms are that either
they allow for analytical solutions in the single-spin case
as shown in Sec.~\ref{sec:2.3} or for comparison with the Geman-Geman
bound mentioned in the previous chapter.

To compare the performance of the two methods QA and  SA,
we calculate the probabilities
$P_{\text{QA}}(t) = 
\left| \langle g |\psi(t) \rangle \right|^2$
for QA and $P_{\text{SA}}(t) = P_g(t)$ for SA, 
where $P_g(t)$ is the probability to find the system
in the ground state at time $t$ in SA
and $| g \rangle$ is the ground-state
wave function of $\mathcal{H}_0$.
Note that we treat only small-size systems (the number of
spins $N=8$) and thus the ground state can be picked
out explicitly.
In the ideal situation $P_{\text{QA}}(t)$ and
$P_{\text{SA}}(t)$ will be very small initially
and increase towards 1 as $t\to \infty$.

It is useful to introduce another set of quantities
$P_{\text{SA}}^{\text{st}}(T)$
and $P_{\text{QA}}^{\text{st}}(\Gamma)$.
The former is the Boltzmann factor
of the ground state of $\mathcal{H}_0$
at temperature $T$ while the latter is defined
as $\left| \langle g |\psi_\Gamma \rangle \right|^2$,
where the wave function $\psi_\Gamma$ is the lowest-energy
stationary state
of the full Hamiltonian (\ref{Hamiltonian})
for a given fixed value of $\Gamma$.
In the quasi-static limit, the system follows equilibrium
in SA and thus
$P_{\text{SA}}(t)
=P_{\text{SA}}^{\text{st}}(T(t))$.
Correspondingly for QA, 
$P_{\text{QA}}(t)=
P_{\text{QA}}^{\text{st}}(\Gamma (t))$
when $\Gamma (t)$ changes sufficiently slowly.
Thus the differences between both sides of these two
equations give measures how closely the system follows
quasi-static states during dynamical process of annealing.

%------------------------------------------------------------------------------
\section{Numerical results}
\label{sec:2.2}

We now present numerical results on
$P_{\text{SA}}$ and $P_{\text{QA}}$
for various types of exchange interactions and transverse fields.
All calculations were performed with a constant longitudinal field $h=0.1$
to remove trivial degeneracy.

%- - - - - - - - - - - - - - - - - - - - - - - - - - - - - - - - - - - - - - -
\subsection{Ferromagnetic Model}
\label{sec:2.2.1}

\begin{figure}[htp]
\scalebox{1}{\includegraphics{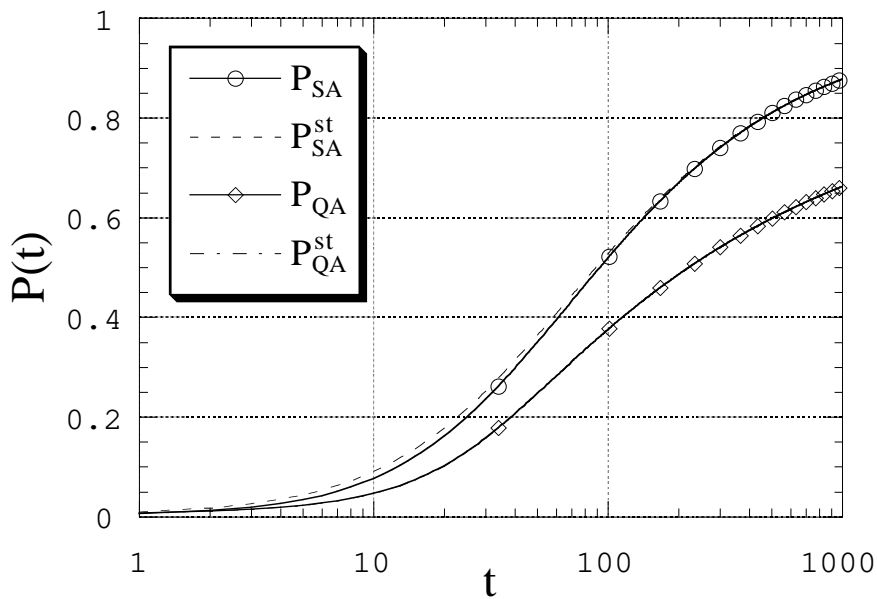}}
\caption{Time dependence of the overlaps $P_{\text{SA}}(t)$,
$P_{\text{QA}}(t)$,
$P_{\text{SA}}^{\text{st}}(T(t))$
and $P_{\text{QA}}^{\text{st}}(\Gamma(t))$
of the ferromagnetic model with $\Gamma (t)=T(t)=3/\log(t+1)$.}
\label{fig:1}
\end{figure}

Let us first discuss the ferromagnetic Ising model
with $J=\text{const}$ for all pairs of spins.
Figure~\ref{fig:1} shows the overlaps for the case
of $\Gamma (t)=T(t)=3/\log(t+1)$.
It is seen that both  QA and SA follow stationary
(equilibrium) states during dynamical processes
rather accurately.
In SA the theorem of Geman and Geman~\cite{Geman} guarantees that
the annealing schedule $T(t)=c/\log (1+t)$ assures convergence
to the ground state ($P_{\text{SA}} \to 1$
in our notation) if $c$ is adjusted appropriately.
Our choice $c=3$ is somewhat arbitrary but the tendency
is clear for $P_{\text{SA}} \to 1$ as $t\to\infty$,
which is also clear from approximate satisfaction
of the quasi-equilibrium condition
$P_{\text{SA}}(t)
=P_{\text{SA}}^{\text{st}}(T(t))$.
Although there are no mathematically rigorous arguments for
QA corresponding to the Geman-Geman bound,
the numerical data indicate convergence to the ground state
under the annealing schedule $\Gamma (t)=3/\log(t+1)$
at least for the ferromagnetic system.
It should be remembered that the unit of time is arbitrary
since we have set $\hbar =1$ in the Schr\"odinger equation
(\ref{Schroedinger}) and the unit of time $\tau =1$
in the master equation (\ref{master}).
Thus the fact that the curves for QA in Fig.~\ref{fig:1}
lie below those for SA at any given time does not
have any positive significance.

\begin{figure}[htp]
\scalebox{1}{\includegraphics{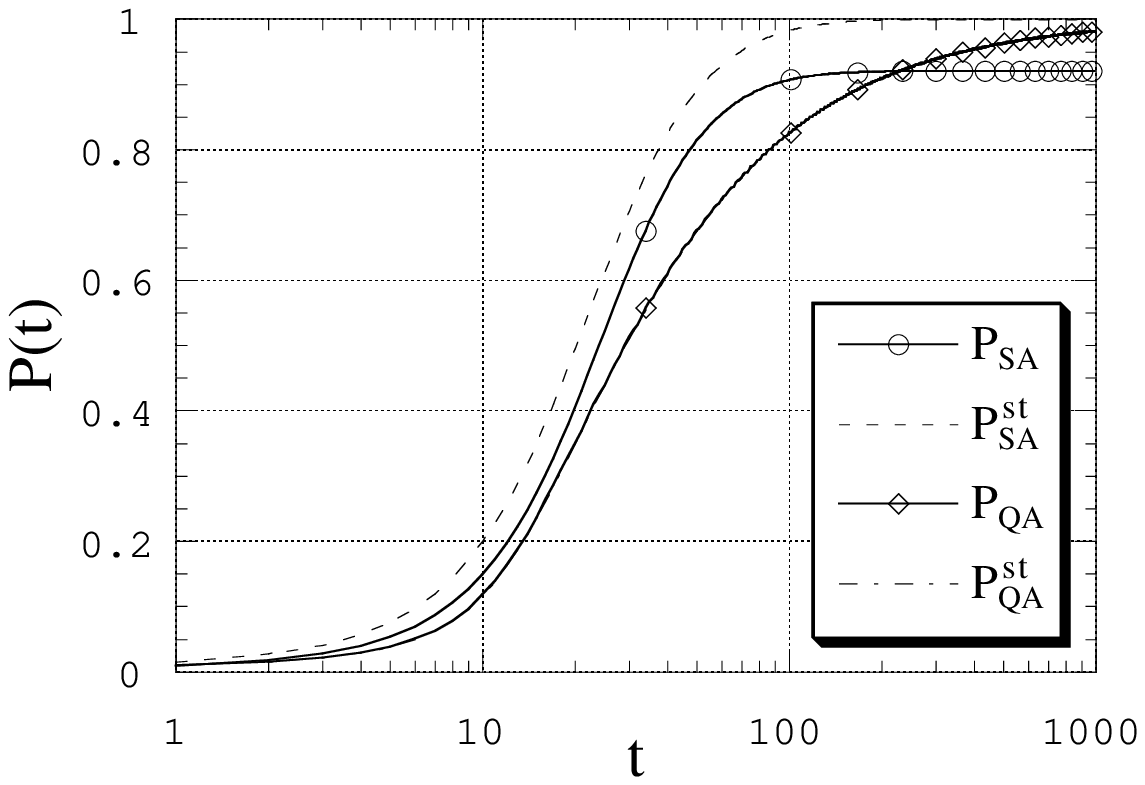}}
\caption{Time dependence of the overlaps
of the ferromagnetic model with $\Gamma (t)=T(t)=3/\sqrt{t}$.}
\label{fig:2}
\end{figure}

\begin{figure}[htp]
\scalebox{1}{\includegraphics{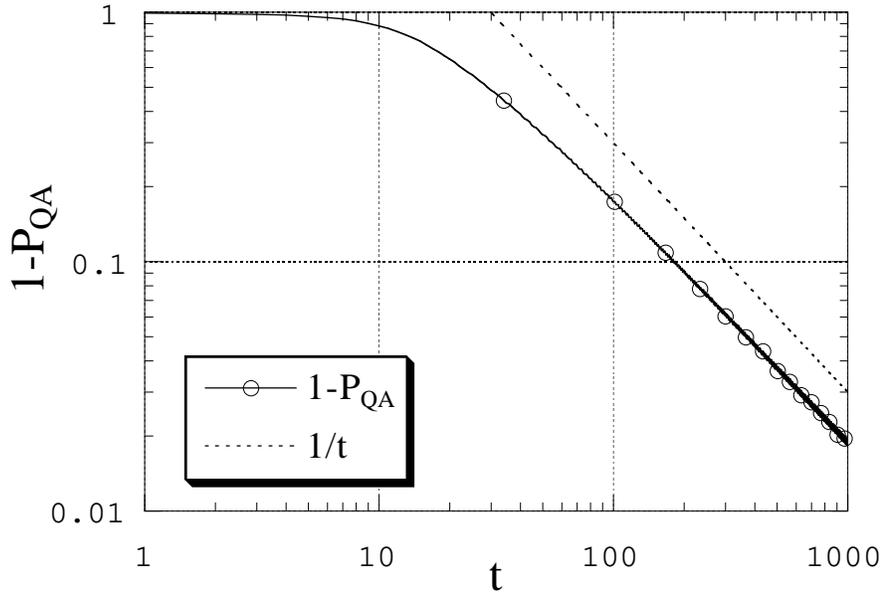}}
\caption{Time dependence of $1-P_{\text{QA}}(t)$
of the ferromagnetic model with $\Gamma (t)=3/\sqrt{t}$.
The dotted line represents $t^{-1}$ to guide the eye.}
\label{fig:3}
\end{figure}

If we decrease the transverse field and the temperature faster,
$\Gamma (t)=T(t)=3/\sqrt {t}$,
there appears a qualitative difference between QA and SA as shown
in Fig.~\ref{fig:2}.
The quantum method clearly gives better convergence to the ground state
while the classical counterpart gets stuck in a local minimum
with a non-negligible probability.
To see the rate of approach of $P_{\text{QA}}$
to 1, we have plotted $1-P_{\text{QA}}$ in a
log-log scale in Fig.~\ref{fig:3}.
It is seen that $1-P_{\text{QA}}$ behaves
as $\text{const}/t$ in the time region between 100 and 1000.

\begin{figure}[htp]
\scalebox{1}{\includegraphics{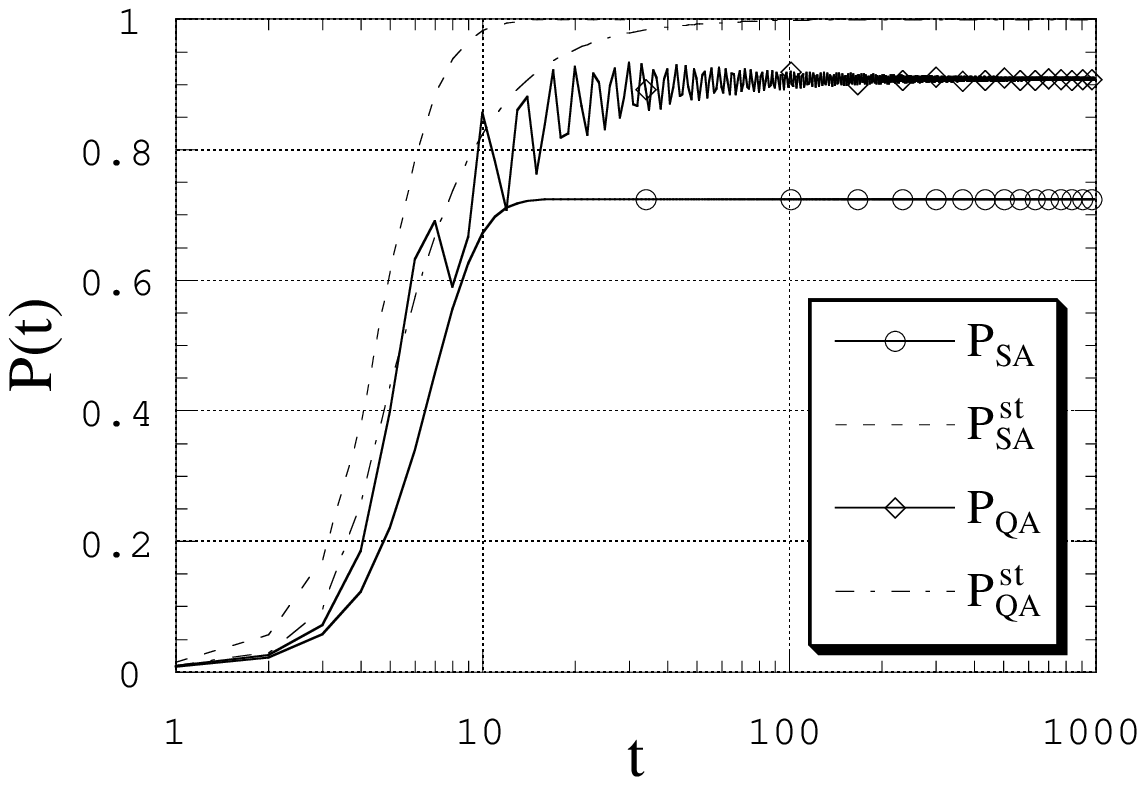}}
\caption{Time dependence of the overlaps
of the ferromagnetic model with $\Gamma (t)=T(t)=3/t$.}
\label{fig:4}
\end{figure}

By a still faster annealing schedule $\Gamma (t)=T(t)=3/t$,
the system becomes trapped in intermediate states
both in QA and SA as seen in Fig.~\ref{fig:4}.

%- - - - - - - - - - - - - - - - - - - - - - - - - - - - - - - - - - - - - - -
\subsection{Frustrated Model}
\label{sec:2.2.2}

\begin{figure}[htp]
\begin{center}
\scalebox{0.8}{\includegraphics{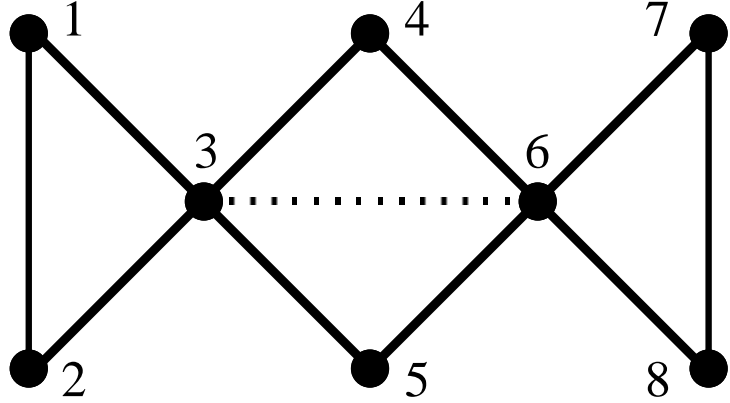}}
\end{center}
\caption{The frustrated model where the
solid lines denote ferromagnetic interactions and the broken line
is for an antiferromagnetic interaction.}
\label{fig:5}
\end{figure}

\begin{figure}[htp]
\scalebox{1}{\includegraphics{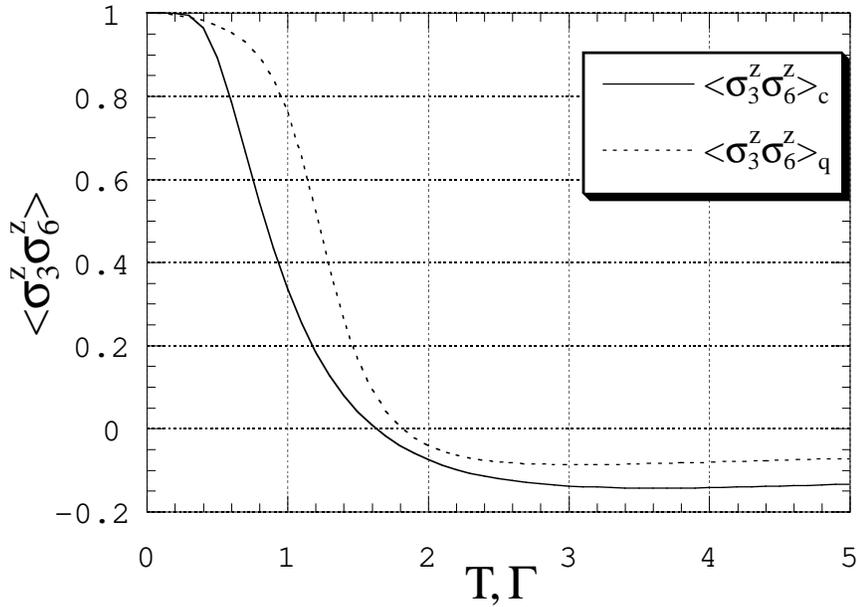}}
\caption{Correlation functions of spins 3 and 6 in Fig.~\ref{fig:5}
for the classical and quantum cases.
In the classical model (full line) the correlation is shown as a function
of temperature while the quantum case (dotted line) is regarded as a function
of the transverse field.}
\label{fig:6}
\end{figure}

We next analyze the interesting case of a frustrated system
shown in Fig.~\ref{fig:5}.
The full lines indicate ferromagnetic interactions while
the broken line is for an antiferromagnetic interaction
with the same absolute value as the ferromagnetic ones.
If the temperature is very high in the classical case,
the spins 4 and 5 are changing their states very rapidly
and hence the effective interaction between spins 3 and 6
via spins 4 and 5 will be negligibly small.
Thus the direct antiferromagnetic interactions between spins
3 and 6 is expected to dominate the correlation of these
spins, which is clearly observed in Fig.~\ref{fig:6}
as the negative value of the thermodynamic correlation function
$\langle \sigma_3^z \sigma_6^z\rangle_c$
in the high-temperature side.
At low temperatures, on the other hand, the spins 4 and 5 tend
to be fixed in some definite direction and consequently
the effective ferromagnetic interactions between spins 3 and 6
are roughly twice as large as the direct antiferromagnetic
interaction.
This argument is justified by the positive value of the correlation
function at low temperatures in Fig.~\ref{fig:6}.
Therefore the spins 3 and 6 must change their relative orientation
at some intermediate temperature.
This means that the free-energy landscape goes under
significant restructuring as the temperature is decreased
and therefore the annealing process should be performed
with sufficient care.

If the transverse field in QA plays a similar role to the
temperature in SA, we expect similar dependence of the
correlation function $\langle \sigma_3^z \sigma_6^z\rangle_q$
on the transverse field $\Gamma$.
Here the expectation value is evaluated by the stationary
eigenfunction of the full Hamiltonian (\ref{Hamiltonian})
with the lowest eigenvalue at a given $\Gamma$.
The broken curve in Fig.~\ref{fig:6} clearly supports
this idea.
We therefore expect that the frustrated system of Fig.~\ref{fig:5}
is a good test ground for comparison of QA and SA in the situation
with a significant change of spin configurations in the dynamical
process of annealing.

\begin{figure}[htp]
\scalebox{1}{\includegraphics{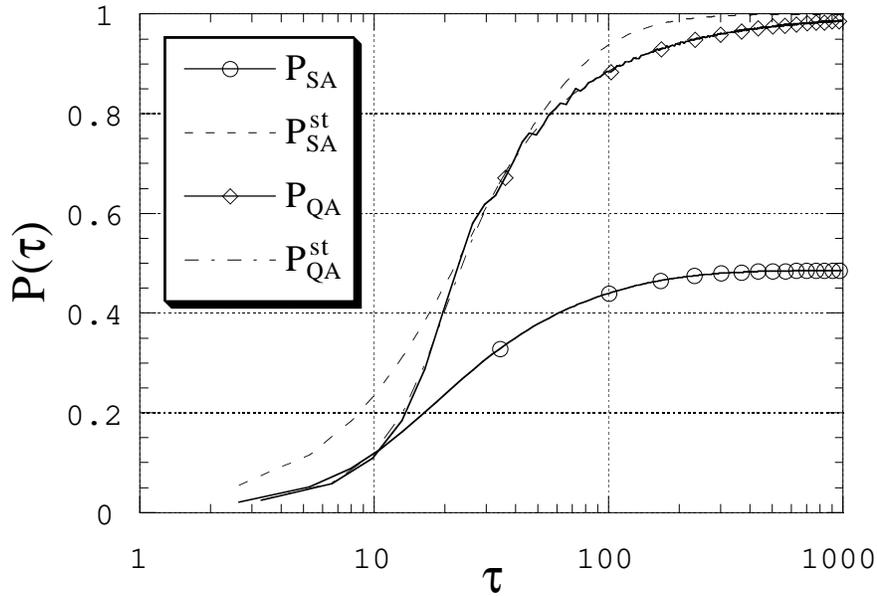}}
\caption{Time dependence of the overlaps
of the frustrated model under
$\Gamma (t)=T(t)=3/\sqrt{t}$.
Here the time scale $\tau$ is normalized by
$\Gamma_{\text{c}}$ and $T_{\text{c}}$
(the points where the correlation functions vanish in Fig.~\ref{fig:6}.)}
\label{fig:7}
\end{figure}

The results are shown in Fig.~\ref{fig:7} for the annealing
schedule $\Gamma (t)=T(t)=3/\sqrt{t}$.
The time scale $\tau$ is normalized as $\tau=tT_{\text{c}}^2$ in SA
and $\tau=t\Gamma_{\text{c}}^2$ in QA.
The values, $T_{\text{c}}$ and $\Gamma_{\text{c}}$, are the points where the
correlation functions vanish in Fig.~\ref{fig:6}.
Thus both classical and quantum correlation functions
vanish at $\tau=1$.
The tendency is clear that QA is better suited for
ground-state search in the present system.

%- - - - - - - - - - - - - - - - - - - - - - - - - - - - - - - - - - - - - - -
\subsection{Random Interaction Model}
\label{sec:2.2.3}

\begin{figure}[htp]
\scalebox{1}{\includegraphics{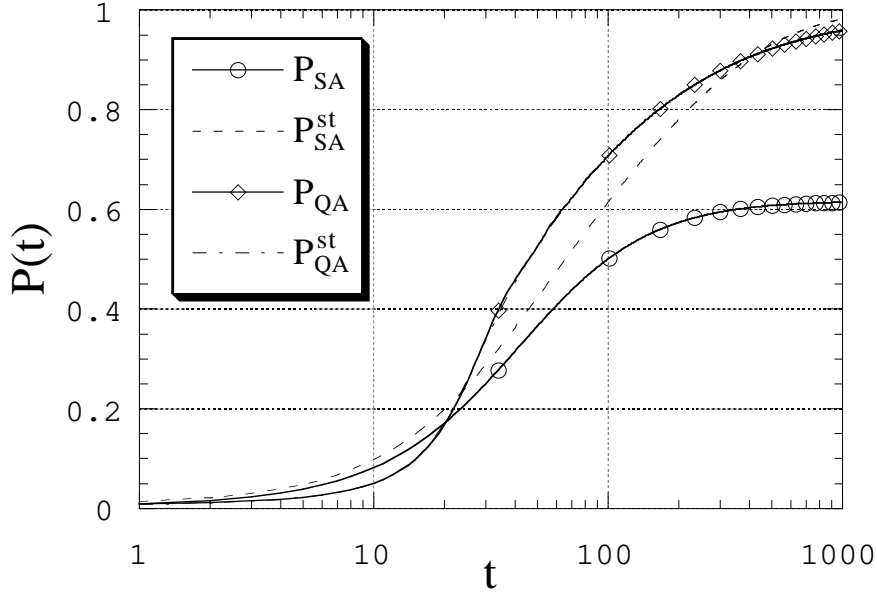}}
\caption{Time dependence of the overlaps for
the SK model with $\Gamma (t)=T(t)=3/\sqrt{t}$.}
\label{fig:8}
\end{figure}

The third and final example is the
Sherrington-Kirkpatrick (SK) model of spin glasses~\cite{SK}.
Interactions exist between all pairs of spins and are chosen
from a Gaussian distribution with vanishing mean and
variance $1/N$ ($N=8$ in our case).
Figure~\ref{fig:8} shows a typical result on the time
evolution of the probabilities under the annealing schedule
$\Gamma(t)=T(t)=3/\sqrt{t}$.
We have checked several realizations of exchange interactions
under the same distribution function and have found that
the results are qualitatively the same.
Figure~\ref{fig:8} again suggests that QA is better suited
than SA for the present optimization problem.

%------------------------------------------------------------------------------
\section{Solution of the single-spin problem}
\label{sec:2.3}

It is possible to solve the time-dependent Schr\"odinger equation
explicitly when the problem involves only a single spin and
the functional form of the transverse field is
$\Gamma (t)=-ct, c/t$ or $c/\sqrt{t}$.
We note that the single-spin problem is trivial
in SA because there are only two states involved (up and down)
and thus there are no local minima.
This does not mean that the same single-spin problem is also
trivial in the quantum mechanical version.
In QA with a single spin, the transition between the two states
is caused by a finite transverse field.
The system goes through tunneling processes
to reach the other state, and an appropriate annealing schedule is
essential to reach the ground state.
On the other hand, in SA, the transition from the higher state
to the lower state takes place even at $T=0$
and thus the system always reaches the ground state.

Let us first discuss the case of $\Gamma (t)=-ct$ with
$t$ changing from $-\infty$ to 0.
This is the well-known Landau-Zener model and the explicit
solution of the time-dependent Schr\"odinger equation
is available in the literature~\cite{Miya1,Zener,Miya2,Miya3,Raedt}.
With the notation $a(t)=\langle +|\psi (t)\rangle$ and
$b(t)=\langle -|\psi (t)\rangle$ and the initial condition
$a(-\infty)=b(-\infty)=1/\sqrt{2}$ (the lowest eigenstate),
the solution for $b(t)$ is found to be (see Appendix)
\begin{equation}
b(t)  =  \frac{h e^{-\pi h^2/8c}}{2\sqrt{c}}
           \biggl\{-\frac{2ct+h}{h}D_{-\lambda-1}(-iz)
          -\frac{ih^2+2c}{\sqrt{2c}h}e^{3/4 \pi i}
            D_{-\lambda-2}(-iz)\biggr\},
\label{eq:lin}
\end{equation}
where $D_{-\lambda -1}, D_{-\lambda -2}$ represent
the parabolic cylinder function (or Weber function) and
$z$ and $\lambda$ are given as
\begin{eqnarray}
 z & = & \sqrt{2c}e^{-\pi i/4}t, \\
 \lambda & = & \frac{ih^2}{2c}.
\end{eqnarray}
The final value of $b(t)$ at $t=0$ is
\begin{eqnarray}
\nonumber
b(0)&=& -\frac{h\sqrt{\pi}2^{-ih^2/4c}e^{-\pi h^2/8c}}{2\sqrt{2c}} \\
 & & \times \biggl\{ \frac{1}{\Gamma(1+ih^2/4c)}
             +\frac{\sqrt{c}e^{3/4\pi i}(1+ih^2/2c)}
            {h\Gamma(3/2+ih^2/4c)}\biggr\}.
\end{eqnarray}
The probability to find the system in the ground state at $t=0$ is,
when $h^2/c \gg 1$,
\begin{equation}
P_{\text{QA}}(0) = |a(0)|^2 =  1-|b(0)|^2 \sim 1-\frac{c^2}{16h^4}.
\end{equation}
Thus the probability $P_{\text{QA}}(t)$ does not approach
1 for finite $c$.

We next present the solution for $\Gamma (t)=c/t$
with $t$ changing from 0 to $\infty$ under the initial
condition $a=b=1/\sqrt{2}$ (see Appendix):
\begin{equation}
b(t) = \frac{1}{\sqrt{2}} e^{iht} t^{ic} F(1+ic,1+2ic;-2iht),
\end{equation}
where $F$ is the confluent hypergeometric function.
The asymptotic form of $b(t)$ as $t\to\infty$ is
\begin{equation}
b(t) \sim \frac{\sqrt{2} (2h)^{-ic} \Gamma(2ic)}{\Gamma(ic)} 
    \times \left\{ e^{-iht-\pi c/2} + c e^{iht+\pi c/2} (2ht)^{-1} \right\}.
\end{equation}
The probability to find the system in the target ground state
behaves asymptotically as
\begin{eqnarray}
P_{\text{QA}}(t) & = & |a(t)|^2 \\
 & = & 1-|b(t)|^2 \\
\nonumber
 & \sim & 1- \frac{\sinh(\pi c)}{\sinh(2\pi c)} \\
 & &       \times \left\{e^{-\pi c} + \frac{c \cos(2ht)}{ht}
                + \frac{c^2 e^{\pi c}}{4h^2 t^2} \right\} \\
  & \sim & 1-e^{-2\pi c},
\end{eqnarray}
the last approximation being valid for $c \gg 1$ after
$t\to\infty$.
The system does not reach the ground state as $t\to\infty$ as
long as $c$ is finite.
Larger $c$ gives more accurate approach to the ground state,
which is reasonable because it takes longer time to reach
a given value of $\Gamma (=c/t)$ for larger $c$, implying
slower annealing.

The final example of the solvable model concerns the annealing
schedule $\Gamma (t)=c/\sqrt{t}$.
The solution for $b(t)$ is derived in Appendix under the
initial condition $a=b=1/\sqrt{2}$ as
\begin{eqnarray}
\nonumber
b(t) & = & \frac{1}{\sqrt{2}} e^{iht}
   F\left(\frac{1}{2}-i\gamma,\frac{1}{2};-2iht \right) \\
& &     + \frac{c}{\sqrt{h}} e^{(3/4)\pi i} e^{iht} (-2iht)^{1/2}
                   F\left( 1-i\gamma,\frac{3}{2};-2iht \right),
\end{eqnarray}
where $\gamma = c^2/2h$.
The large-$t$ behavior is found to be
%
%\begin{equation}
%\label{asym}
%b(t) \sim \sqrt{\pi} e^{-\pi c^2/4h} \Biggl[
%        e^{-iht} (2ht)^{-i\gamma} \left\{
%        \frac{1}{\sqrt{2}\Gamma(\frac{1}{2}-i\gamma)}
%        + \frac{\sqrt{h} e^{(5/4)\pi i}}{c\Gamma(-i\gamma)} \right\}
%      + e^{iht} (2ht)^{-1/2+i\gamma} \left\{
%        \frac{e^{-(1/4)\pi i}}{\sqrt{2}\Gamma(i\gamma)}
%        + \frac{c}{2\sqrt{h}\Gamma(\frac{1}{2}+i\gamma)} \right\} \Biggr],
%\end{equation}
\begin{eqnarray}
\nonumber
b(t) & \sim & \sqrt{\pi} e^{-\pi c^2/4h} \Biggl[
        e^{-iht} (2ht)^{-i\gamma} \left\{
        \frac{1}{\sqrt{2}\Gamma(\frac{1}{2}-i\gamma)}
        + \frac{\sqrt{h} e^{(5/4)\pi i}}{c\Gamma(-i\gamma)} \right\} \\
\label{asym}
 & &  + e^{iht} (2ht)^{-1/2+i\gamma} \left\{
        \frac{e^{-(1/4)\pi i}}{\sqrt{2}\Gamma(i\gamma)}
        + \frac{c}{2\sqrt{h}\Gamma(\frac{1}{2}+i\gamma)} \right\} \Biggr],
\end{eqnarray}
\noindent
and the probability $P_{\text{QA}}(\infty)$ for $c^2/h\gg 1$
is obtained as
\begin{equation}
P_{\text{QA}}(\infty) = 1-|b(\infty)|^2 \sim 1-\frac{h^2}{64c^4}.
\end{equation}
This equation indicates that the single-spin system does not reach
the ground state under the present annealing schedule
$\Gamma (t)=c/\sqrt{t}$ for which the numerical data in the
previous section suggested an accurate approach.
We therefore conclude that the asymptotic value of
$P_{\text{QA}}(t)$ in the previous section may not
be exactly equal to 1 for $\Gamma(t)=3/\sqrt{t}$ although
it is very close to 1.

The annealing schedule $\Gamma (t)=c/\sqrt{t}$ has a feature
which distinguishes this function from the other ones
$-ct$ and $c/t$.
As we saw in the previous discussion, the final asymptotic value
of $P_{\text{QA}}(t)$ is not 1 if the initial
condition corresponds to the ground state for $\Gamma\to\infty$,
$a=b=1/\sqrt{2}$.
However, as shown in Appendix, by an appropriate choice of the initial
condition, it is possible to drive the system to the
ground state if $\Gamma (t)=c/\sqrt{t}$.
This is not possible for any initial conditions in the case
of $\Gamma (t)=-ct$ or $c/t$.

%------------------------------------------------------------------------------
\section{Summary}
\label{sec:2.4}

In this chapter we have started the study of quantum annealing (QA)
from the transverse Ising model obeying the time-dependent
Schr\"odinger equation.
The transverse field term was controlled so that the system
approaches the ground state.
The numerical results on the probability to find the system
in the ground state were compared with the corresponding
probability derived from the numerical solution of the 
master equation representing the SA processes.
We have found that QA shows convergence to the optimal (ground)
state with larger probability than SA in all cases if the
same annealing schedule is used.
The system approaches the ground state rather accurately
in QA for the annealing schedule $\Gamma =c/\sqrt{t}$ but
not for faster decrease of the transverse field.

We have also solved the single-spin model exactly for QA
in the cases of $\Gamma (t)=-ct, c/t$ and $c/\sqrt{t}$.
The results showed that the ground state is not reached
perfectly for all these annealing schedules.
Therefore the asymptotic values of $P_{\text{QA}}(t)$
in numerical calculations are probably not exactly 1
although they seem to be quite close to the optimal value 1.

The rate of approach to the asymptotic value close to 1,
$1-P_{\text{QA}}(t)$, was found to
be proportional to $1/t$ in Fig.~\ref{fig:3} for the ferromagnetic
model.
On the other hand, the single-spin solution shows the existence
of a term proportional to $1/\sqrt{t}$, see Eq. (\ref{asym}).
Probably the coefficient of the $1/\sqrt{t}$-term is very small
in the situation of Fig.~\ref{fig:3}
and the next-order contribution dominates in the time region
shown in Fig.~\ref{fig:3}.

A simple argument using perturbation theory yields useful
information about the asymptotic form of the probability
function if we assume that the system follows quasi-static
states during dynamical processes.
The probability to find the system in the ground state is expressed
using the perturbation in terms of $\Gamma (\ll 1)$ as
\begin{equation}
 P_{\text{QA}}(\Gamma)
  \sim  1-\Gamma^2\sum_{i\neq 0}\frac{1}{(E_0^{(0)}-E_i^{(0)})^2},
\end{equation}
where $E_i^{(0)}$ is the energy of the $i$th state of
the non-perturbed (classical) system
and $E_0^{(0)}$ is the ground-state energy.
If we set $\Gamma=c/\sqrt{t}$, we have
\begin{equation}
 P_{\text{QA}}(\Gamma) \sim
   1-\frac{1}{t}\sum_{i\neq 0}\left(\frac{c}{E_0^{(0)}-E_i^{(0)}}\right)^2.
\end{equation}
Thus the approach to the asymptotic value is proportional to $1/t$
as long as the system stays in quasi-static states.
The corresponding probability for SA is
\begin{equation}
 P_{\text{SA}}(T) \sim \frac{e^{-E_0/T}}{\sum_i e^{-E_i/T}}
  \sim 1-\sum_{i\neq 0} e^{-(E_i-E_0)/T},
\end{equation}
which shows absence of universal ($1/t$-like) dependence on time.

The present method of QA bears some similarity to the approach by
the generalized transition probability in which
the dynamics is described by the master equation but the transition
probability has power-law dependence on the temperature in contrast
to the usual exponential form of the Boltzmann factor~\cite{Tsallis}.
This power-law dependence on the temperature allows the system
to search for a wider region in the phase space
because of larger probabilities of transition to higher-energy
states at a given $T(t)$, which may be the reason of faster
convergence to the optimal states~\cite{Tsallis,Nishi}.
The transverse field term $\Gamma$ in our QA represents the rate of
transition between states which is larger than the transition rate
in SA (see (\ref{rate})) at a given small value of
the control parameter $\Gamma (t)=T(t)$.
This larger transition probability may lead to a more active search
in wider regions of the phase space, leading to better
convergence similarly to the case of the generalized transition
probability.

We have solved the Schr\"odinger equation and the master equation directly
by numerical methods for the purpose of comparison of QA and SA.
This method faces difficulties for larger $N$ because the number
of states increases exponentially as $2^N$.
The classical SA solves this problem by exploiting stochastic processes,
Monte Carlo simulations, which have the computational complexity
growing as a power of $N$.
The corresponding reduction of the computational complexity is lacking
in QA, because the Schr\"odinger equation is not replaced directly
by the stochastic processes.
While the quantum Monte Carlo is not equivalent to the Schr\"odinger
equation, we will conduct QA in the quantum Monte Carlo framework
in the next chapter.
Another problem is to devise implementations of QA in other
optimization problems such as the traveling salesman problem or
the Hamilton problem.
The implementation of QA to the traveling salesman problem is explained
as an example in Chapter~\ref{chap:4}.

\chapter[Monte Carlo Analysis]{Monte Carlo Analysis of Larger Systems}
\label{chap:3}

We will apply the Monte Carlo method to QA in this chapter.
Two possible ways to solve QA are considered.
Comparison between SA and QA by the Monte Carlo method is also performed.

We desire to solve SA and QA by another method, because the calculations
of the master equation for SA and the Schr\"odinger equation for QA
become difficult for large-size systems.
As shown in the previous chapter, when we search the ground state by SA
and QA, the operation of $2^N$-dimensional matrices is needed to solve
the differential equations in the Ising-spin systems.
For example, real and complex $1024 \times 1024$ matrixes are needed for SA
and QA with ten spins, respectively.
To avoid such a difficulty of solving the master equation, the
calculation is performed by the Monte Carlo method in the original
definition of SA.
Solving SA by the master equation is the equivalent to the Monte Carlo
method.
From the relation of the master equation and the Monte Carlo method in
SA, we consider QA by the Monte Carlo method.

Another motivation of considering the Monte Carlo method is the
following:
In the actual procedure of QA, all the elements of the Hamiltonian matrix
are calculated to solve the the Schr\"odinger equation.
If the matrix is expressed in the $z$-representation, the diagonal elements
are the classical-term energy of each configuration.
This implies that there is no advantage to the enumeration method
(to enumerate all the possible configurations), because we already
calculated all the energy of possible configurations.
Therefore, we have to solve QA by another method which requires less
calculations than to solve the differential equation, like the Monte
Carlo method.

We consider the two methods, the path-integral Monte Carlo and the
quantum Monte Carlo methods, for QA.
The results of the two methods are not the same as the result of the
Schr\"odinger equation, but the ground state can be found by both of the
methods.
We adopt the quantum Monte Carlo method for QA and compare the results
of SA and QA by the Monte Carlo simulation.

\section{Monte Carlo Method}
\label{sec:3.1}

The Monte Carlo method is powerful to analyze various problems of
statistical mechanics.
In classical statistical mechanics, we obtain the expectation value
of a quantity $A$ as a function of the temperature $T=1/\beta$ in the
following form:
\begin{equation}
\langle A \rangle = \frac{\sum A e^{-\beta E}}{\sum e^{-\beta E}},
\label{expectation}
\end{equation}
where the summation runs over all the possible configurations of the
system and $E$ corresponds to the energy of each configuration.
The number of all possible configurations is $S^N$, if the system
includes $N$ sites and each site takes $S$ states independently.
The number of the terms in the summation increases exponentially as a
function of the system size $N$.
It is difficult to calculate the whole summation for large-size systems.

We consider replacing the summation with the weighted sampling.
Equation (\ref{expectation}) is rewritten as
\begin{equation}
\langle A \rangle = \sum_i P_i A_i,
\end{equation}
where $P_{i} = e^{-\beta E_i} / \sum_j e^{-\beta E_j}$.
$P_i$ is the Gibbs distribution, the probability of the configuration $i$.
The sampling configurations are generated to obey the Gibbs distribution.

The expression of the expectation values by the weighted sampling is given
by
\begin{equation}
\langle A \rangle_{\text{sampling}} = \frac{1}{M} \sum_{i=1}^{M} A(i),
\end{equation}
where the summation runs over the sampled configurations.
In the large $M$ limit, $\langle A \rangle_{\text{sampling}}$ converges
to $\langle A \rangle$.

The configuration obeying the Gibbs distribution can be obtained from the
Markov chain which comes from the master equation as
\begin{equation}
\frac{d\boldsymbol{P}(t)}{dt}=\boldsymbol{\mathcal{L}P}(t),
\end{equation}
where $\boldsymbol{\mathcal{L}}$ is the transition matrix defined
in the previous chapter.
The matrix element which is the transition probability from $i$th state
to $j$th state has a non-zero value, when the configurations of $i$th
and $j$th states differ only by one spin. 
Therefore, the $i$th state can be modified to $N$ states.
We create a new configuration from the $i$th state by the following steps:
Choose a site randomly and determine the direction of the spin on the
site by the transition ratio from the configuration before flip to the
configuration after flip.
Next, repeat this procedure $N$ times.
These $N$ trials of the spin flip are called ``one Monte Carlo step''.
The Markov chain is obtained from the calculation of this Monte Carlo steps.
In principle, every configuration can change to any configuration in one
Monte Carlo step, because all spins can flip once.
However, the distance between two configurations is close, if one
configuration is obtained from the other configuration by one Monte
Carlo step.
For the precise calculation, the interval of the sampling has to be long
enough.

\section{Quantum Monte Carlo Method}
\label{sec:3.2}

In quantum systems, the definition of the expectation value is
not the same as in the classical systems.
The expectation value of a quantity $A$ is given by
\begin{equation}
\langle A \rangle = \frac{\text{Tr} A e^{-\beta \mathcal{H}}}
                         {\text{Tr} e^{-\beta \mathcal{H}}},
\end{equation}
where $\mathcal{H}$ is the Hamiltonian.
To calculate the operator $e^{-\beta \mathcal{H}}$ is difficult
for large-size systems because of non-diagonal terms
of the Hamiltonian $\mathcal{H}$.
The same procedure as in the classical Monte Carlo method can not be
applied to quantum systems, if we can not diagonalize the Hamiltonian.

The following procedure avoids the above difficulty.
Now we consider the the Ising system with transverse field $\Gamma$
as a quantum system,
\begin{equation}
\mathcal{H} =  - \sum_{ij} J_{ij} \sigma^z_i \sigma^z_j
            - \Gamma \sum_i \sigma^x_i .
\end{equation}
This Hamiltonian is not diagonal in the $z$-representation.
By the Trotter formula,
the operator $e^{-\beta \mathcal{H}}$ can be described
as a product of many operators diagonalized in $z$- or $x$-representation:
\begin{equation}
e^{-\beta \mathcal{H}} = \lim_{M \rightarrow \infty}
( e^{A/M} e^{B/M} )^M ,
\end{equation}
where $A=\sum_{ij} K_{ij} \sigma^z_i \sigma^z_j$
($K_{ij} = \beta J_{ij}$) and $B=\gamma \sum_i \sigma^x_i$
($\gamma = \beta \Gamma$). 
This decomposition is called the Suzuki-Trotter
decomposition~\cite{Trotter,Suzuki-Trotter}.
Using this decomposition,
we obtain the partition function of the $M$th decomposition $Z_M$ as
\begin{eqnarray}
Z_M & = & \text{Tr} ( e^{A/M} e^{B/M} )^M \\
& = & \sum_{\{\sigma_{jk} = \pm 1\}}
       \langle \{\sigma_{j1}\}|e^{A/M}|\{\sigma_{j1}^{\prime}\} \rangle
       \langle \{\sigma_{j1}^{\prime}\}|e^{B/M}|\{\sigma_{j2}\} \rangle
\nonumber \\
& & \hspace{3em} \times 
       \langle \{\sigma_{j2}\}|e^{A/M}|\{\sigma_{j2}^{\prime}\} \rangle
       \langle \{\sigma_{j2}^{\prime}\}|e^{B/M}|\{\sigma_{j3}\} \rangle
\nonumber \\
& & \hspace{3em} \times \cdots\cdots \nonumber \\
& & \hspace{3em} \times 
       \langle \{\sigma_{jM}\}|e^{A/M}|\{\sigma_{jM}^{\prime}\} \rangle
       \langle \{\sigma_{jM}^{\prime}\}|e^{B/M}|\{\sigma_{j1}\} \rangle,
\end{eqnarray}
where $|\{\sigma_{jk}\} \rangle$ is the $M$th direct product of eigenstates
$\{\sigma_{j} \}$ defined as
\begin{equation}
|\{\sigma_{jk}\} \rangle = |\sigma_{j1} \rangle \otimes
                           |\sigma_{j2} \rangle \otimes \cdots \otimes
                           |\sigma_{jM} \rangle .
\end{equation}
From this decomposition, the partition function is expressed
as a trace of products of diagonalized matrices.
Each part of the product is rewritten as follows:
\begin{equation}
\langle \{\sigma_{jk}\}|e^{A/M}|\{\sigma_{jk}^{\prime}\} \rangle
= \exp \left(\frac{1}{M} \sum_{ij} K_{ij} \sigma_{ik} \sigma_{jk} \right)
  \prod_{j}^{N}\delta(\sigma_{jk}, \sigma_{jk}^{\prime}).
\end{equation}

\begin{equation}
\langle \{\sigma_{jk}\}|e^{B/M}|\{\sigma_{j,k+1}\} \rangle
= a_M^N \exp \left(\gamma_M \sum_{j=1}^{N} \sigma_{jk} \sigma_{j,k+1} \right),
\end{equation}
where
\[
a_M = \left \{ \frac{1}{2} \sinh\left( \frac{2 \gamma}{M} \right)
      \right\}^{1/2}, \ \ \
\gamma_M = \frac{1}{2} \ln \coth \left(\frac{\gamma}{M} \right).
\]

\begin{figure}[t]
\scalebox{1}{\includegraphics{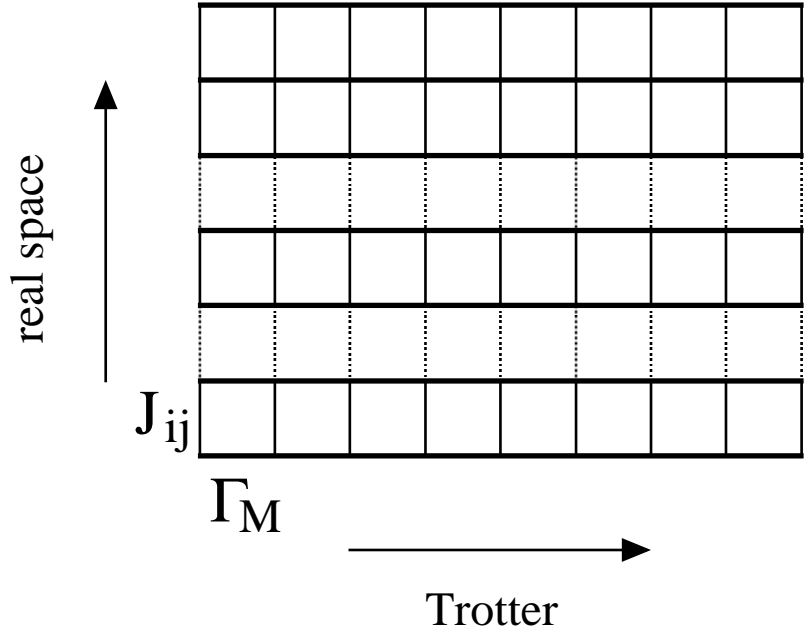}}
\caption{
The $d$-dimensional quantum system is mapped to a $(d+1)$-dimensional
classical system at the inverse temperature $\beta_{\text{eff}}$.
The original $d$-dimensional space is expressed as the ``real space'' and
the additional space is expressed as the ``Trotter''.
Copies of the original system without the quantum term are placed in
the Trotter direction.
The real space interactions $J_{ij}$ are random variables in the real space
direction but uniform in the Trotter direction.
The interactions between the Trotter slices, $\Gamma_M$, are uniform and
depend on the amplitude of the transverse field $\Gamma$.
}
\label{fig_qmc}
\end{figure}

The partition function $Z$ is represented as
\begin{eqnarray}
Z & = & \lim_{M \rightarrow \infty} a_M^{NM} \sum_{\{\sigma_{jk}=\pm 1\}}
\nonumber \\
& & \hspace{3em} \exp\left \{ \sum_{k=1}^M \left(
   \frac{1}{M} \sum_{ij} K_{ij} \sigma_{ik}\sigma_{jk}
 + \gamma_M \sum_{j=1}^M \sigma_{jk} \sigma_{j,k+1}
\right) \right \} \nonumber \\
& = & \lim_{M \rightarrow \infty} a_M^{NM} \sum_{\{\sigma_{jk}=\pm 1\}}
\nonumber \\
& & \hspace{3em} \exp\left \{ \beta_{\text{eff}} \sum_{k=1}^M \left(
   \sum_{ij} J_{ij} \sigma_{ik}\sigma_{jk}
 + \Gamma_M \sum_{j=1}^N \sigma_{jk} \sigma_{j,k+1}
\right) \right \},
\end{eqnarray}
where $\beta_{\text{eff}} = \beta/M$ and
$\Gamma_M = \gamma_M/\beta_{\text{eff}}$.
This is the  partition function of a $(d+1)$-dimensional classical spin
system at the effective inverse temperature $\beta_{\text{eff}}$.
The quantum $d$-dimensional partition function is mapped to a
$(d+1)$-dimensional classical partition function.
The representation of the classical $(d+1)$-dimension system thus can be
obtained by the usual Monte Carlo method (see Fig.~\ref{fig_qmc}).

The framework does not change, when the longitudinal field is applied.
It produces the term $-h \sum_i \sigma_i^z$ in the Hamiltonian.
We can deal with this term similarly to the term
$- \sum_{ij} J_{ij} \sigma_i^z \sigma_j^z$.

\section{Path-Integral Monte Carlo}
\label{sec:3.3}

\begin{figure}[p]
\scalebox{1}{\includegraphics{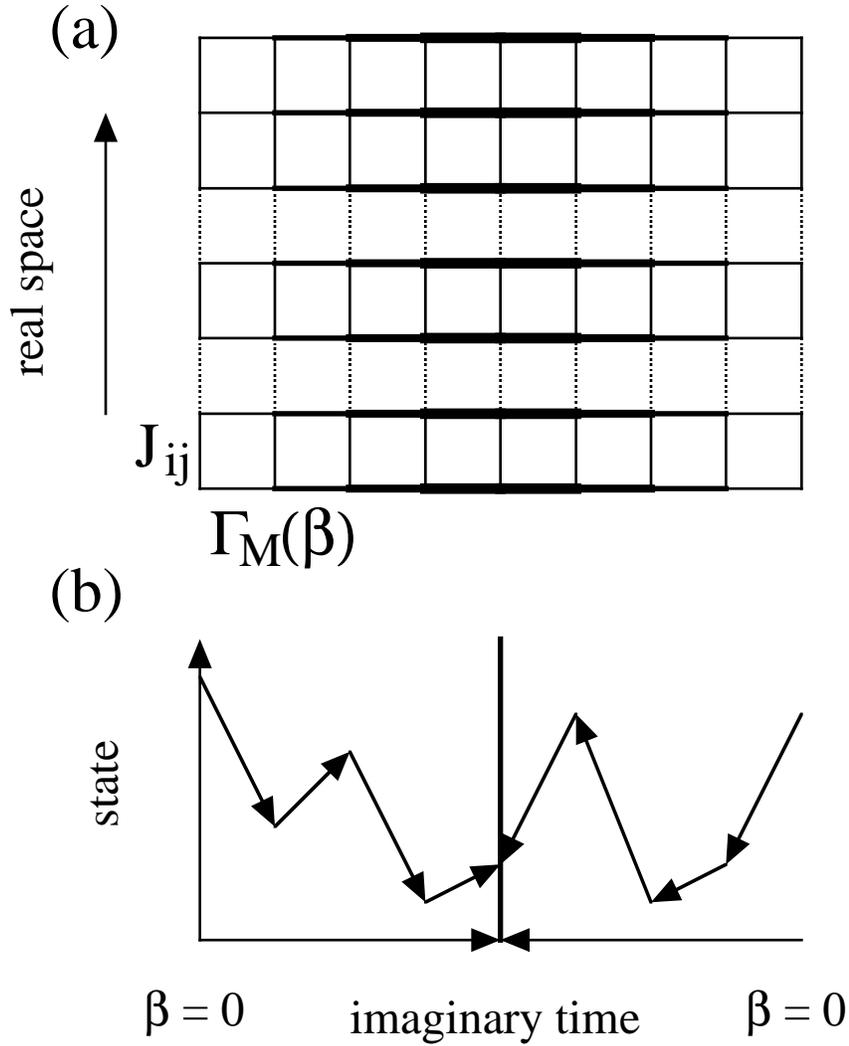}}
\caption{
(a) A path of the imaginary time evolution is expressed as
a configuration on a $(d+1)$-dimensional space.
As the transverse field becomes weak, the interaction between
the Trotter slices, $\Gamma_M(\beta)$, becomes strong.
This fact is illustrated as the thickness of bonds.
(b) The initial states $|\Phi(0)\rangle$ are located at each end
and  evolute toward the center.
}
\label{fig_pimc}
\end{figure}

The time development of the wave function in quantum systems is given as
\begin{equation}
|\Psi(T) \rangle = \mathrm{T} e^{-i\int_0^T dt \mathcal{H}(t)}
                              |\Psi(0) \rangle ,
\end{equation}
where the symbol $\mathrm{T}$ means that the operator acts on the ket
vectors with time ordering.
Decomposing the product of the short time $(\delta t=T/M)$ operators,
we have 
\begin{equation}
|\Psi(T) \rangle
 =  \lim_{M \rightarrow \infty}
    e^{-i \frac{T \mathcal{H}_{M-1}}{M}}
    e^{-i \frac{T \mathcal{H}_{M-2}}{M}}
      \cdots
    e^{-i \frac{T \mathcal{H}_1}{M}}
    e^{-i \frac{T \mathcal{H}_0}{M}}
      |\Psi(0) \rangle ,
\end{equation}
where $\mathcal{H}_j=\mathcal{H}(Tj/M)$.
By replacing the time with the imaginary time ($i T \rightarrow \beta$),
we obtain
\begin{equation}
|\Psi(-i\beta) \rangle =  \lim_{M \rightarrow \infty}
       e^{-\frac{\beta \mathcal{H}_{M-1}}{M}} \cdots
       e^{-\frac{\beta \mathcal{H}_0}{M}} |\Psi(0) \rangle.
\end{equation}

The expectation value of $A$ is obtained as,
\begin{eqnarray}
\langle A(T) \rangle & = &\langle A(-i\beta) \rangle \\
& = & \langle \Psi(-i\beta) | A | \Psi(-i\beta) \rangle \\
& = & \lim_{M \rightarrow \infty} 
      \langle \Psi(0) | e^{-\beta/M \ \mathcal{H}_0} \cdots
                        e^{-\beta/M \ \mathcal{H}_{M-1}} A \nonumber \\
& &   \times            e^{-\beta/M \ \mathcal{H}_{M-1}} \cdots
                        e^{-\beta/M \ \mathcal{H}_0} | \Psi(0) \rangle.
\label{eq_pimc}
\end{eqnarray}

If the imaginary time $\beta$ is a real number, this formulation is
similar to the quantum Monte Carlo method.
Dividing the Hamiltonian into two parts, the terms including $\sigma^z$
or $\sigma^x$, and inserting the complete sets between each product,
we can map the equation (\ref{eq_pimc}) to the weighted sampling of the
Monte Carlo simulation.
One configuration in the Monte Carlo simulation corresponds to one
possible path of the imaginary time dynamics in the original system.
The summation of the samplings is the discrete version of the imaginary
time path integral.
The system simulated in the Monte Carlo is shown in Fig.~\ref{fig_pimc}.
The initial state $|\Psi(0)\rangle$ is located at the right and the left
sides.
The evolved state $|\Psi(\beta)\rangle$ locates at the center.
A path of the sampling is obtained from the snapshot of the thermal
equilibrium state of a $(d+1)$-dimensional space.
The horizontal interaction comes from the transverse field.
If the field is scheduled (decreases as a function of time),
the interaction changes stronger toward the center.
The same systems are located in the horizontal direction.
These are similar to the Trotter slices in the quantum Monte Carlo
method.

\section[Applying Monte Carlo to Annealing Process]
{Applying Monte Carlo Method to Annealing Process}
\label{sec:3.4}

\begin{figure}[t]
\scalebox{1}{\includegraphics{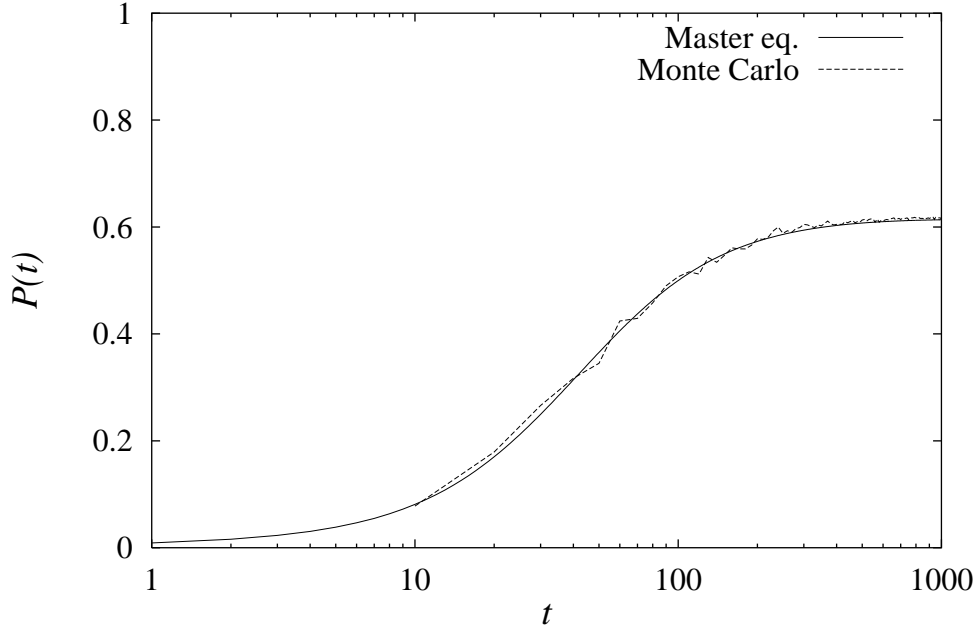}}
\caption{
The results of the master equation and the Monte Carlo simulation.
Two curves show good agreement.
}
\label{fig_sa_dm}
\end{figure}

In statistical mechanics, Monte Carlo simulation and the master equation
of the classical system give the same result except for statistical
errors.
The results in the dynamics are also the same.
(For the dynamics of the SK model, see
\cite{Coolen,Laughton,Nishi_Yama,Yamana})
From this equivalence, we regard the Monte Carlo step as the time
of the master equation.
Figure~\ref{fig_sa_dm} shows the probability to find the ground
state of the SK model with $N=8$ for the master equation and the Monte
Carlo simulation.
The probability for the Monte Carlo simulation is obtained from the
ratio of the number of the ground state configurations for each time in
one hundred independent runs.
The temperature schedule of both cases is $T=3/\sqrt{t}$.
Two curves show good agreement.

\begin{figure}[pt]
\scalebox{1}{\includegraphics{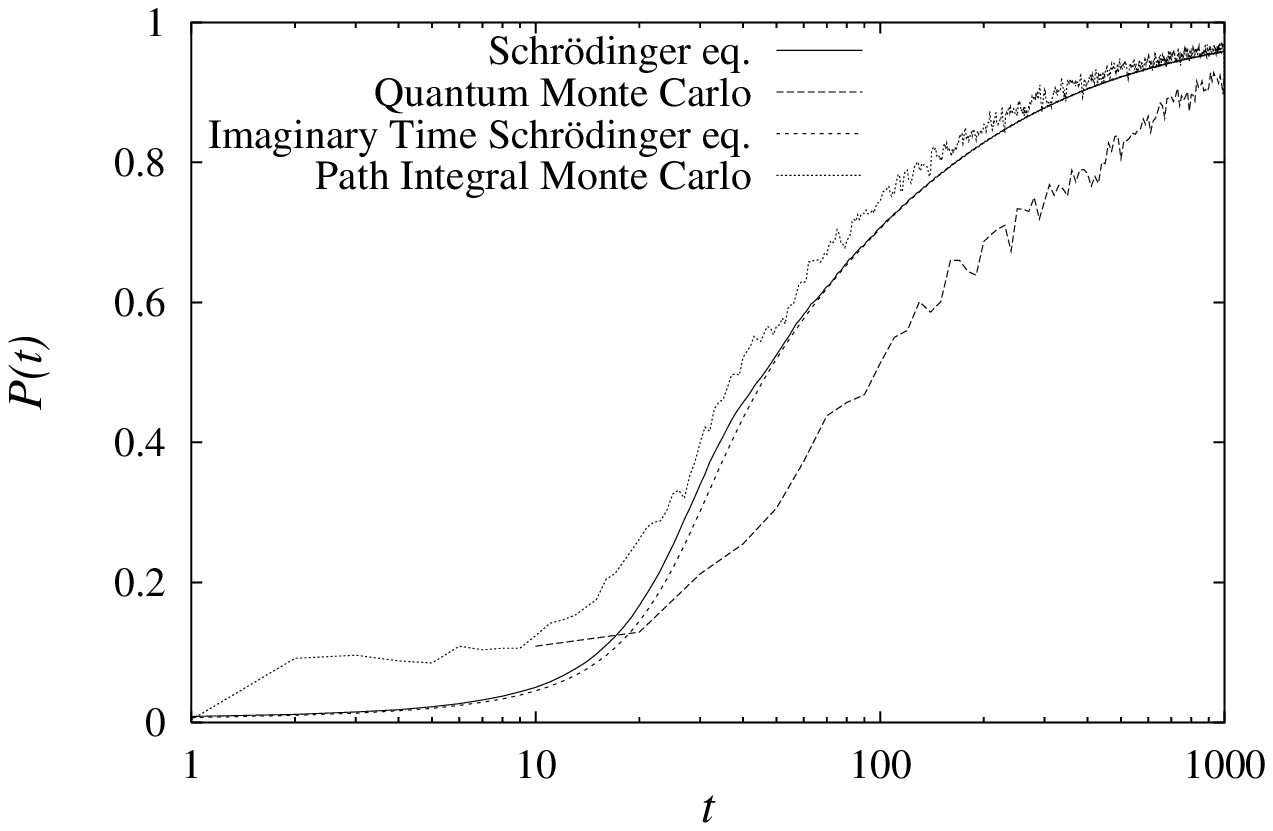}}
\caption{
The results for the Schr\"odinger equation,
the quantum Monte Carlo method, the imaginary time Schr\"odinger equation
and the path-integral Monte Carlo method.
Imaginary time Schr\"odinger equation means the results of imaginary time
dynamics of the Schr\"odinger equation.
The schedule of $\Gamma(t)$ is $3/\sqrt(t)$.
}
\label{fig_qa_dm_1}
%\end{figure}

%\begin{figure}[hb]
\scalebox{1}{\includegraphics{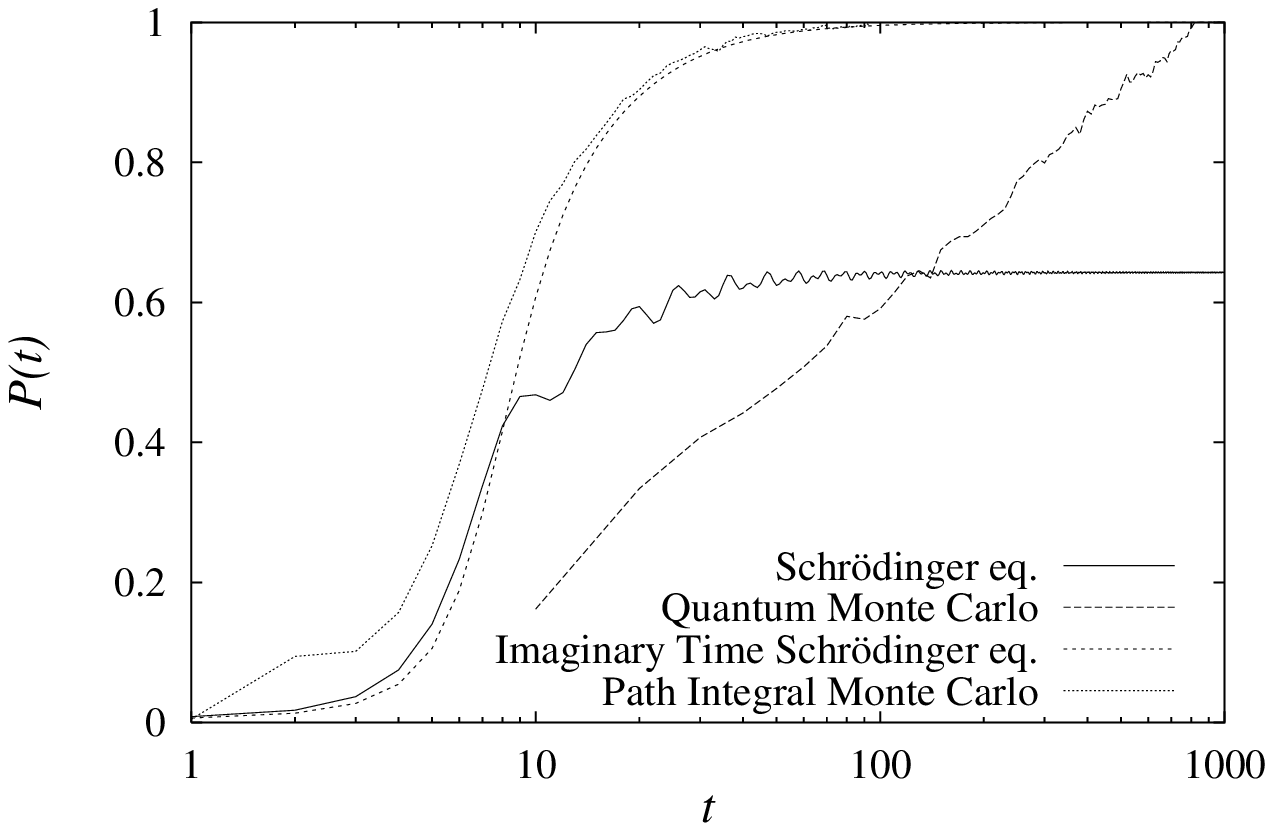}}
\caption{
The results for the Schr\"odinger equation,
the quantum Monte Carlo method, the imaginary time Schr\"odinger equation
and the path-integral Monte Carlo method.
The schedule of $\Gamma(t)$ is $3/t$.
Except for the result for the Schr\"odinger equation,
the probabilities converge to one.
}
\label{fig_qa_dm_2}
\end{figure}

On the other hand,
there is no reason that the quantum Monte Carlo simulation
is equivalent to the solution of the Schr\"odinger equation.
However, we regard that the Monte Carlo step in the quantum Monte Carlo
simulation is also equivalent or similar to the time in the Schr\"odinger
equation.
How different between the two results is shown later.

Besides the above approach, the path-integral Monte Carlo also describes
the dynamics of the quantum system.
The imaginary time path integral of the quantum system and the partition
function at the temperature $1/\beta$, where $\beta$ is the imaginary
time, have the same expression.
The time developing operator $e^{-\beta \mathcal{H}}$ can
be decomposed like the Suzuki-Trotter decomposition in the
quantum Monte Carlo method.
Therefore, the imaginary time path integral is evaluated numerically.

In this formulation, the path integral is replaced with the Monte Carlo
simulation.
However, we have three difficulties about this simulation.
The Hamiltonian with the imaginary time $\mathcal{H}(-i\beta j/M)$ is no
longer a Hermitian matrix,
because the amplitude of the transverse field $\Gamma(t)$ is a function
of the time and generally becomes a complex number when $\beta$ is a
real number.
We can avoid this difficulty by changing the definition of the time in
$\Gamma(t)$ as $t\rightarrow \sqrt{tt^{\ast}}$, where $t^{\ast}$ is the
complex of $t$ conjugate of $t$, because $\sqrt{tt^{\ast}}$ is a real
number for any complex number $t$.
Secondly, the simulation is performed by replacing imaginary numbers
with real numbers, so that we have to evaluate $A(T)$ from $A(\beta)$
by the process of analytic continuation.
The analytic continuation for the function whose values are given
as a numerical number is a great problem itself
and needs large computational power like the simulated annealing.
Finally, the computational time of the path-integral Monte Carlo method
may be longer than the quantum Monte Carlo method.
The imaginary time evolution is mapped to the spatial axis, so that
we have to wait until the system becomes in equilibrium.
We can decrease the transverse field in the quantum Monte Carlo method
without care that the system is trapped at a local minimum or not.
In the path-integral Monte Carlo, we care about that the system is
trapped at a local minimum or not, because the expectation value
is obtained from the average of various paths whose distribution obeys
the Gibbs distribution of the $(d+1)$-dimensional system.

If we accept that the imaginary time expectation value $A(\beta)$ is
regarded as a real time one $A(T)$, the final problem is still a
difficult problem.
It takes long time to wait until the system becomes in equilibrium,
if the landscape of the system is complicated.

We compared the above two approaches,
the quantum and the path-integral Monte Carlo methods,
with the results of the direct solutions of the Schr\"odinger equation.
The system is the SK model with $N=8$
and the schedule is $\Gamma=3/\sqrt{t}$.
The probability to find the ground state is plotted in Fig.~\ref{fig_qa_dm_1}.
The quantum Monte Carlo is performed with $M=1000$.
We wait $5000$ Monte Carlo steps for the initial relaxation and
take average over the next $5000$ steps in the path-integral Monte Carlo
simulation.
We put the inverse temperature $\beta=1000$ for each calculation.
The result of the imaginary time dynamics of the Schr\"odinger equation
is also plotted.

In Fig.~\ref{fig_qa_dm_2}, we plot the solution whose schedule is
$\Gamma = 3/t$.
In this schedule, the solution of the Schr\"odinger equation does not
converge to the ground state with probability one,
but the others converge to one.
The curve of the path-integral Monte Carlo agrees with the curve of the
imaginary time Schr\"odinger equation.

From these results, we can conclude that the quantum Monte Carlo and
the path-integral Monte Carlo simulations are not the same as the
Schr\"odinger equation.
We regard that the assumptions of the two Monte Carlo simulations are
not correct.
The assumptions are the following:
For the quantum Monte Carlo simulation, we assume that the Monte Carlo
step equals to the real time.
For the path-integral Monte Carlo simulation, we assume that the
imaginary time dynamics equals to the real time one.

However, the goal is to find the ground state as fast as possible.
The two Monte Carlo methods find the ground state configuration with
probability one, even if the schedule is too fast to solve for QA by the
Schr\"odinger equation.
For this purpose, we can adopt one of the Monte Carlo methods.
We consider that the quantum Monte Carlo method has the advantage.
The quantum Monte Carlo method is similar to SA, because the control
parameter ($T$ for SA and $\Gamma$ for QA) decreases as a function of
the Monte Carlo step.
Thus, the longer we perform the simulation, the larger the probability
to find the ground state becomes.
On the other hand, a long simulation makes the statistical errors small
for the path-integral Monte Carlo method, but the probability does not
increase.
For these reasons hereafter we adopt the quantum Monte Carlo method for
QA of the large-size systems.

\begin{figure}[t]
\scalebox{1}{\includegraphics{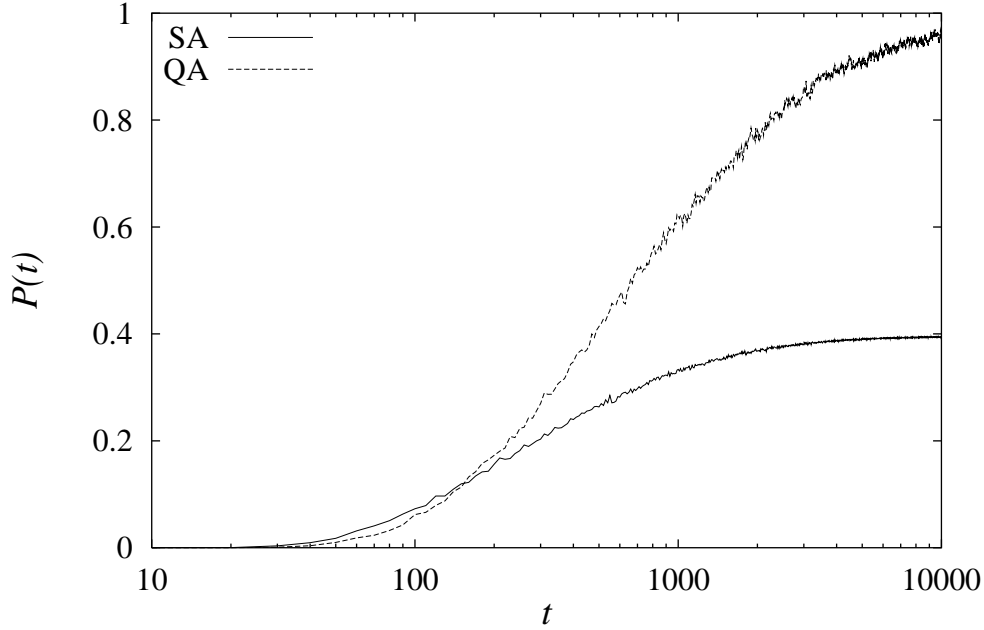}}
\caption{
Overlaps with the ground state are plotted.
The system is the SK model with $N=51$ and $3/\sqrt{t}$ schedule.
}
\label{fig_sa_qa_sk_p}
\end{figure}

\begin{figure}[htp]
\scalebox{1}{\includegraphics{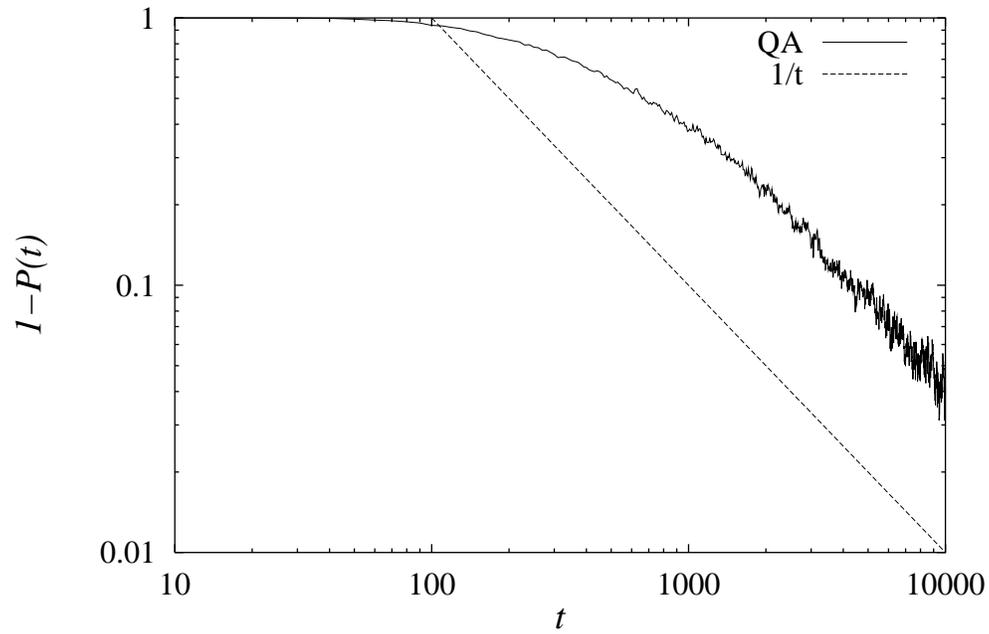}}
\caption{
The quantity $1-P(t)$ is plotted.
The conditions are the same as in Fig.~\ref{fig_sa_qa_sk_p}.
}
\label{fig_sa_qa_sk_1-p}
\end{figure}

We compared the probabilities to find the ground state in the Monte
Carlo version of SA and QA for the SK model with $N=51$ and
$T=\Gamma=3/\sqrt{t}$ in Fig.~\ref{fig_sa_qa_sk_p}.
In this calculation, we perform $100$ independent runs of long-time
($t=100000$) SA with the sufficiently slow schedule $T=3/\ln(t+1)$
and determine the ground state energy beforehand.
Using this ground state energy, we can calculate the probability to
find the ground state.
The probability to find the ground state is the ratio of the number of
the ground state configurations for each time in ten thousand
independent runs of SA and Trotter slices of QA.
The probability in QA is larger than SA.
This result means that the QA in the quantum Monte Carlo method also
improves the performance in finding ground state.
The quantity $1-P(t)$ is plotted in Fig.~\ref{fig_sa_qa_sk_1-p}
under the log-log scale.
The curve converges to zero asymptotically as $1/t$.
This implies that the system almost follows the stationary state and the
annealing schedule is sufficiently slow.
Moreover, we apply fast schedule to the same system and the probability
converges to one asymptotically as $1/t^2$.
In this fast schedule, system also follow the stationary state.
\begin{figure}[htp]
\scalebox{1}{\includegraphics{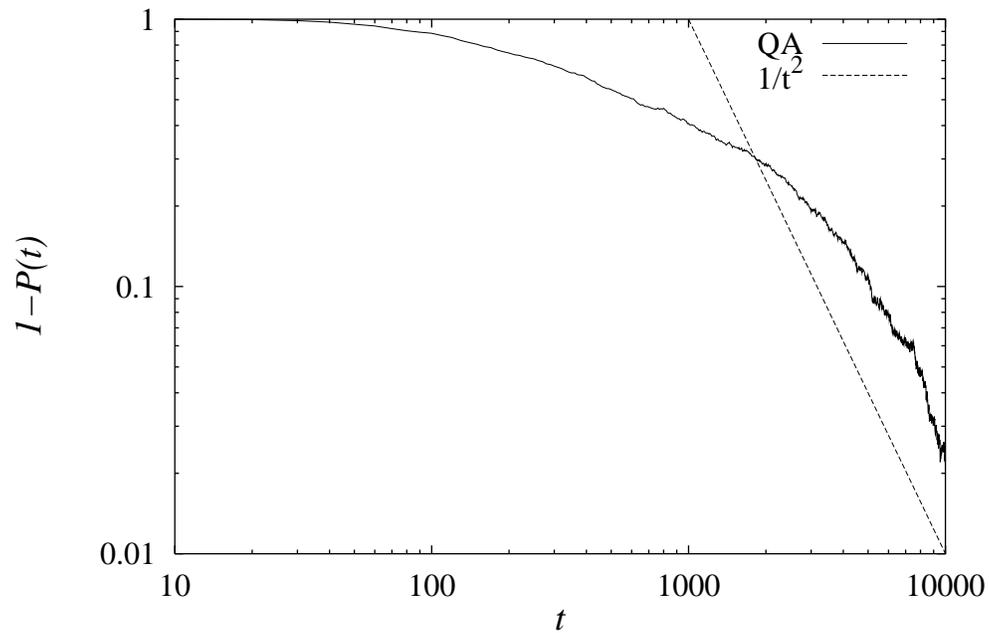}}
\caption{
The quantity $1-P(t)$ is plotted.
The schedule is $\Gamma=3/t$.
}
\label{fig_sa_qa_sk_1-p_2}
\end{figure}

\begin{figure}[t]
\scalebox{1}{\includegraphics{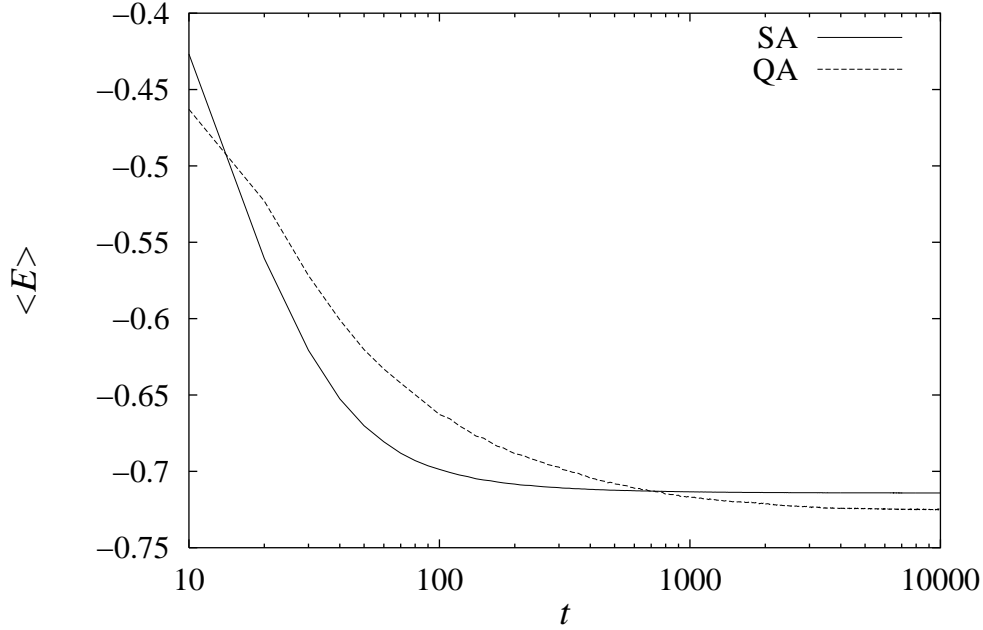}}
\caption{
The average energy is plotted for SA and QA.
The system is the SK model with $N=51$ and $3/\sqrt{t}$ schedule.
}
\label{fig_sa_qa_sk_e}
\end{figure}

We also plot the average energy of SA and QA under the same condition
in Fig.~\ref{fig_sa_qa_sk_e}.
As we need the ground state of the classical term of the Hamiltonian,
we neglect the contribution of the transverse field term in the
calculation of the energy.
We take the average energy which does not contain the interactions
between the Trotter slices for each snapshot of the Trotter slices in QA.
The average of the energy per spin is calculated by $M$ independent runs
for SA and $M$ Trotter slices for QA, respectively.
In spite of the large difference between SA and QA in $P(t)$,
the energy of QA is lower than SA, but the difference is not so large.
This can be understood that if the system is trapped in a local minimum,
the trapped system also lowers the average energy but does not count in
$P(t)$.

Calculating the probability to find the ground state is a useful index
to check which method has a good performance.
However it is too difficult to perform this calculation for large-size
systems, because the calculation needs the ground state configuration,
but we do not know it.
In this case, we regard that the method whose average energy is lower
has a better performance.
If the average energy of QA is lower than SA, QA may find the ground
state more efficiently than SA.

On the other hand, another index can be considered.
That is the lowest energy in the $M$ independent runs of SA or in $M$
Trotter slices of QA.
However, comparing the performance by the average energy gives more
precise conclusion than by the lowest energy, because the lowest energy
depends on the initial state more than the average energy.
If the initial state is in the basin of the ground state,
the system goes to the ground state with large probability.
We will calculate SA and QA by the Monte Carlo method for the large-size
systems up to $N=10000$ and compare their performance by the average energy.

\section{Results of the Monte Carlo simulation}
\label{sec:3.5}

Two additional parameters which do not appear in the Schr\"odinger
equation are needed, when the calculation of QA is performed by the
quantum Monte Carlo simulation.
The first one is the (inverse) temperature $T(=1/\beta)$,
and second one is the Trotter number $M$.
The original dynamics, the Schr\"odinger equation, does not contain
thermal fluctuations, so that we have to take the limit $\beta
\rightarrow \infty$.
The Trotter number $M$ should also be infinity to take into account
the quantum effect correctly.

Each Trotter slice is the classical system with the effective inverse
temperature $\beta_{\text{eff}} (=\beta/M)$ except for the interactions
between Trotter slices.
We consider the parameters $\beta_{\text{eff}}$ and $M$ instead of the
parameters $\beta$ and $M$, because it is natural to regard that we
simulate the classical $(d+1)$-dimensional system at the effective
inverse temperature.
The infinity limit of the Trotter number also means the infinity limit
of the inverse temperature, because the ratio $\beta/M$ is fixed.

\subsection{Dependence on the temperature}

First, we consider the effective inverse temperature $\beta_{\text{eff}}$.
At the initial time, the strength of the interaction between Trotter
slices is small and each slice is almost an independent classical system.
At the end, the strength goes to infinity.
The $M$ spins in the Trotter direction take the same value because of the
interactions of the infinite strength and the freedom of this direction
vanish.
The $(d+1)$-dimensional system which is equivalent to the
$d$-dimensional quantum system reduced to the $d$-dimensional classical
system.
This can be understood naturally, because the transverse field
$\Gamma(t)$ vanishes and the Hamiltonian goes to the classical Hamiltonian.

The spins in the Trotter direction take same value when the interactions
between the Trotter slices is stronger than the thermal fluctuation.
If the temperature is high, we need long time until the interaction
becomes large enough.
From this, the temperature should be low to save the time for the
calculation.
On the other hand, the temperature is required to be high enough for
searching the configuration space widely at the initial time.
If the temperature is low, the system can not visit various
configurations and trapped at a local minimum.

To satisfy the above two conflicting conditions, we have to choose moderate
values of temperature.
The system can seek all possible configurations when the temperature is
above the critical temperature of the order-disorder transition of the
classical system.
Thus, we choose $\beta_{\text{eff}}=1$ for the SK model.

\begin{figure}[t]
\scalebox{1}{\includegraphics{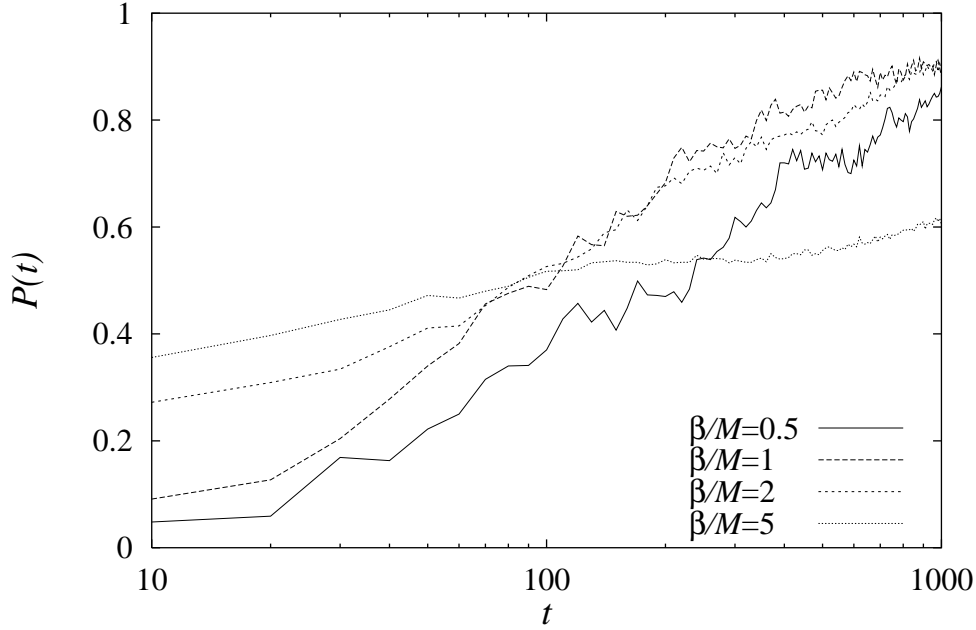}}
\caption{
The probability dependence on the ratio $\beta/M$.
The simulation with $\beta_{\text{eff}}=1$ has a best performance among
the four.
}
\label{fig_dep_bm}
\end{figure}

The dependence of the probability on $\beta_{\text{eff}}$ is shown in
Fig.~\ref{fig_dep_bm}.
The condition is $\beta_{\text{eff}}= 1/2, 1, 2, 5$, $N=8$ and $M=1000$
for the SK model in the $3/\sqrt{t}$ schedule.
The simulation with $\beta_{\text{eff}}=1$ has a best performance among
the four.

\subsection{Dependence on the number of Trotter slices}

Secondly, we consider the Trotter number $M$.
A simulation with large number of Trotter slices gives precise
results, but the computational power for the calculation becomes large.
The computational time for the quantum Monte Carlo simulation is $M$
times longer than the conventional simulated annealing.
We have to estimate the reasonable Trotter number of the correct
simulation.

\begin{figure}[t]
\scalebox{1}{\includegraphics{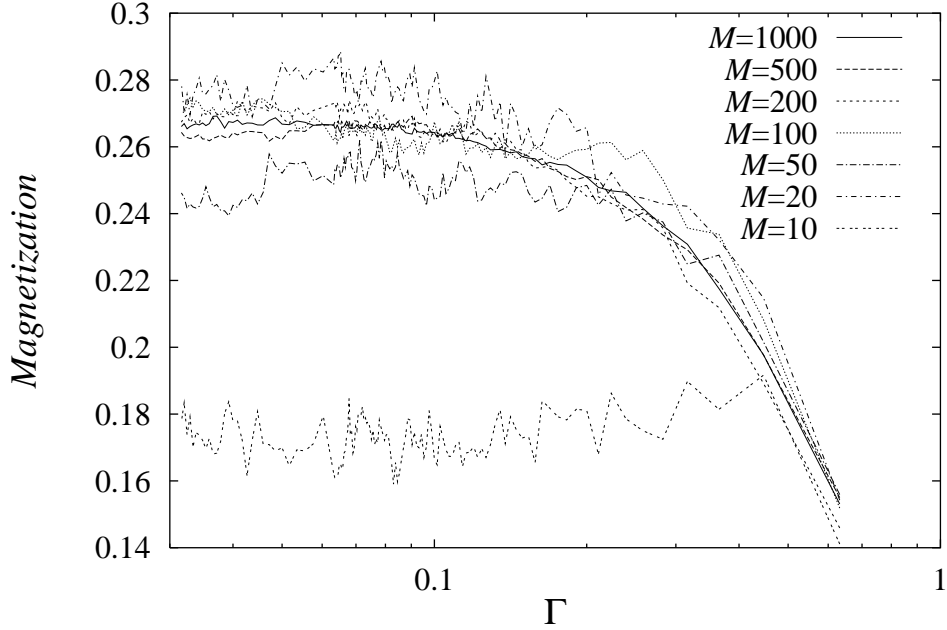}}
\caption{
The magnetization dependence on the Trotter number $M$.
Curves of over $M=100$ are large enough.
}
\label{fig_dep_tn}
\end{figure}

We check the dependence of the simulation on the Trotter number.
The probabilities to find the ground state versus the transverse field
are plotted for the cases of $M=10, 20, 50, 100, 200, 500, 1000$ in
Fig.~\ref{fig_dep_tn}.
The condition is $\Gamma=2/\sqrt{t}$, $N=625$ and $\beta_{\text{eff}}=1$
for the two-dimensional Edward-Anderson model.
We find that $M=100$ is large enough.

From the above two results, we regard that the conditions of
$\beta_{\text{eff}}=1$ and $M=100$ give a reasonable solution.
Hereafter we will use this condition.

\subsection{Results of large size system for the EA model}

Next, we consider systems of larger size than the previous calculations
to demonstrate that our method can be applied to the actual problems.
The model is the two-dimensional Edward-Anderson (EA) model.
The calculation per Monte Carlo step for the EA model is less than
the SK model.
The spin flip is determined by the calculation of the energy gap of the
spin flip.
We have to sum up $N-1$ interactions for the SK model, because all spins
interact with each other in such an infinite range model.
For the two-dimensional EA model on the square lattice the number of the
interactions is four.
Therefore, we can accelerate the summation in the EA model $(N-1)/4$ times
faster than the SK model.
The acceleration affects both of the methods SA and QA.

We choose the standard condition of the calculation of this model as
follows;
$N=625 (25^2)$ with periodic boundary conditions, $\beta_{\text{eff}}=1$,
$M=100$ and the schedule $T=\Gamma=c/\sqrt{t}$.
The critical temperature $T_c$ is considered zero in this model.
This is not the same value of the SK model.
However, the zero temperature dynamics only seeks the configuration
whose energy is lower than the present configuration and tends to be
trapped at local minima, so that we use the same condition as
$\beta_{\text{eff}}=1$.

We check how the system follows the stationary (equilibrium) state
which means equilibrium state at fixed $T$ for SA
and ground state at fixed $\Gamma$ for QA.
For small-size systems, we can calculate the probabilities to find
the ground state $P_{\text{SA}}(t)$ and $P_{\text{QA}}(t)$ and compare
with the probabilities in the stationary state
$P_{\text{SA}}^{\text{st}}(T(t))$ and
$P_{\text{QA}}^{\text{st}}(\Gamma(t))$.
If the two curves, the results of the dynamics and the statics, are the
same, the system follows the stationary state.
For large-size systems we cannot calculate the probability to find the
ground state, because the calculation of the probability needs the
ground state configuration beforehand and that is difficult.
Nevertheless, we can check whether the system follows the stationary
state or not by another quantity.
In our case, a longitudinal magnetic field is applied to the system,
and magnetic order grows as a consequence of this field.
If the system is in the stationary state, the magnetization is
characterized by the temperature for SA or the transverse field for QA
and does not depend on the speed of the schedule.
The functional form of the schedule is $c/\sqrt{t}$, thus the speed of
our schedule is controlled by $c$
(slow for large $c$ and fast for small $c$).
The system can not follow the stationary state for the rapid schedules,
the cases of small $c$.
We check the dependence of magnetization on $c$.

\begin{figure}[p]
\scalebox{1}{\includegraphics{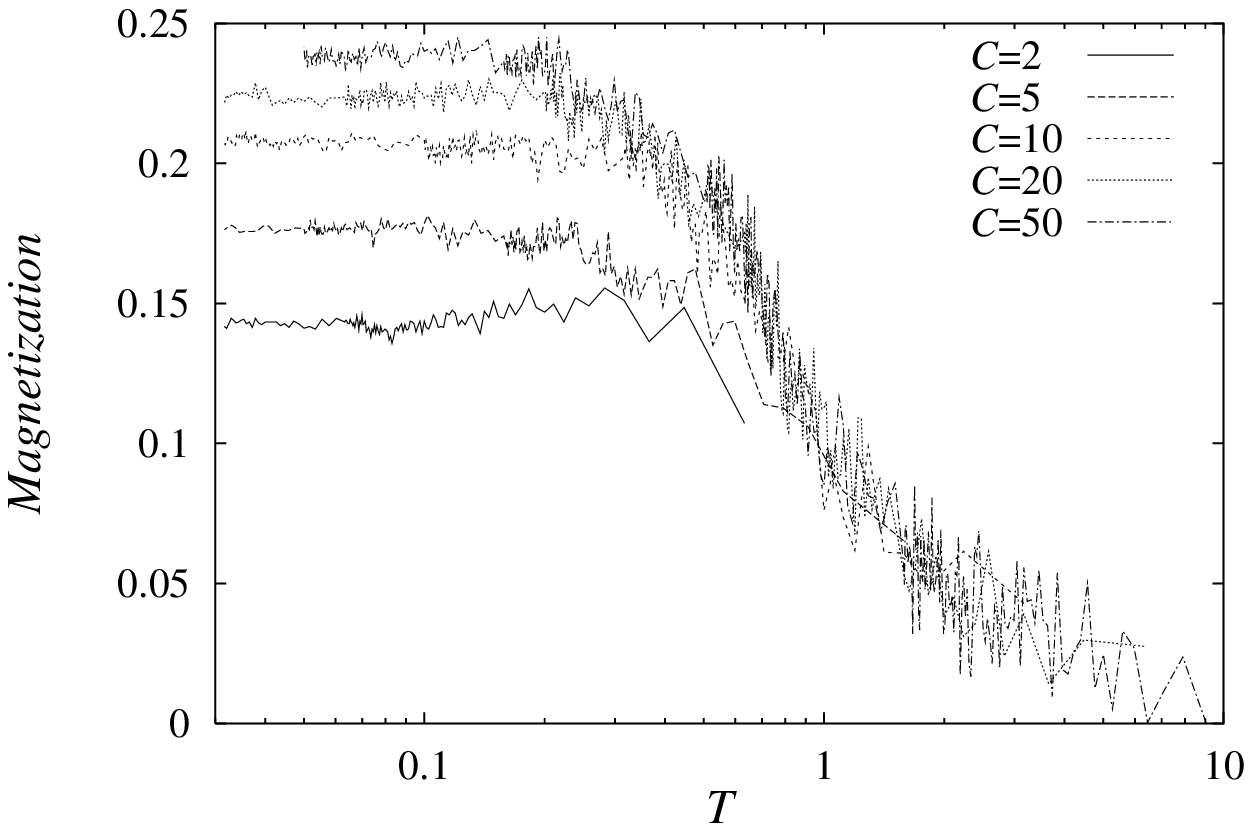}}
\caption{
The dependence of magnetization on $c$ for the conventional simulated
annealing (SA) 
}
\label{fig_sa_c}
%\end{figure}

%\begin{figure}[t]
\scalebox{1}{\includegraphics{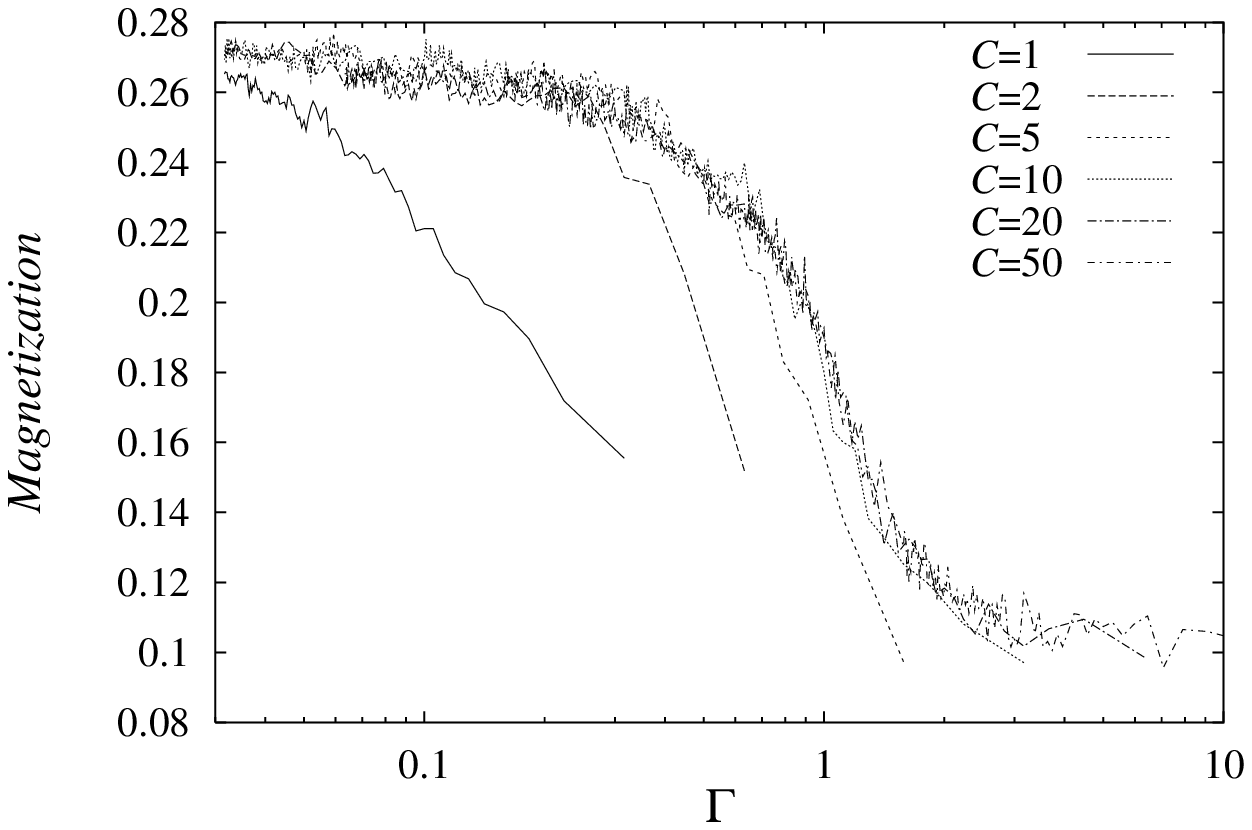}}
\caption{
The dependence of magnetization on $c$ for quantum annealing (QA)
}
\label{fig_qa_c}
\end{figure}

The magnetization versus the temperature for SA and the transverse field
for QA is plotted in Figs.~\ref{fig_sa_c} and~\ref{fig_qa_c}.
For SA, the final value of the magnetization (the left end of the
curves) depends on $c$.
This means that the schedule is too fast to follow the stationary state.
This result implies that the system tends to be trapped at local minima
in this schedule up to $c=50$.
For QA, the final value of the magnetization does not depend on $c$
beyond $c=10$.
Thus, we can regard that the schedule $\Gamma=c/\sqrt{t}$ with $c=10$
is slow enough to search the ground state of the two-dimensional EA model.
The conclusion of these two results for SA and QA is that SA does not
follow the stationary state and QA does in the $c/\sqrt{t}$ schedule.
This is consistent with small-size analysis by differential equations
(the master equation for SA and the Schr\"odinger equation for QA).

\begin{figure}[t]
\scalebox{1}{\includegraphics{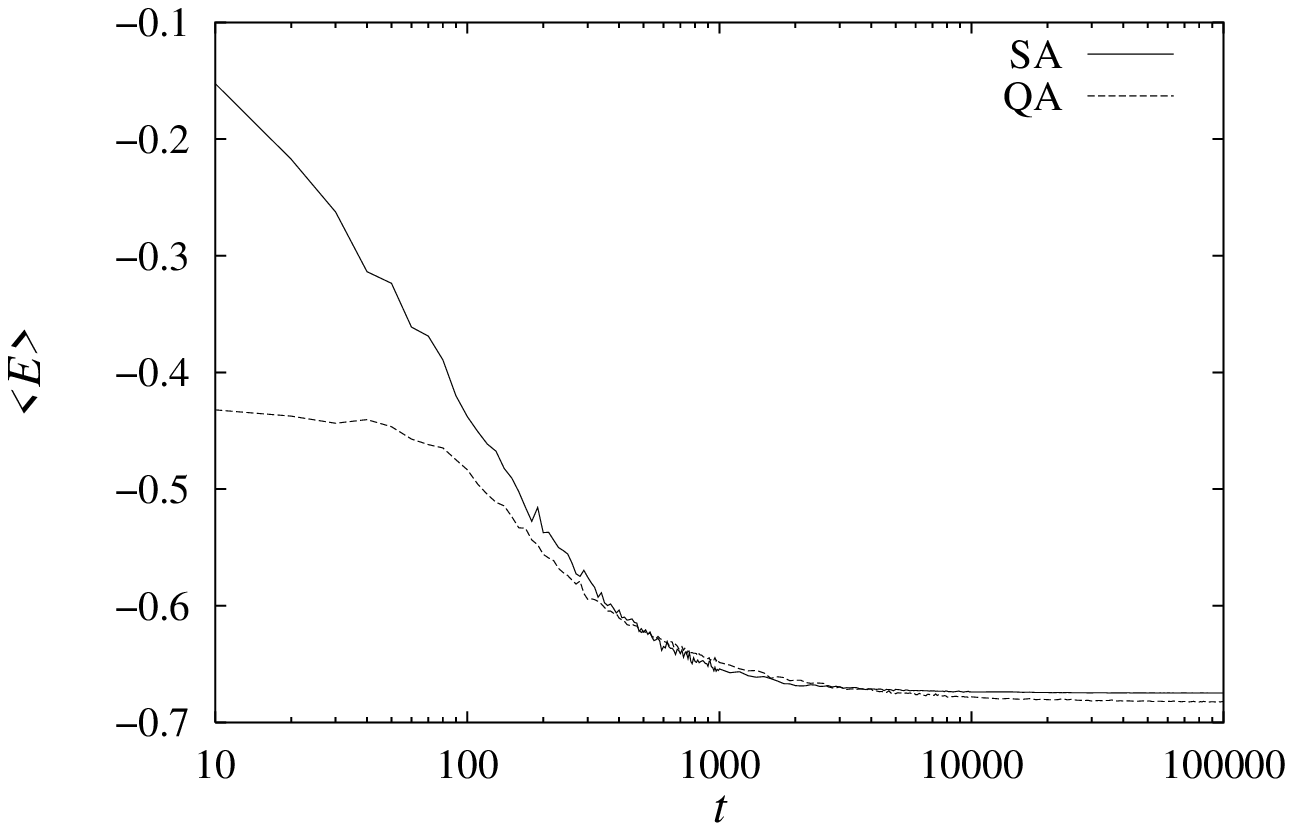}}
\caption{
Average energy for SA and QA are plotted.
The schedule is $T=\Gamma=10/\sqrt{t}$ and the system size is $N=625$.
}
\label{fig_sa_qa_1}
\end{figure}

We average out the energy for SA with $M(=100)$ independent runs
and QA with $M$ Trotter slices.
The results on the average energy are plotted in Fig.~\ref{fig_sa_qa_1}.
Only in the middle region of the time, the energy in QA is higher than
in SA, but finally the energy in QA goes lower than in SA.

As shown in the previous chapter, the performance in finding the ground
state is improved by the quantum effect.
However, the quantum Monte Carlo uses the Suzuki-Trotter decomposition
by definition.
The calculation of the quantum Monte Carlo needs Trotter-number $M$ times
longer than the calculation of the classical system.
Therefore, the comparison between SA and QA is still incomplete.
The time scales of the two computations are not the same.
One Monte Carlo step in QA takes about $M$ times evaluations of spin
flips than in SA.
(To be precise, the calculation in QA is $M(z+2)/z$ times longer than the
calculation in SA, where $z$ is the number of nearest neighbors and
``2'' is the number of the interactions in the Trotter direction.
It is exactly ``$M$ times'' for the infinite range model in the
thermodynamic limit.)
Taking into account this disadvantage gives a reasonable comparison
between SA and QA.
We provide a reasonable definition of the time $t^{\prime}$ of which the
quantity is plotted and compared as a function as,
\begin{gather}
 t^{\prime} = t \ \ \text{(for SA)} \ \ \ \ \ \ \
 t^{\prime} = M t \ \ \text{(for QA)},
\end{gather}
The Trotter number is $100$ in our case, so that the time is
$t^{\prime} = 100 t$ for QA.

\begin{figure}[t]
\scalebox{1}{\includegraphics{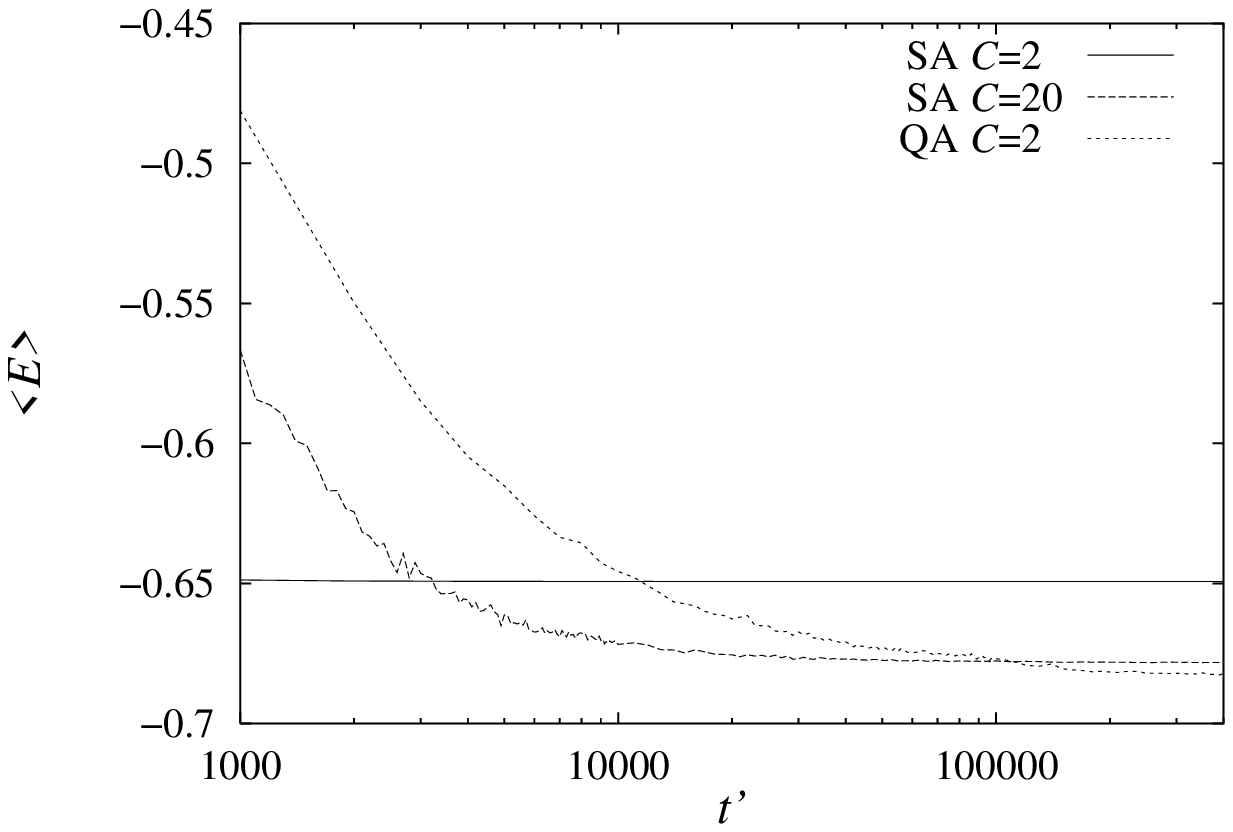}}
\caption{
Time evolution of the average energy for QA and SA.
The system size is $N=625$.
One hundred Trotter slices are averaged for QA
and $10$ runs are averaged for SA.
For ``SA $C=2$'' and ``QA $C=2$'' the schedules of the temperature and
the transverse field are the same function.
``SA $C=20$'' and ``QA $C=2$'' take the same value
of the temperature and the transverse field
$T=\Gamma=0.1/\sqrt{10}=0.0316\cdots$ at the end of the simulations.
}
\label{fig_sa_qa_2}
%\end{figure}

%\begin{figure}[t]
\scalebox{1}{\includegraphics{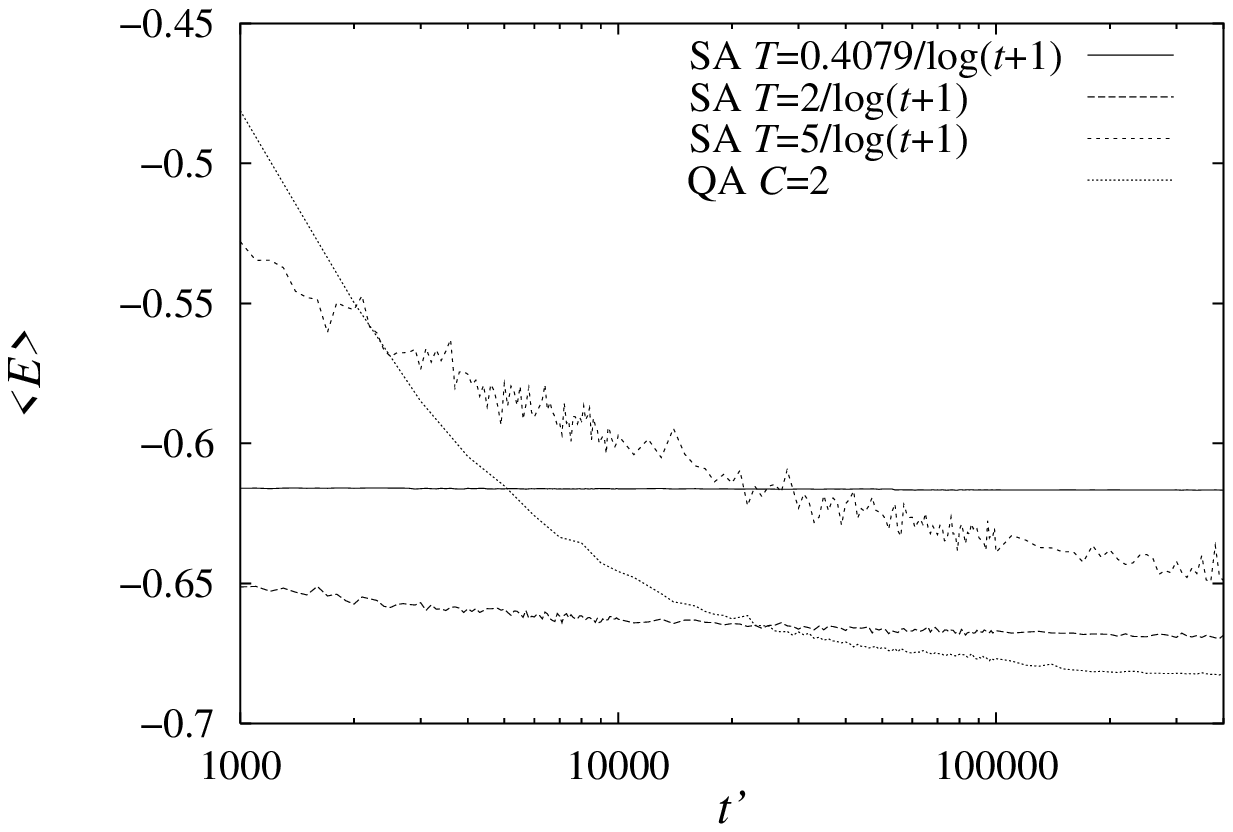}}
\caption{
Time evolution of the average energy for QA and SA.
The system size is $N=625$.
One hundred Trotter slices are averaged for QA
and $10$ runs are averaged for SA.
``SA $T=0.4079...$'' and ``QA $C=2$'' take the same values
$T=\Gamma=0.1/\sqrt{10}=0.0316\cdots$ at the end of the simulations.
}
\label{fig_sa_qa_3}
\end{figure}

\begin{figure}[t]
\scalebox{1}{\includegraphics{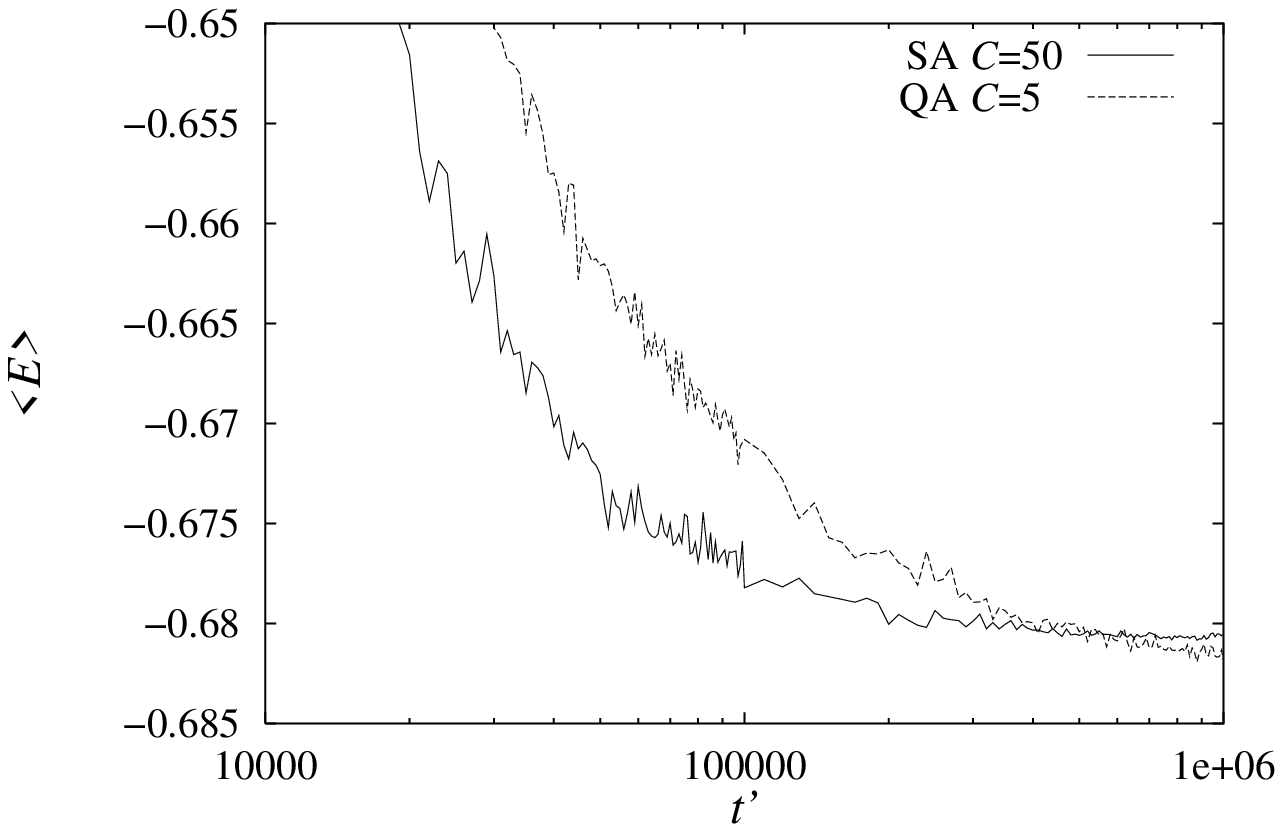}}
\caption{
Time evolution of the average energy for QA and SA.
The system size is $N=625$ and the schedule is $T=50/\sqrt{t}$
and $\Gamma=5/\sqrt{t}$.
One hundred Trotter slices are averaged for QA
and $10$ runs are averaged for SA.
The temperature and the transverse field at the end of the simulations
are the same value $T=\Gamma=0.05$.
}
\label{fig_sa_qa_4}
%\end{figure}

%\begin{figure}[t]
\scalebox{1}{\includegraphics{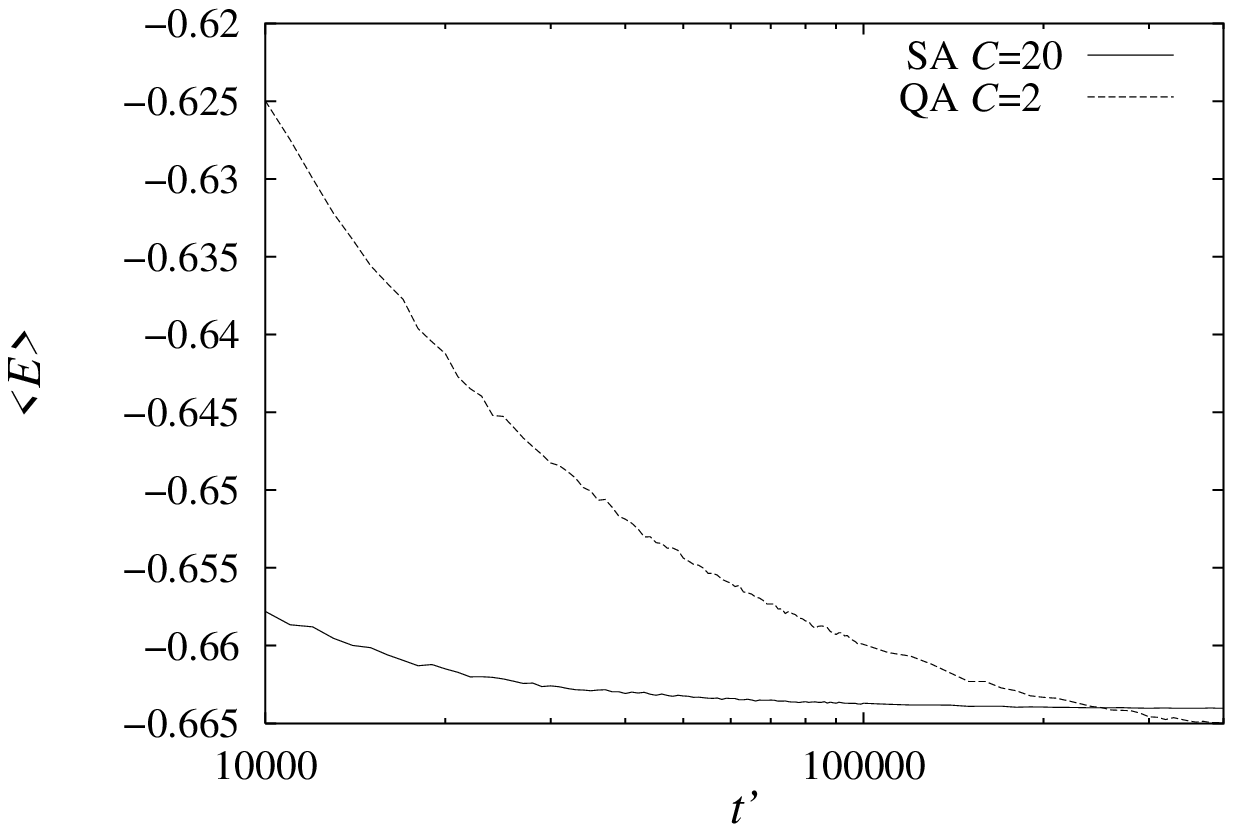}}
\caption{
Time evolition of the average energy for QA and SA.
The system size is $N=10000$ and the schedule is $T=20/\sqrt{t}$
and $\Gamma=2/\sqrt{t}$.
One hundred Trotter slices are averaged for QA
and $10$ runs are averaged for SA.
The temperature and the transverse field at the end of the simulations
are the same value $T=\Gamma=0.1/\sqrt{10}=0.0316\cdots$.
}
\label{fig_sa_qa_5}
\end{figure}

The average energy for SA and QA is plotted in Fig.~\ref{fig_sa_qa_2}
under the condition of such a scaling of the time.
QA improves the performance in finding the ground state.
In the calculation, we adopt that the schedule in SA is $100$ times slower
than QA, because SA could not follow the stationary state in the
$c/\sqrt{t}$ schedule (see Fig.~\ref{fig_sa_c}).
The total Monte Carlo step of SA is $100$ times longer than QA, because
we have to use same amount of the computational power to compare the two
methods (QA needs $100$ times calculations than SA).
The values of the temperature and the transverse field
at the end of the simulations are the same.
We also plot the cases of various schedules for the temperature
and the transverse field in Fig.~\ref{fig_sa_qa_3} and
Fig.~\ref{fig_sa_qa_4}.
QA's performance is also good in the various schedules.
For another system size, $N=10000$, the average energy in QA is also lower
than SA as seen in Fig.~\ref{fig_sa_qa_5}.

At the end of simulations, we cannot reach the zero temperature or zero
transverse field.
The values are quite small $(\sim 0.03)$ but not equal to zero.
For example, the strength of the interacrion between the Trotter slices
is $1.73$, when $\Gamma=0.1/\sqrt{10}$.
It is not so strong compared with the effective inverse temperature
$\beta_{\text{eff}}=1$.
The thermal fluctuation still remains.
To remove this fluctuation, we execute zero temperature dynamics after
the annealing.
Then, almost all the Trotter slices (more than 90\% for the situation
of the calculation for Fig.~\ref{fig_sa_qa_2}) converge to the same
configuration in QA, while the independent runs in SA do not.

\subsection{Difference between the annealing and the quenching processes}

\begin{figure}[p]
\scalebox{1}{\includegraphics{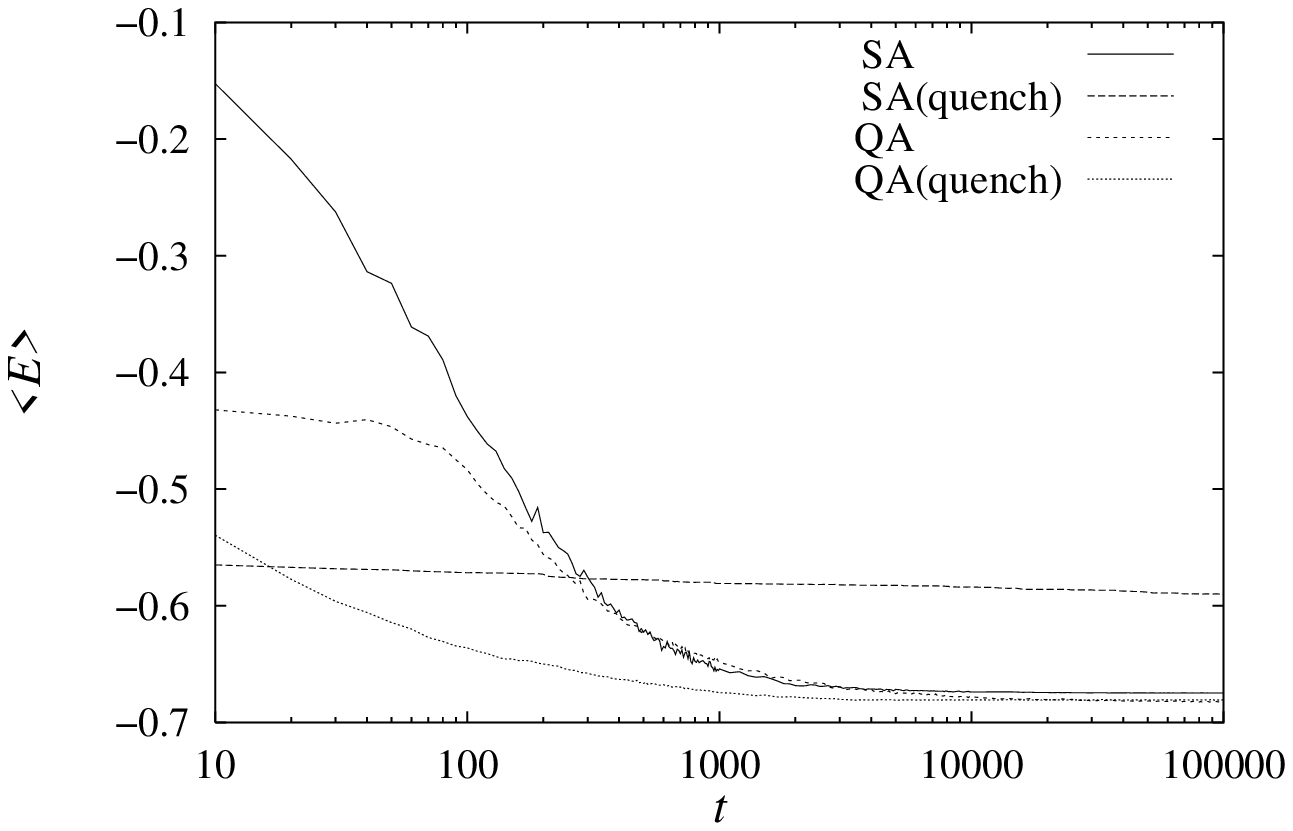}}
\caption{
Solutions for annealing and quenching processes are plotted.
The annealing schedule is $T=\Gamma=10/\sqrt{t}$ and
the quenched value is $T=\Gamma=0.1/\sqrt{10}=0.0316\cdots$
which is the final value of the schedule.
}
\label{fig_quench_1}
%\end{figure}

%\begin{figure}[t]
\scalebox{1}{\includegraphics{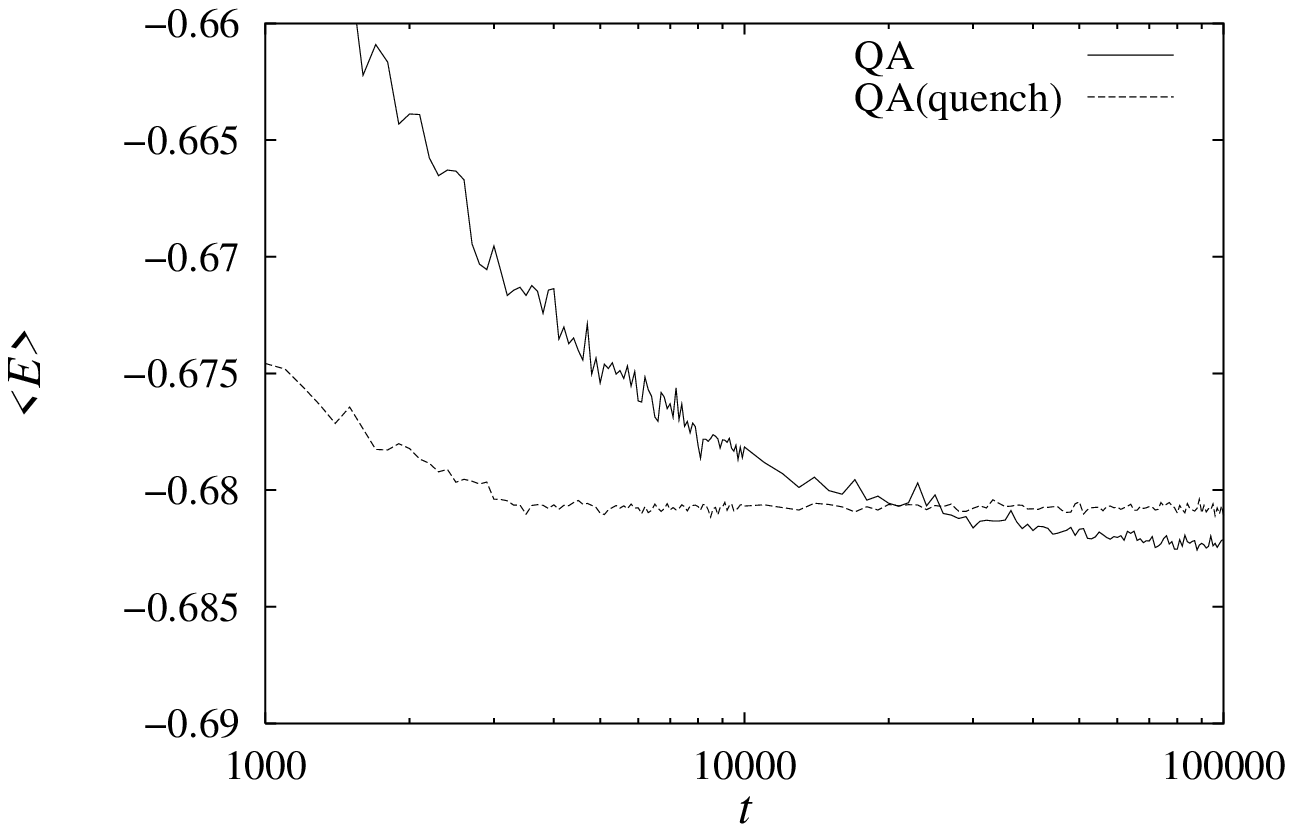}}
\caption{
Solutions for annealing and quench process for QA are plotted.
This is the enlargement of Fig.~\ref{fig_quench_1}.
}
\label{fig_quench_2}
\end{figure}

Let us take another approach to investigate the property of QA.
How efficiently does the annealing process find the ground state
in comparison with the quenching process?
We calculate the average energy of the two processes, the annealing and
the quenching, for SA and QA.
The temperature and the transverse field are quenched at
$T=\Gamma=0.1/\sqrt{10}=0.0316\cdots$, and scheduled from infinity to zero by a
function $10/\sqrt{t}$.
The quenching process of the quantum system is almost equivalent to the
case of the work by Sato {\it et al.}~\cite{Sato}
They considered the dynamics of the quenched system with
$\Gamma \rightarrow \infty$.
The results are shown in Fig.~\ref{fig_quench_1}.
The temperature $T$ and the transverse field $\Gamma$ for annealed
and quenched systems are the same at the end of the simulations.
If the systems stay in the stationary state, the curves arrive at the same
point at the end.
The average values of energy of SA, QA and $\Gamma$-quenched systems
converge to almost the same value,
but the $T$-quenched system is trapped in a high-energy state.
Of course the energy of the $\Gamma$-quenched system is higher than that of
QA (see Fig.~\ref{fig_quench_2}. It is an enlargement of
Fig.~\ref{fig_quench_1}).
The behavior of the two quenched systems are quite different,
though both of the controlled parameters $T$ and $\Gamma$ are quenched.
The quantum system relaxes faster than the classical system,
when the transverse field takes the same value as the temperature
of the classical system.
This is consistent with the result that QA reaches the stationary state
in magnetization, but SA does not when the coefficient $c$ changes
from $2$ to $50$.

\section{Summary}

We applied the Monte Carlo method to QA.
There are two ideas to translate the time in the Schr\"odinger equation
to the Monte Carlo method.
One is the path-integral Monte Carlo which regards the Trotter direction
as the time (more precise, the imaginary time).
The other is the quantum Monte Carlo with ``the Monte Carlo
step''-dependent inter-Trotter interaction.
We regard the Monte Carlo step as the time in the Schr\"odinger equation.

The path-integral Monte Carlo is performed in the imaginary time
dynamics.
This results in the different behavior from the solution of the
Schr\"odinger equation.
For the quantum Monte Carlo, the result also differs from the result of
the Schr\"odinger equation.
Two reasons are mentioned.
The energy dissipation occurs in the Monte Carlo simulation,
while it does not occur in the original Schr\"odinger equation.
Beside this, the time in the path-integral and the quantum Monte Carlo
methods are not same as the time in the Schr\"odinger equation.
However, these methods have good performance and the ground state is found,
even if the schedule is too fast to find the ground state for the
Schr\"odinger equation.

We adopt the quantum Monte Carlo for large-size QA to reduce the
computational time.
We have to wait until the system is in equilibrium in the path-integral
Monte Carlo and summing up sampling paths.
This takes large computational time.

The Trotter number we adopt is $M=100$, which is large enough for our
calculations (two-dimensional EA model with $N=625$).
The results are compared with SA.
We found that QA improves a performance in finding the ground state even
if we take into account that the QA need $M$ times more calculations than
SA.
Moreover, almost all the configurations of the Trotter slices
converge to the same configuration.

We also checked the difference between annealed and quenched systems.
The relaxation time of the classical quenched system is quite long.
On the other hand, the quantum quenched system does not so differ from the
annealed system QA.
This implies that the quantum effect accelerates relaxations.
Moreover, it is remarkable that the average energy of the quantum
quenched system is lower than SA.
The short relaxation time makes the dependence on the schedule small.
This can be seen in Figs.~\ref{fig_sa_c} and~\ref{fig_qa_c}.
A rapid decrease in the characteristic relaxation times for the random
Ising magnet, $\text{LiHo}_{0.167}\text{Y}_{0.833}\text{F}_4$, with the
transverse field is observed~\cite{Wu_Ellman_etc}.
We consider that the results may come from the same mechanism.

From the application side, the optimal schedule is not known in general.
If the dependence of the performance to find the ground state on the
annealing speed is small, we can decrease the rate not to reach the
ground state or the optimal configuration.
This is an attractive property for the actual applications.

\chapter[Application to TSP]{Application to the Traveling Salesman Problem}
\label{chap:4}

All the problems we calculated up to the previous chapter were on the
Ising-spin systems.
However, the combinatorial optimization problem is not limited to
the problems based on physics.
One of the major problems is the traveling salesman problem
(TSP)~\cite{TSP1,TSP2}.
We will apply our method to the TSP in this chapter.
The TSP is a hard problem to solve by a computer, say an $NP$-hard
problem.
Therefore, comparison between SA and QA in the TSP may provide a good
measure of their performances in the problems of other areas not
limited to physics.
In other words, we will check generality of QA for combinatorial problems.

\section{The Traveling Salesman Problem}
\label{sec:4.1}

\begin{figure}[htb]

\scalebox{0.8}{\includegraphics{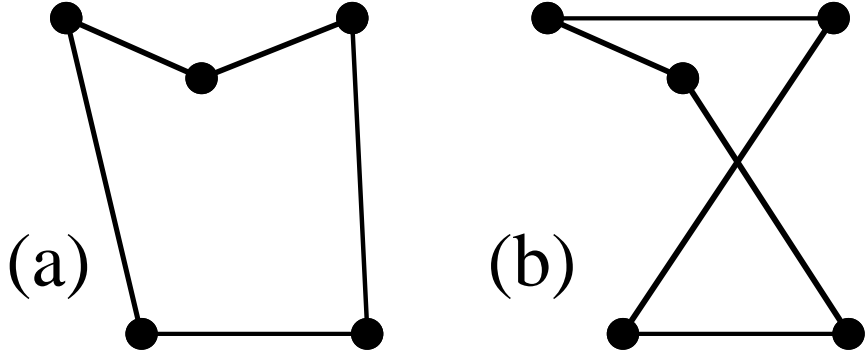}}
\caption{
The traveling salesman problem, showing a good (a) and a bad (b)
solution to the same problem.
} 
\label{fig_tsp_sample}
\end{figure}

We have $N$ points or cities in a space with distances $d_{ij}$ between
them.
The task is to find the minimum-length closed tour that visits each city
once and returns to its starting point.
Figure~\ref{fig_tsp_sample} shows an example.
Practical applications include scheduling of truck deliveries,
airline crews, and the movements of an automatic drill-press or robot
arm.
In many applications the ``distances'' $d_{ij}$ are abstract quantities
not related to a Euclidean distance between points in a real space;
they may not even satisfy the triangle inequality
$d_{ij} \leq d_{ik}+d_{kj}$.
For the example of truck deliveries, the cost to transport
freight from city $i$ to $j$ is affected by many factors, amount of the
freight to city $j$, labor, gas, highway charge, the weather, etc.
The length of the tour is defined by
\begin{equation}
L = \sum_{m=1}^N d_{i(m),i(m+1)},
\end{equation}
where $i(m)$ means the city of the $m$th stop, the tour is closed as
$i(N+1)=i(1)$ and $N$ is the number of cities.

One can think of a simple approach, listing all possible tours of the
problem to search the optimal tour.
The number of all the possible tours is $(N-1)!/2$.
This number increases rapidly and becomes over the limit of the computer
power.
Thus, a faster algorithm is needed to solve the actual problems.
Some algorithms are optimized for TSP~\cite{TSP1,TSP2} and
others are general algorithms which can be applied to TSP~\cite{Kirk}.
The simulated annealing and the quantum annealing belong to the latter.

\section{Simulated Annealing on TSP}
\label{sec:4.2}

\begin{figure}[htb]
\scalebox{0.8}{\includegraphics{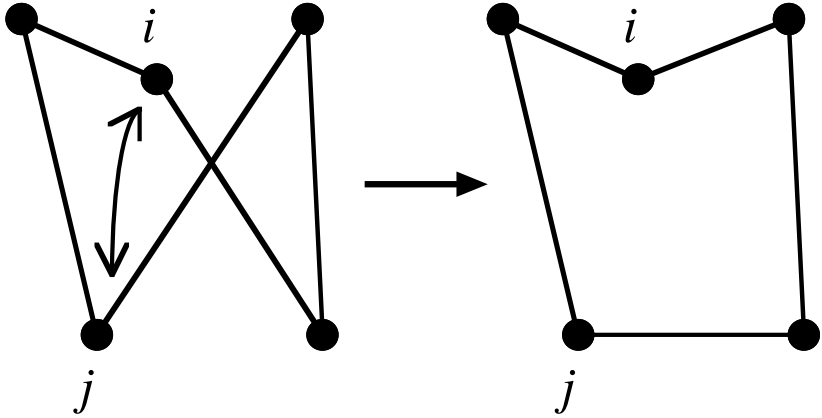}}
\caption{
The simplest rearrangement of the tour.
This rearrangement is expressed in four spin-flip dynamics.
} 
\label{fig_tsp_exchange}
\end{figure}

The simulated annealing is based on the Monte Carlo simulation, so that
we have to determine the dynamics, how to rearrange of the tour.
We choose the simplest rearrangement of the tour like one-spin flip
in the Ising-spin system.
That is to exchange two stops of cities $i$ and $j$, which is
illustrated in Fig.~\ref{fig_tsp_exchange}.

This rearrangement of the tour is realized in the following way:
(1) Choose two cities $i$ and $j$.
(2) Exchange stops of cities $i$ and $j$ with the probability
\begin{equation}
\omega_{A\rightarrow B} =
           \frac{\exp(-\beta L_B)}{\exp(-\beta L_A)+\exp(-\beta L_B)},
\end{equation}
where $A$ is the route before exchange and $B$ is the route after exchange.

One Monte Carlo step consists of $N(N-1)/2$ exchange trials of all
pairs.
We fix the starting point and reduce the trials to $(N-1)(N-2)/2$,
because the tour is closed.
The reversibility of the tour still remains in our calculation.
The temperature $T$ is scheduled in a form $c/\sqrt{t}$,
where $t$ is the Monte Carlo step.

\section{Quantum Annealing on TSP}
\label{sec:4.3}

We have to introduce a quantum fluctuation to the TSP, but that is not
trivial.
The system is not the physical system.
We cannot easily imagine how to introduce a quantum fluctuation.
A straightforward way is that we express TSP as a binary system
namely, the Ising system, so that we can introduce
the quantum effect.
The idea to express TSP as a binary system was proposed by Hopfield and
Tank~\cite{Hopfield_Tank1,Hopfield_Tank2}.

\begin{figure}[htb]
\scalebox{1}{\includegraphics{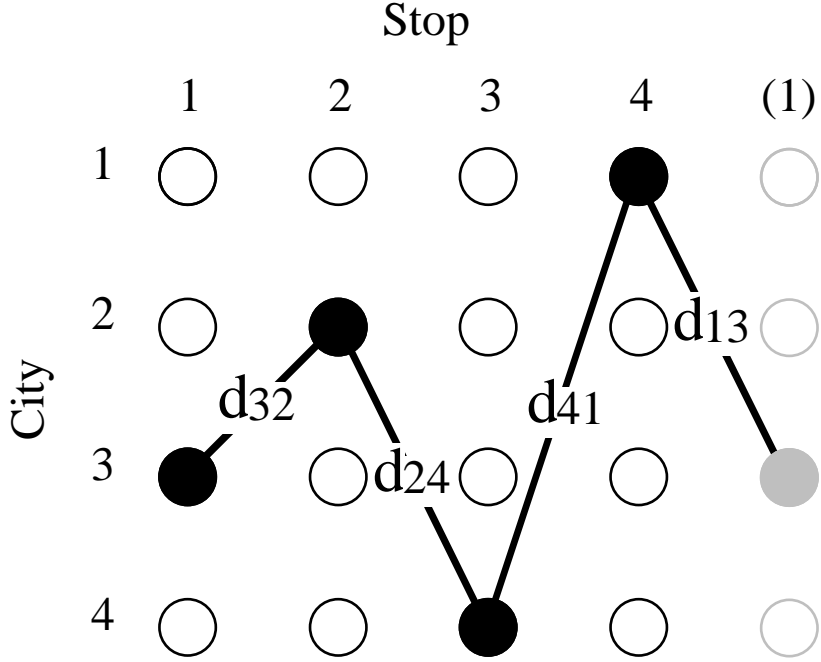}}
\caption{
The illustrated tour is $3 \rightarrow 2 \rightarrow 4 \rightarrow 1$
with $N=4$ TSP.
Solid and open circles denote units that are ``1'' and ``0''
respectively.
The condition is open boundary.
} 
\label{fig_tsp_illust}
\end{figure}

We choose stochastic binary units $n_{ia}$ to represent possible
solutions:
$n_{ia}=1$ if and only if city $i$ is the $a$th stop on the tour
and $n_{ia}=0$ for otherwise.
There are $N^2$ units altogether.
The spin representation can be obtained by the transformation
$n_{ia} \rightarrow (\sigma_{ia}+1)/2$.
The total length of the tour is
\begin{eqnarray}
L & = & \frac{1}{2} \sum_{ij,a} d_{ij} n_{ia}(n_{j,a+1}+n_{j,a-1}) \\
& = & \frac{1}{8} \sum_{ij,a} d_{ij} \sigma_{ia}
          (\sigma_{j,a+1}+\sigma_{j,a-1}) + \text{const.}\ ,
\end{eqnarray}
and there are two constraints:
\begin{eqnarray}
\sum_a n_{ia} & = & 1 \ \ \ \text{(for every city $i$)}, 
\label{constraint1} \\
\sum_i n_{ia} & = & 1 \ \ \ \text{(for every stop $a$)}.
\label{constraint2}
\end{eqnarray}
The first constraint says that each city appears only once on the tour,
while the second says that each stop on the tour is at just one city.
The tour is expressed as a configuration of $N \times N$ units as
illustrated in Fig.~\ref{fig_tsp_illust}

The simplest rearrangement of the route of the traveling salesman
is exchanging the stop of cities $i$ and $j$.
In the spin expression, four spins are flipped at a time.
The salesman stops at cities $i$ and $j$ for the $a$th and the $b$th stops
respectively.
On-route spins take $\sigma_{ia} = \sigma_{jb} = 1$,
and off-route spins take $\sigma_{ib} = \sigma_{ja} = -1$.
One up spin and one down spin in the same row are flipped at a time,
so that the summation which runs in the horizontal direction does not
change.
In the vertical direction, the same situation occurs.
The constraints (\ref{constraint1}) and (\ref{constraint2}) are kept in
the four-spin flip.
The Monte Carlo simulation in the spin expression is performed by this
procedure.

We fix $\sigma_{11}=1$ which means that the tour starts from city $1$.
Possible exchanges with any two stops of cities are the combination
number of choosing two from $N-1$ as $(N-1)(N-2)/2$.
One Monte Carlo step must contain $(N-1)(N-2)/2$ spin-flip trials.
This number of trials is suitable that the system has $N^2$ spins and
$\mathcal{O}(N^2)$ trials are performed in one Monte Carlo step
like the usual one-spin flip procedure.

Let us consider to introduce quantum effects to the classical system.
The dynamics of SA was described by one spin-flip procedure in the
preceding chapters, and we chose the quantum effect which flips one
spin at a time.
That is the transverse field term, $-\Gamma \sum_i \sigma_i^x$.
The linear combination of the off-diagonal operator $\sigma_i^x$
has a non-zero value, when a bra state differ from a ket state only by
one spin.
If the two spin-flip dynamics is applied to SA, we have to choose the square
term of $\sigma_i^x$ as $-\Gamma \sum_{ij} \sigma_i^x \sigma_j^x$.
As known in the Suzuki-Trotter decomposition of the XY or Heisenberg
models, the two-spin interaction is mapped to the four-spin interaction
in the classical Ising model of the $(d+1)$-dimensional
system~\cite{Suzuki}.

\begin{figure}[htb]
\begin{center}
\scalebox{0.8}{\includegraphics{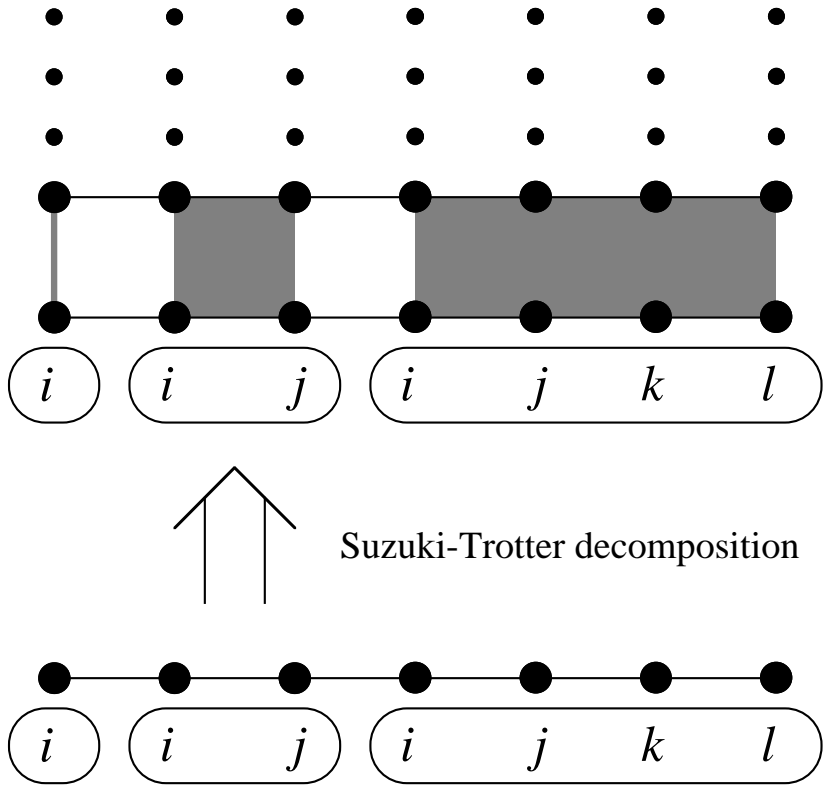}}
\end{center}
\caption{
By the Suzuki-Trotter decomposition
one-, two- and four-spin interactions transform to interactions
which contain spins twice as many --- two, four and eight spins.
} 
\label{fig_4spin_Trotter}
\end{figure}

On the other hand, the traveling salesman problem mapped to the Ising
system has to flip four spins at a time.
The quantum fluctuation term should be the four-spin interaction term as
$-\Gamma \sum_{ijkl} \sigma_i^x \sigma_j^x \sigma_k^x \sigma_l^x$.
We have an eight-spin interaction term by the Suzuki-Trotter
decomposition of the Hamiltonian with this term.
Figure~\ref{fig_4spin_Trotter} shows the interactions in a decomposed
classical system of one-, two- and four-spin interactions of the quantum
system.
Spins which contact through the gray domain in the figure interact with
each other.
The computational cost is expected to increase when the number of spins
which interact each other increases.
Moreover, various types of spin-flip dynamics have to be introduced
to recover ergodicity for such a system.

\begin{figure}[htb]
\begin{center}
\scalebox{0.8}{\includegraphics{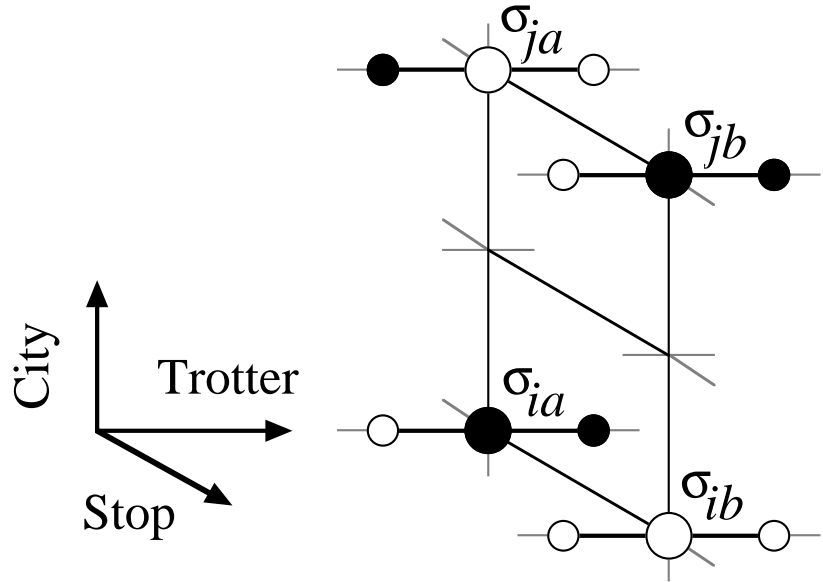}}
\end{center}
\caption{
The four spins (big circles) are flipped at a time.
The transition probability is computed from the length of the tour and
the two-spin interactions of the nearest neighbors in the Trotter
directions.
(the interactions between big and small circles)
} 
\label{fig_tsp_4spin}
\end{figure}

To avoid such difficulties, we adopt the same quantum fluctuation
we used before as $-\Gamma \sum_i \sigma_i^x$.
The same approach as in the previous chapter can be applied.
The system is mapped to a $(d+1)$-dimensional system by the
Suzuki-Trotter decomposition.
However, the spin-flip procedure of QA for TSP differs from the procedure
of QA in the previous chapter.
The four spins in a Trotter slice are flipped at a time to keep the
constraints, though one spin was flipped so far.
The spins are flipped by obeying the transition probability which is obtained
from the energy gap between the states before flip and after flip.
We computed the energy gap from the classical energy term of the
Hamiltonian (the tour length) and the interactions between the Trotter
slices for the four spins.
That is shown in Fig.~\ref{fig_tsp_4spin}. 
We can perform QA for the traveling salesman problem in this procedure.

\section{Results of Monte Carlo Simulations}
\label{sec:4.4}

\begin{figure}[htb]
\scalebox{0.5}{\includegraphics{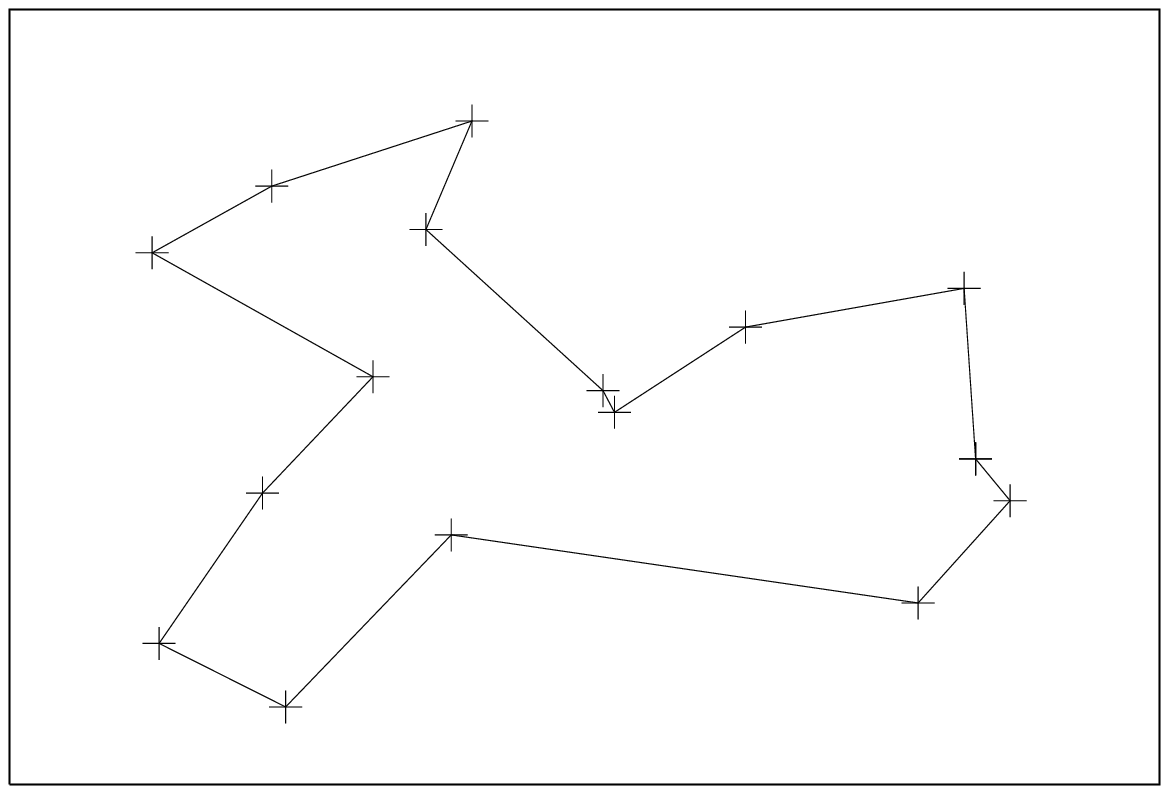}}
\scalebox{0.5}{\includegraphics{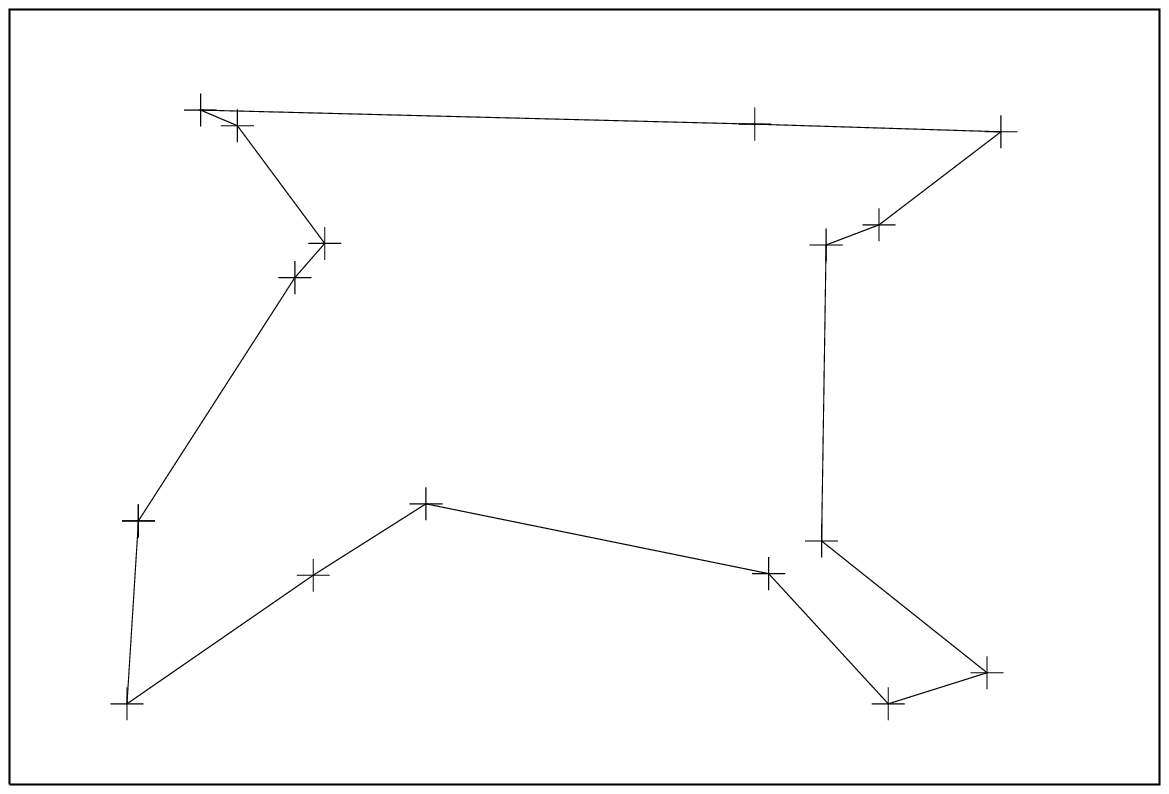}}
\\
\scalebox{0.5}{\includegraphics{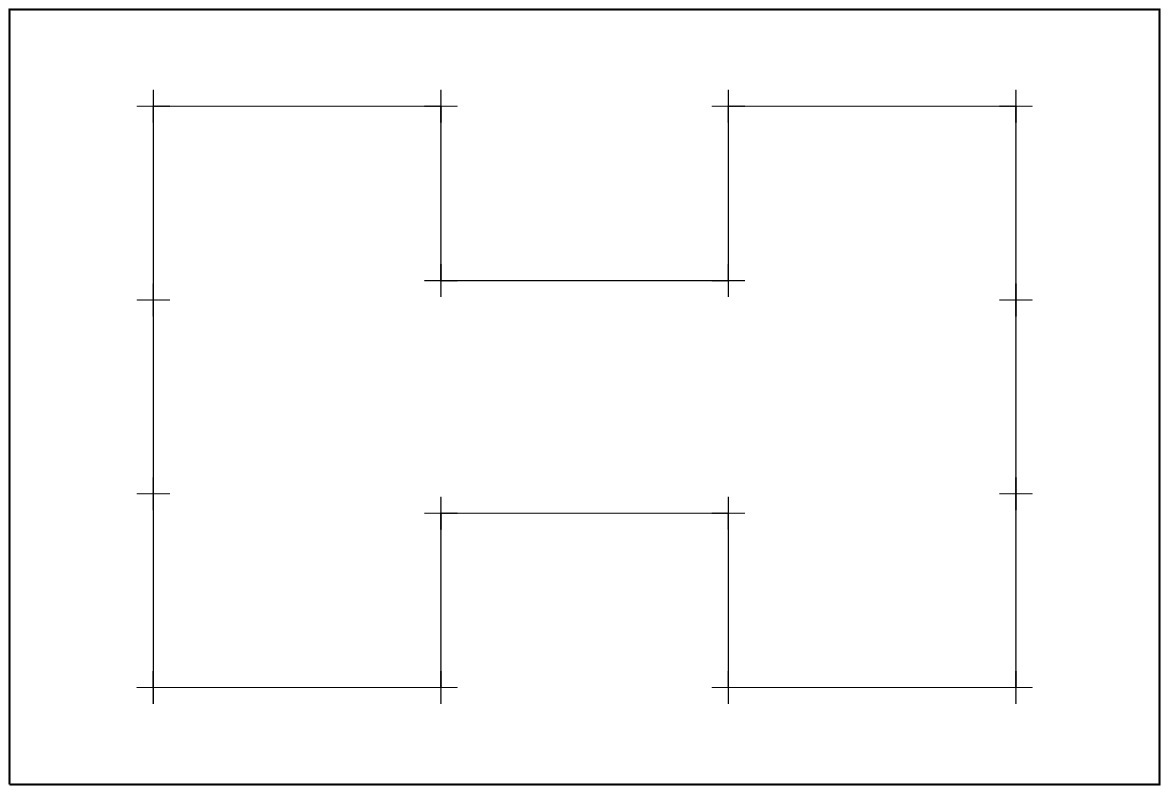}}
\scalebox{0.5}{\includegraphics{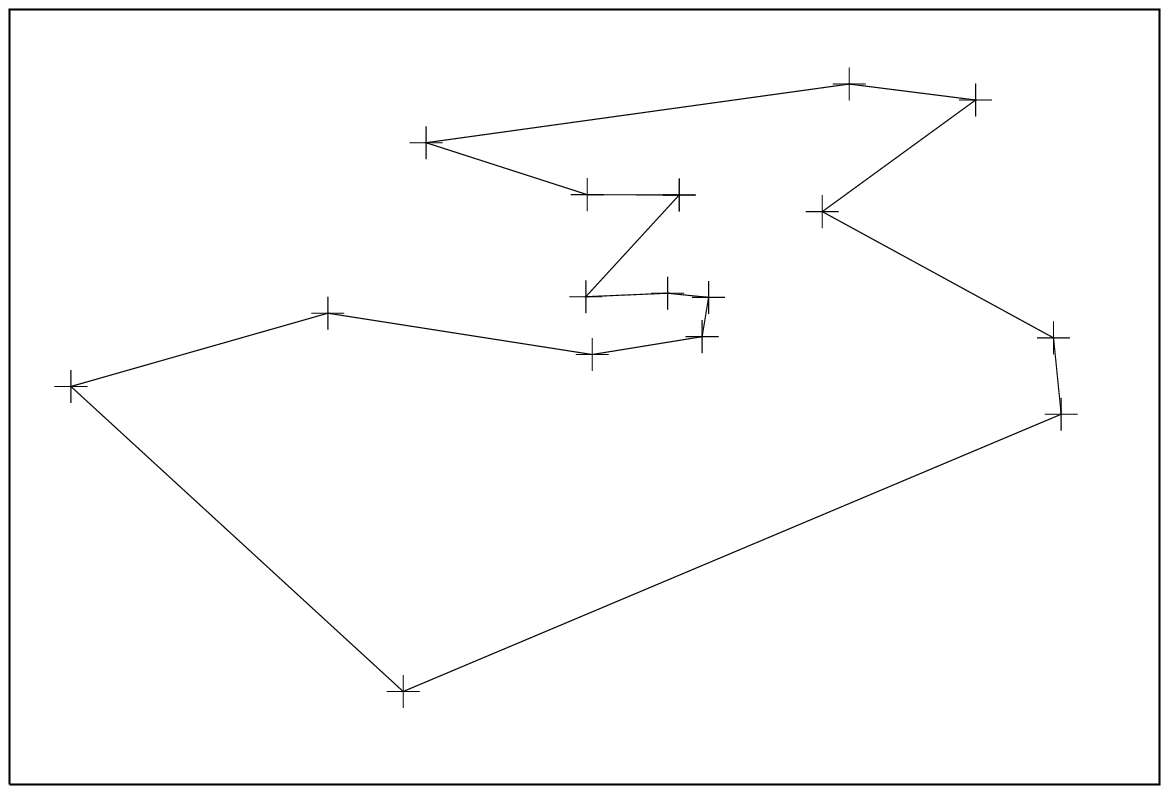}}
\caption{
Optimal tours of the four problems.
} 
\label{fig_tsp_opt}
\end{figure}

We perform SA and QA for the four cases of TSP whose optimal tours
are shown in Fig.~\ref{fig_tsp_opt}.
The first problem ``random'' is the case in which the cities are located at
random.
Cities are located at random but clumped with dense and spares regions
for the second problem ``semi-random''.
The optimal route of the third problem has an ``H'' shape.
(we call this problem ``H-character'' hereafter.)
The fourth problem is ``ulysses16'' of TSPLIB~\cite{TSPLIB}.
The number of cities is $N=16$ for all the problems.

The cities are located on a square with the side length $\sqrt{N}$ to make the
length of the tour extensive for ``random'', ``semi-random'' and
``H-character''.
For ``ulysses16'', we re-scale $d_{ij}$ and set the average to $2.2$.
The average, the dispersion and the ratio of the dispersion and the
average are shown in Table~\ref{table_1}.

\begin{table}
\begin{center}
\begin{tabular}[tb]{|l||r|r|r|r|} \hline
               & random & semi-random & H-character & ulysses16 \\
 \hline \hline
average (a)    & 2.195  & 2.642       & 2.155       & 2.200 \\
 \hline
dispersion (b) & 0.973  & 1.152       & 0.867       & 1.556 \\
 \hline
b/a            & 0.443  & 0.436       & 0.402       & 0.707 \\
 \hline
\end{tabular}
\end{center}
\caption{
The average, the dispersion and the ratio of the dispersion and the
average for the four problems.
}
\label{table_1}
\end{table}

\begin{figure}[p]
\scalebox{0.5}{\includegraphics{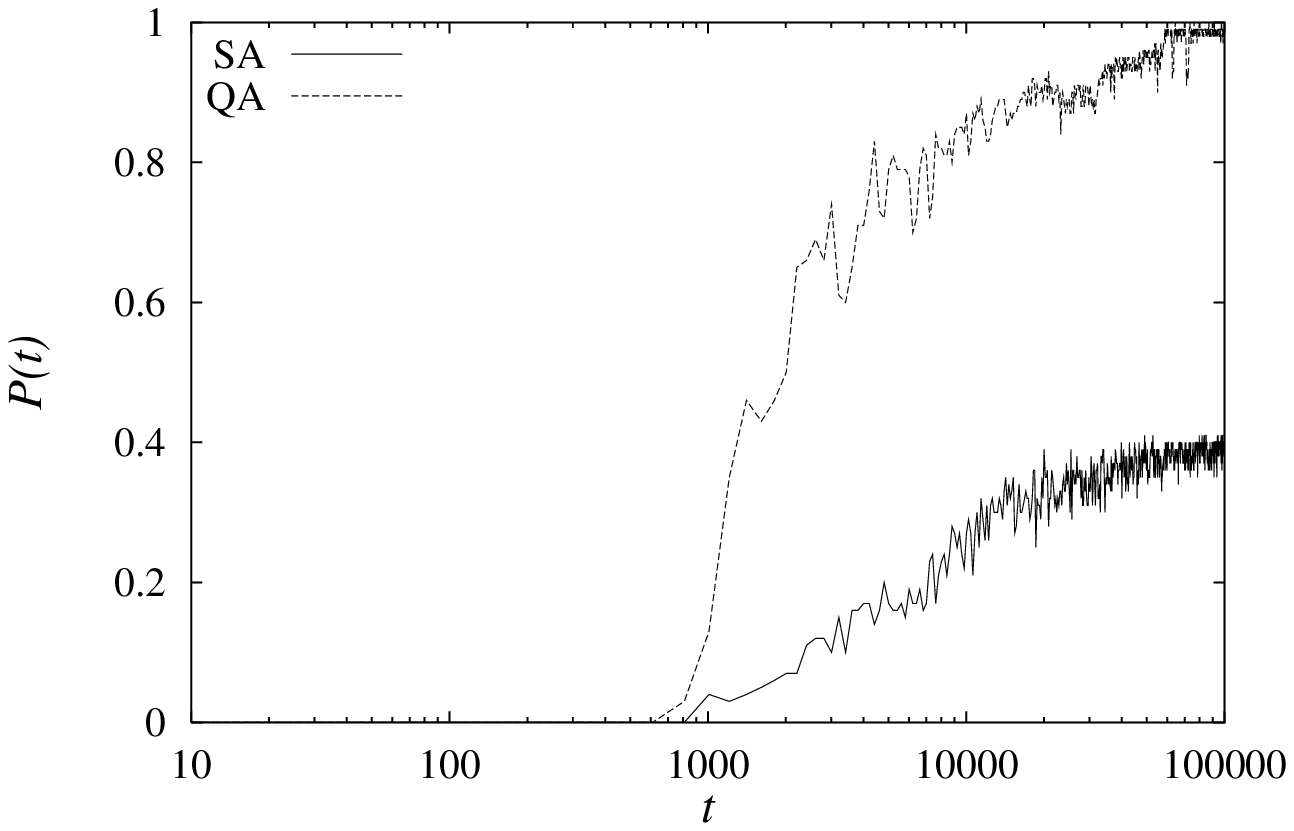}}
\scalebox{0.5}{\includegraphics{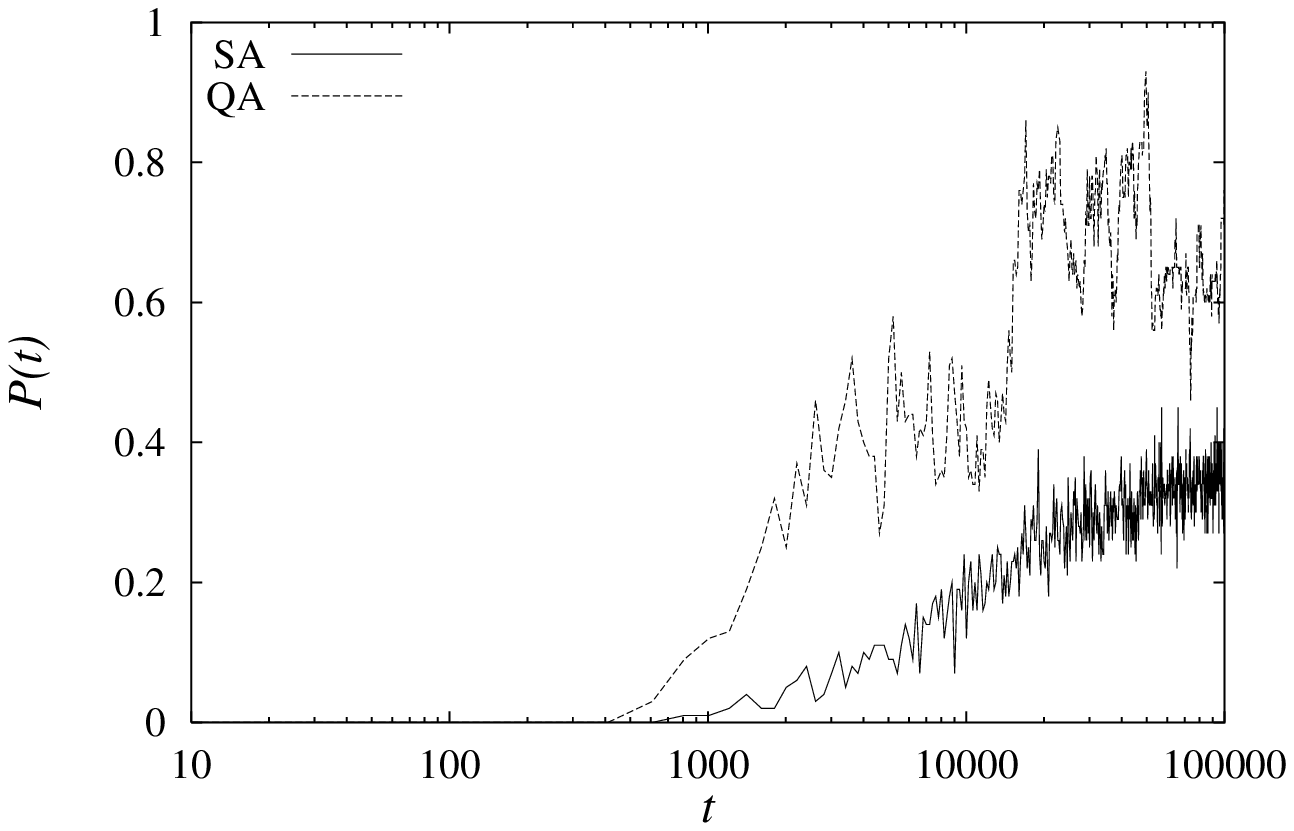}}
\\
\scalebox{0.5}{\includegraphics{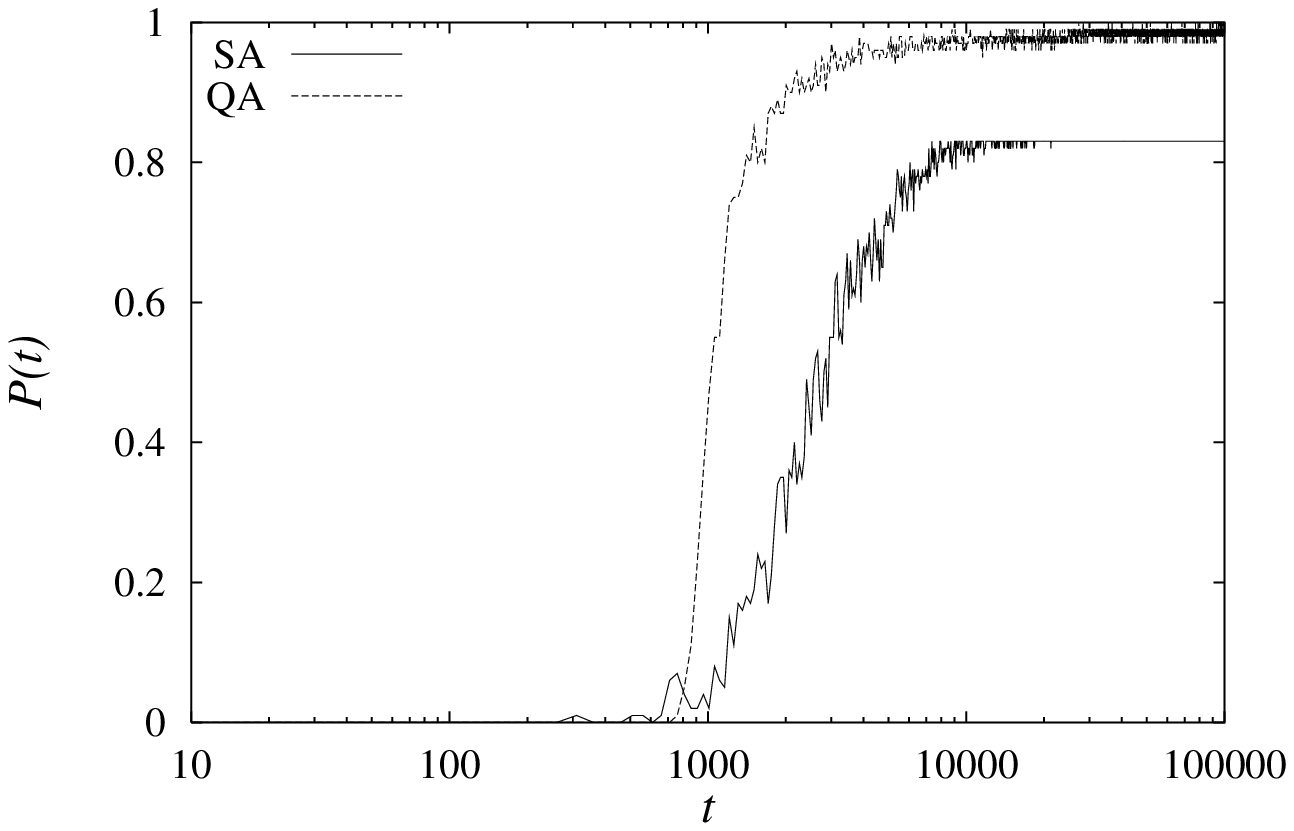}}
\scalebox{0.5}{\includegraphics{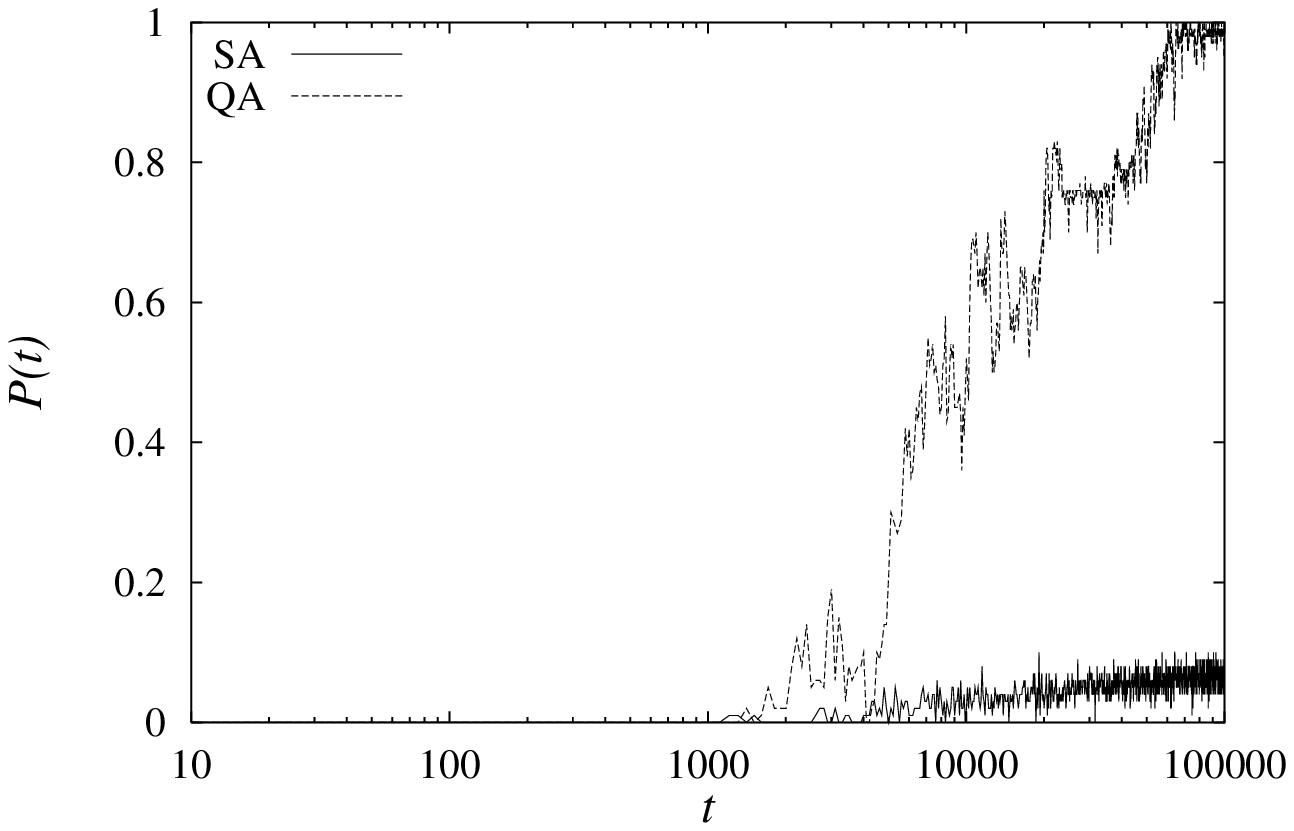}}
\caption{
Probability to find the minimum-length.
The scheduling is $10/\sqrt{t}$.
} 
\label{fig_tsp_p_1}
\end{figure}

\begin{figure}[p]
\scalebox{0.5}{\includegraphics{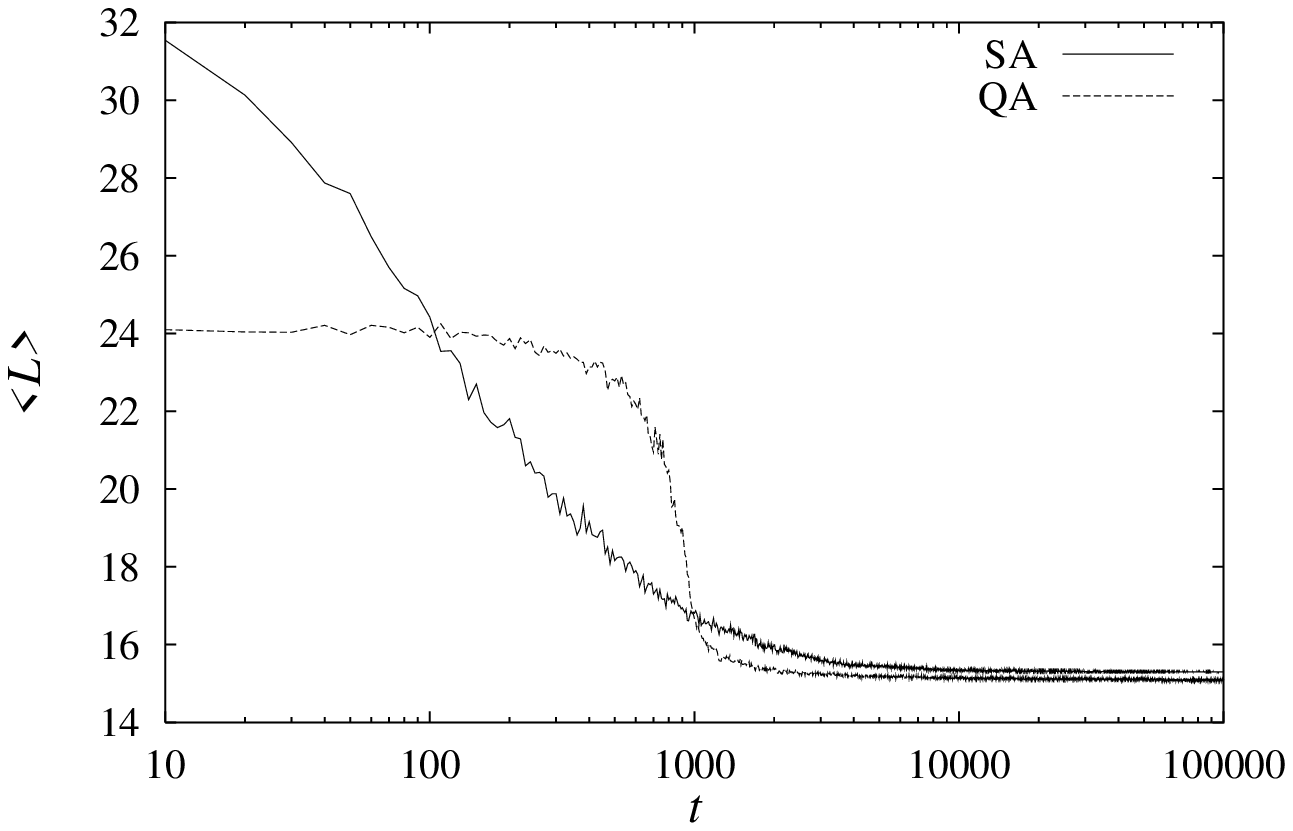}}
\scalebox{0.5}{\includegraphics{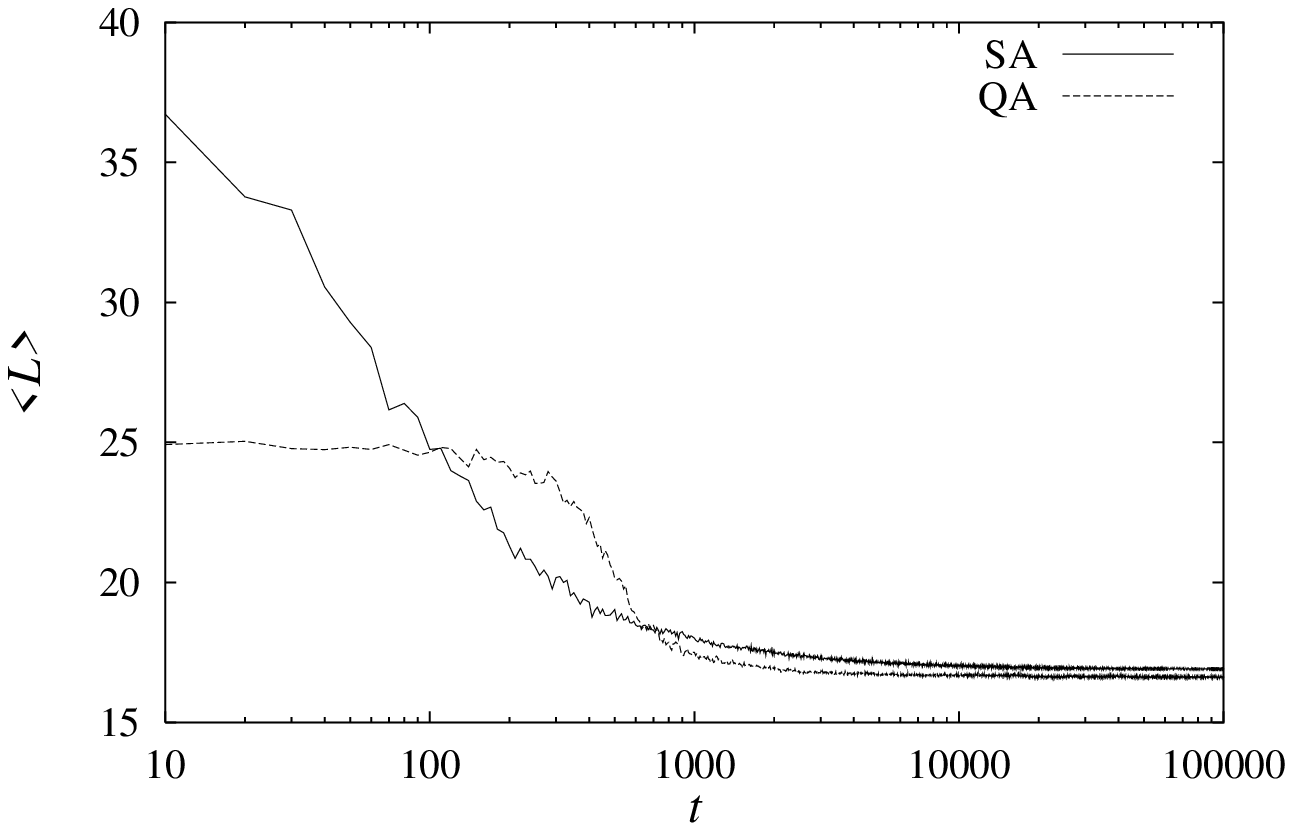}}
\\
\scalebox{0.5}{\includegraphics{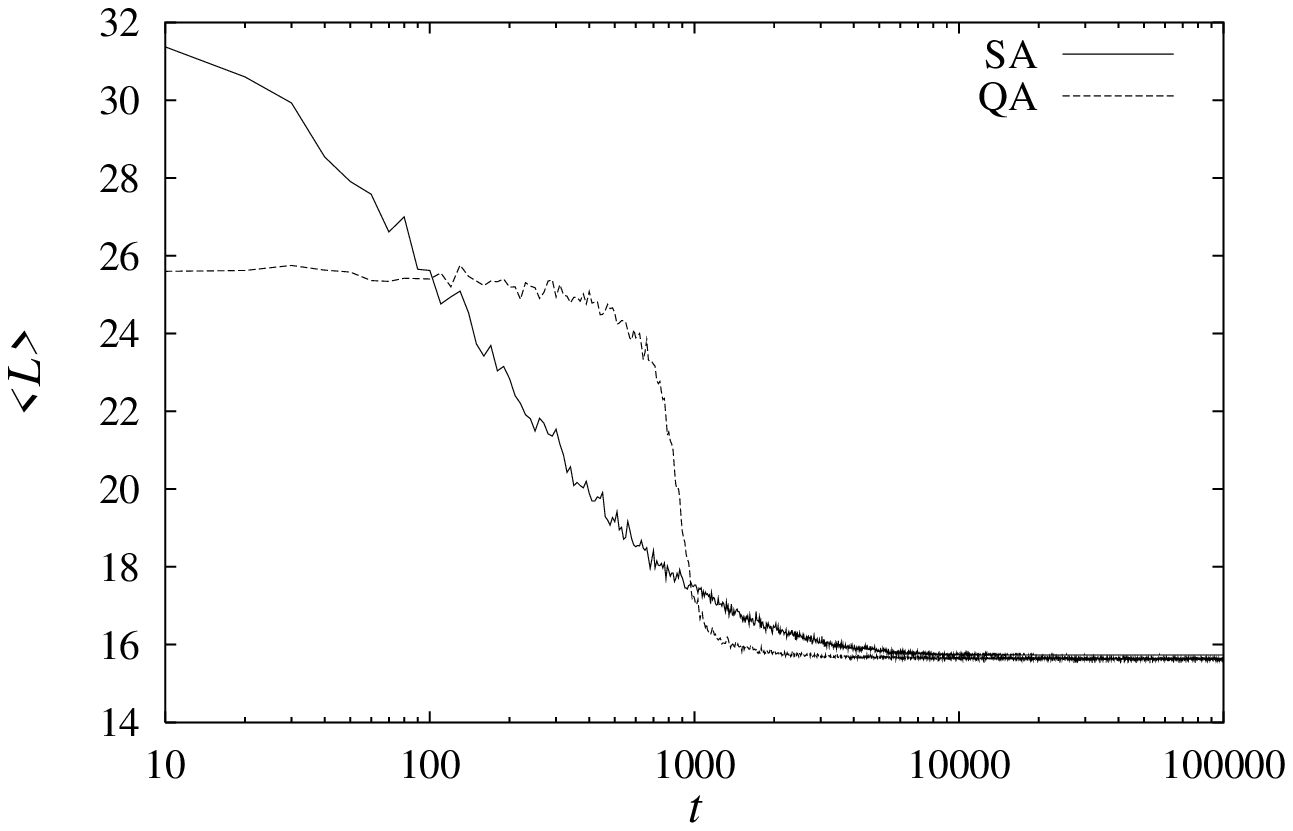}}
\scalebox{0.5}{\includegraphics{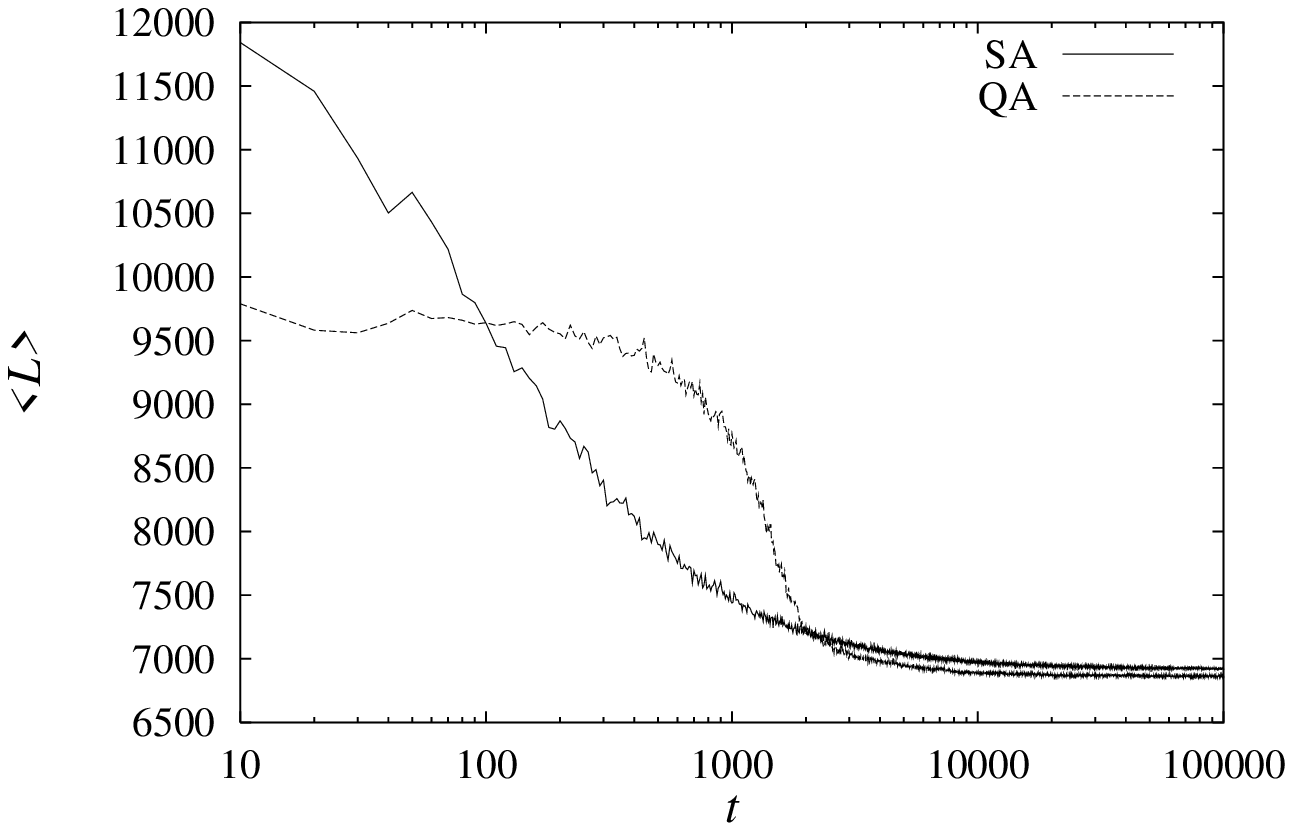}}
\caption{
Average of the length.
The scheduling is $10/\sqrt{t}$.
} 
\label{fig_tsp_l_1}
\end{figure}

\begin{figure}[p]
\scalebox{0.5}{\includegraphics{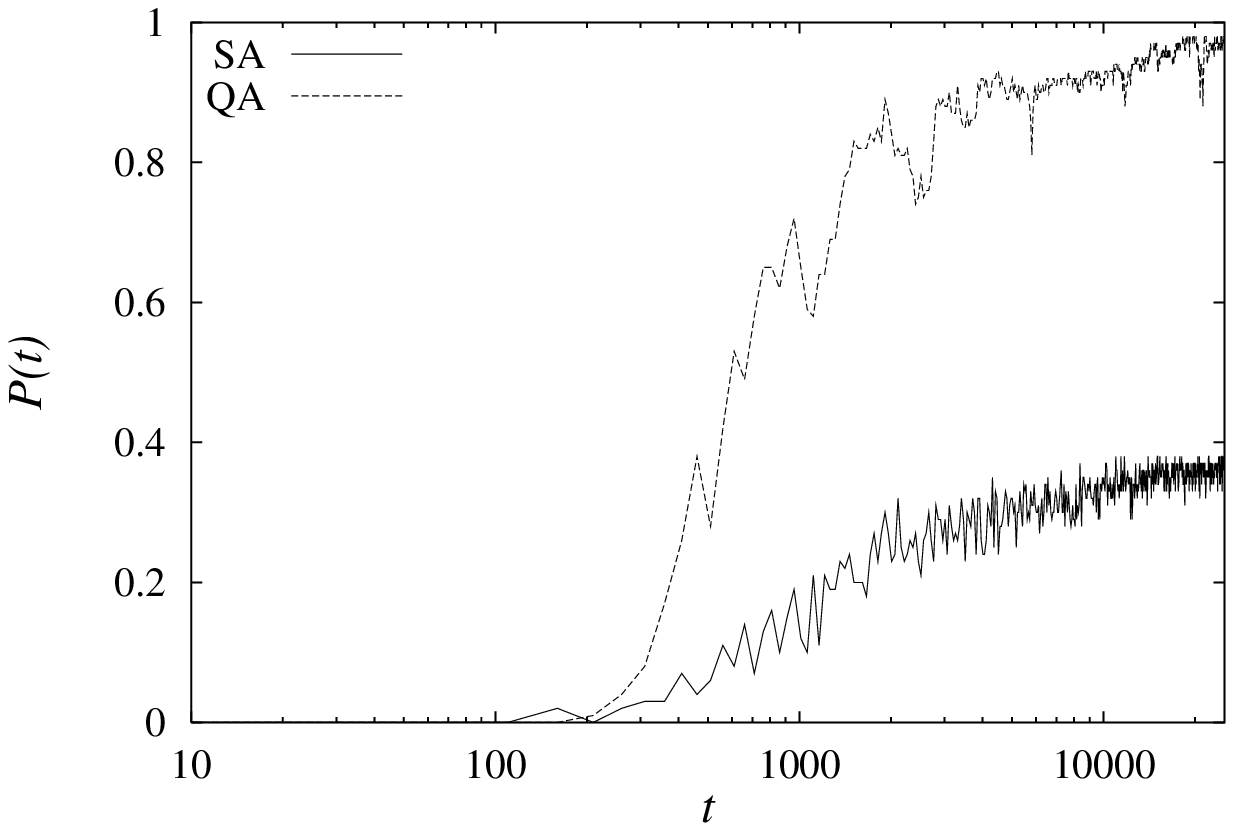}}
\scalebox{0.5}{\includegraphics{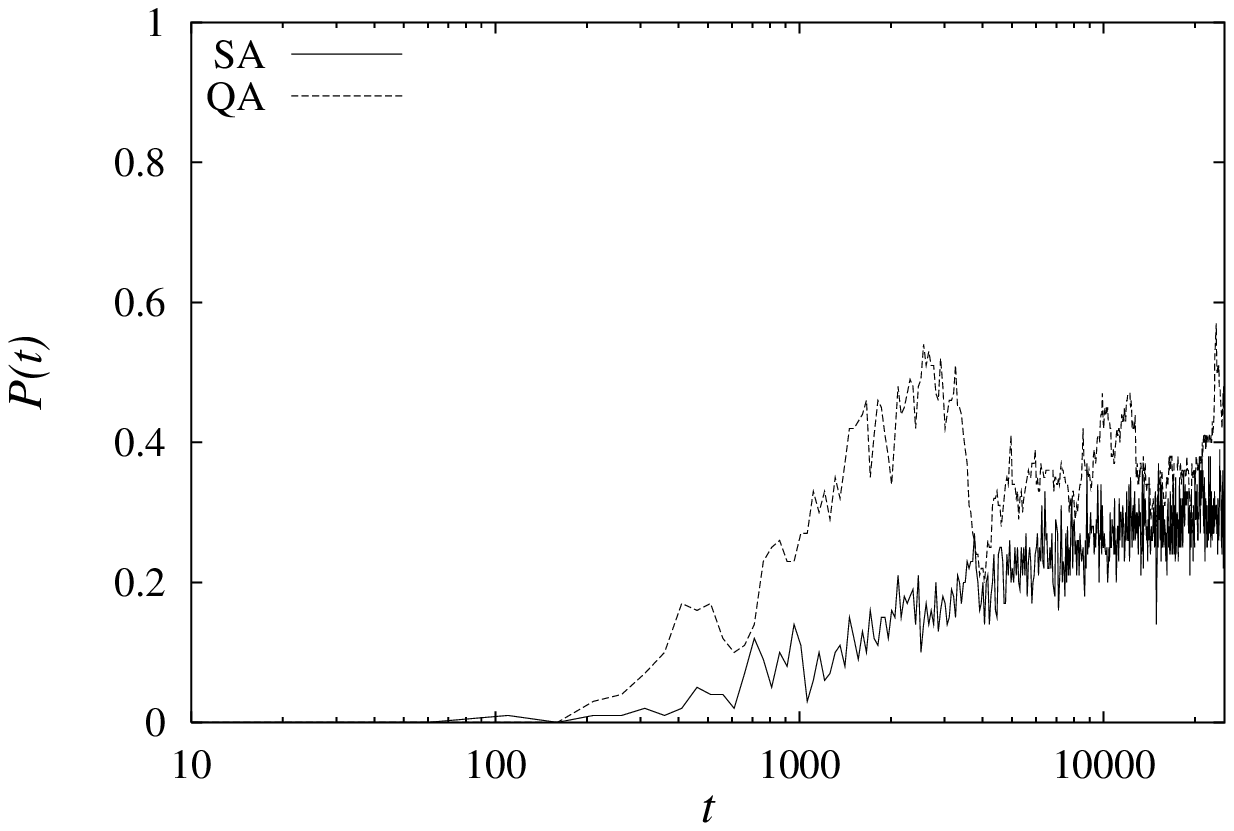}}
\\
\scalebox{0.5}{\includegraphics{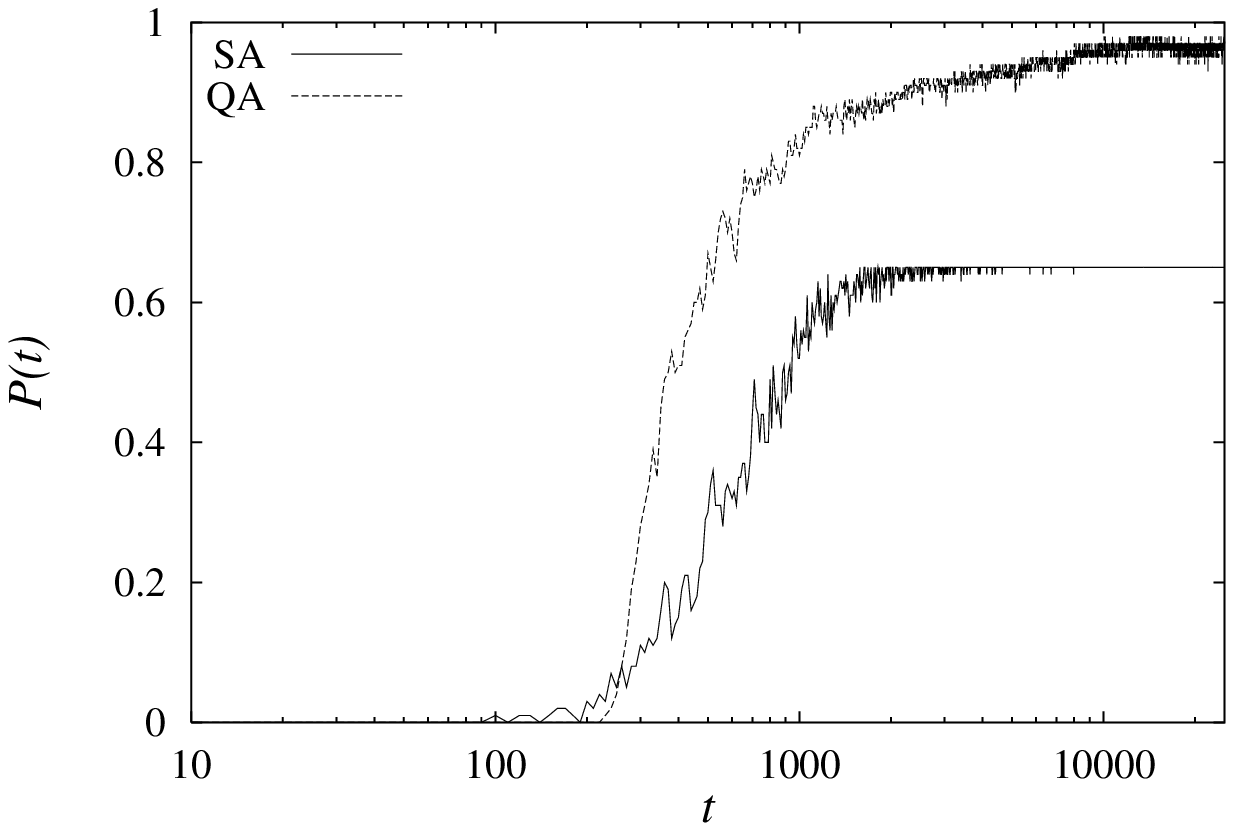}}
\scalebox{0.5}{\includegraphics{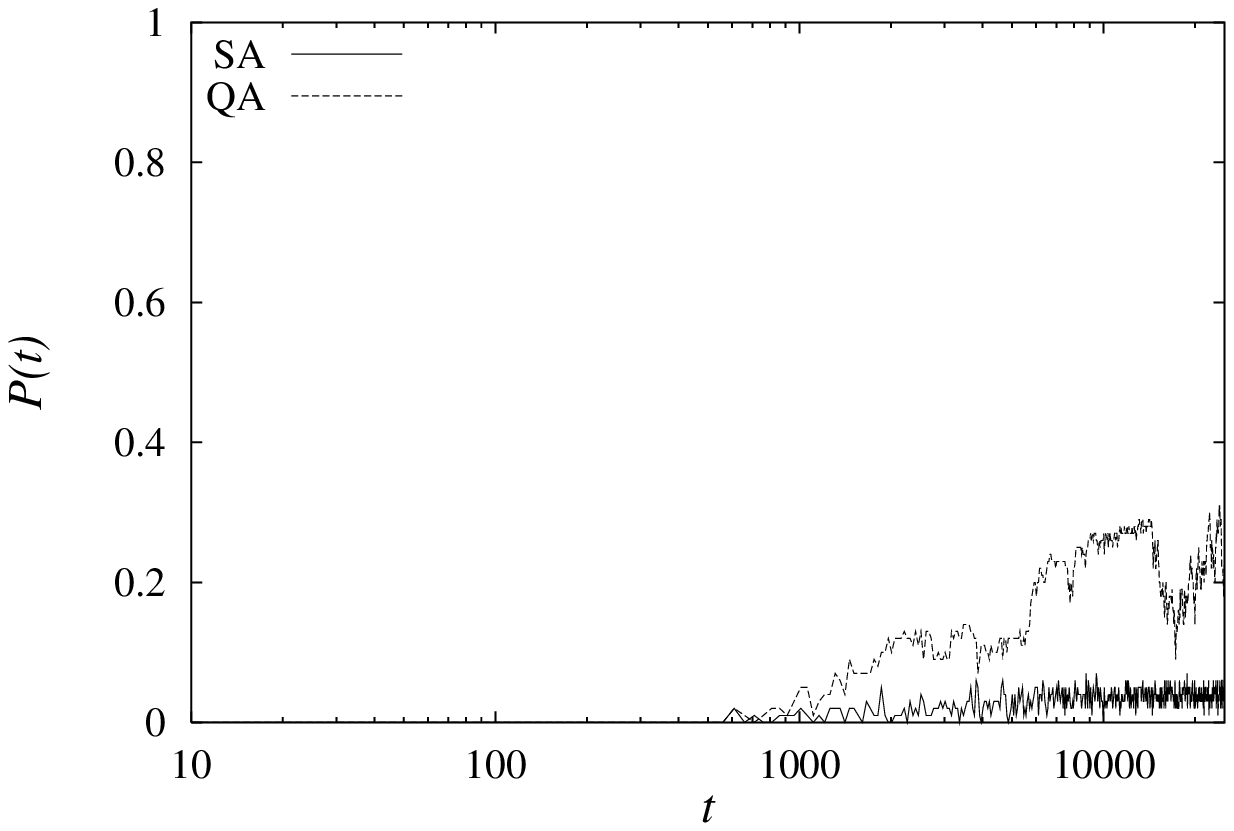}}
\caption{
Probability to find the minimum-length.
The scheduling is $5/\sqrt{t}$.
} 
\label{fig_tsp_p_2}
\end{figure}

\begin{figure}[p]
\scalebox{0.5}{\includegraphics{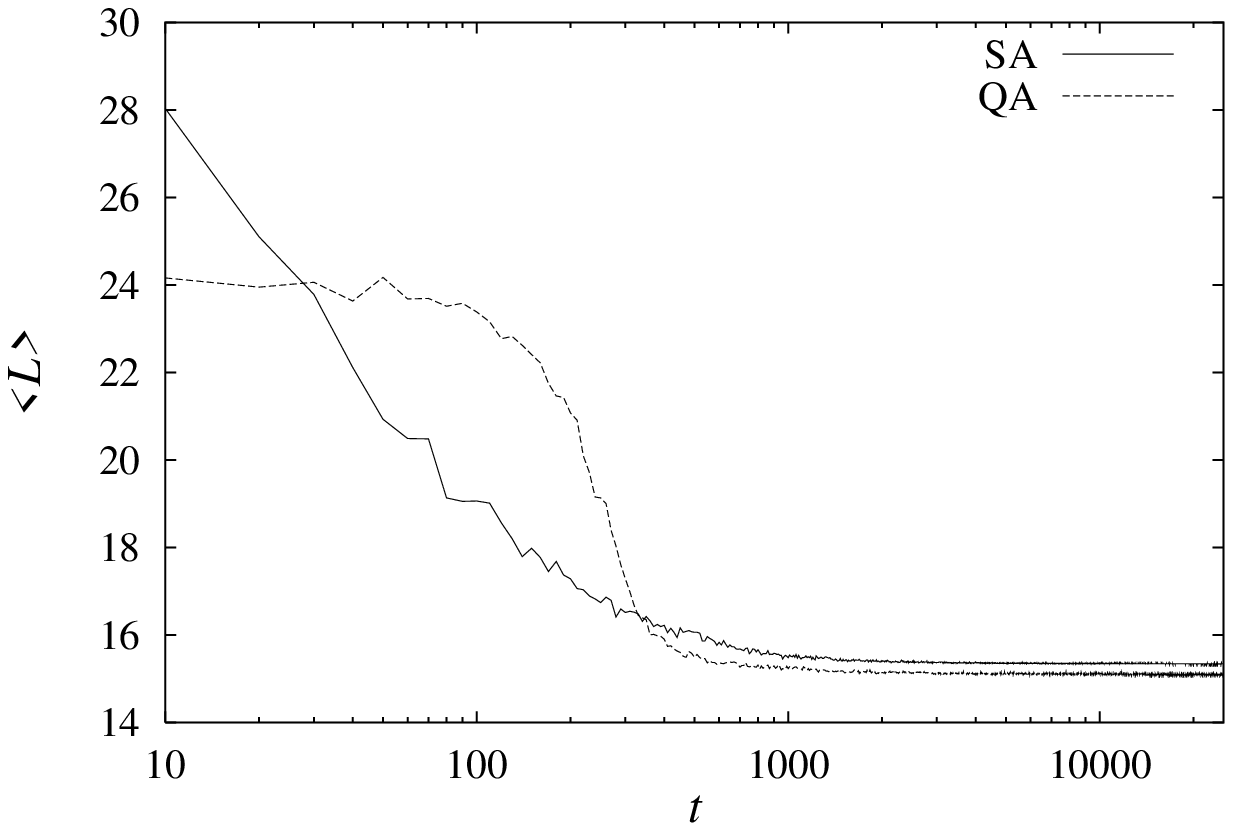}}
\scalebox{0.5}{\includegraphics{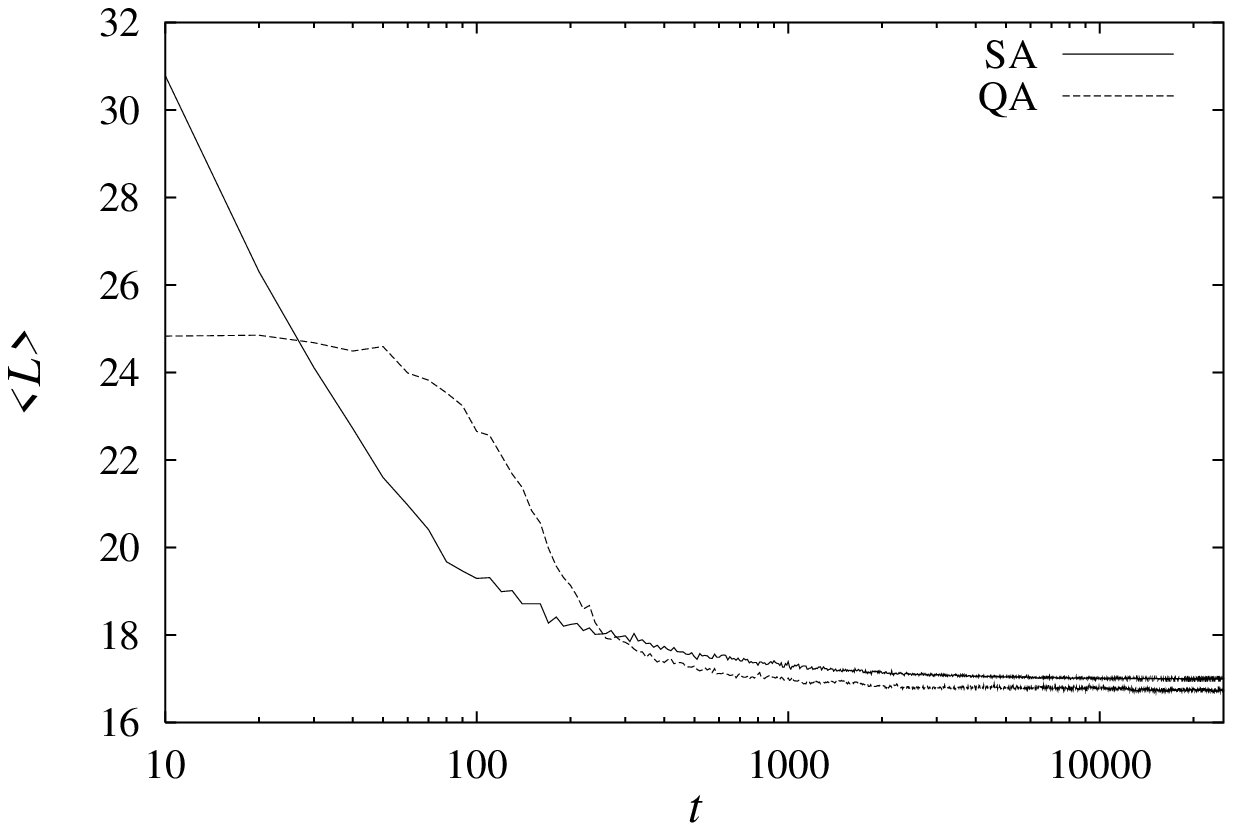}}
\\
\scalebox{0.5}{\includegraphics{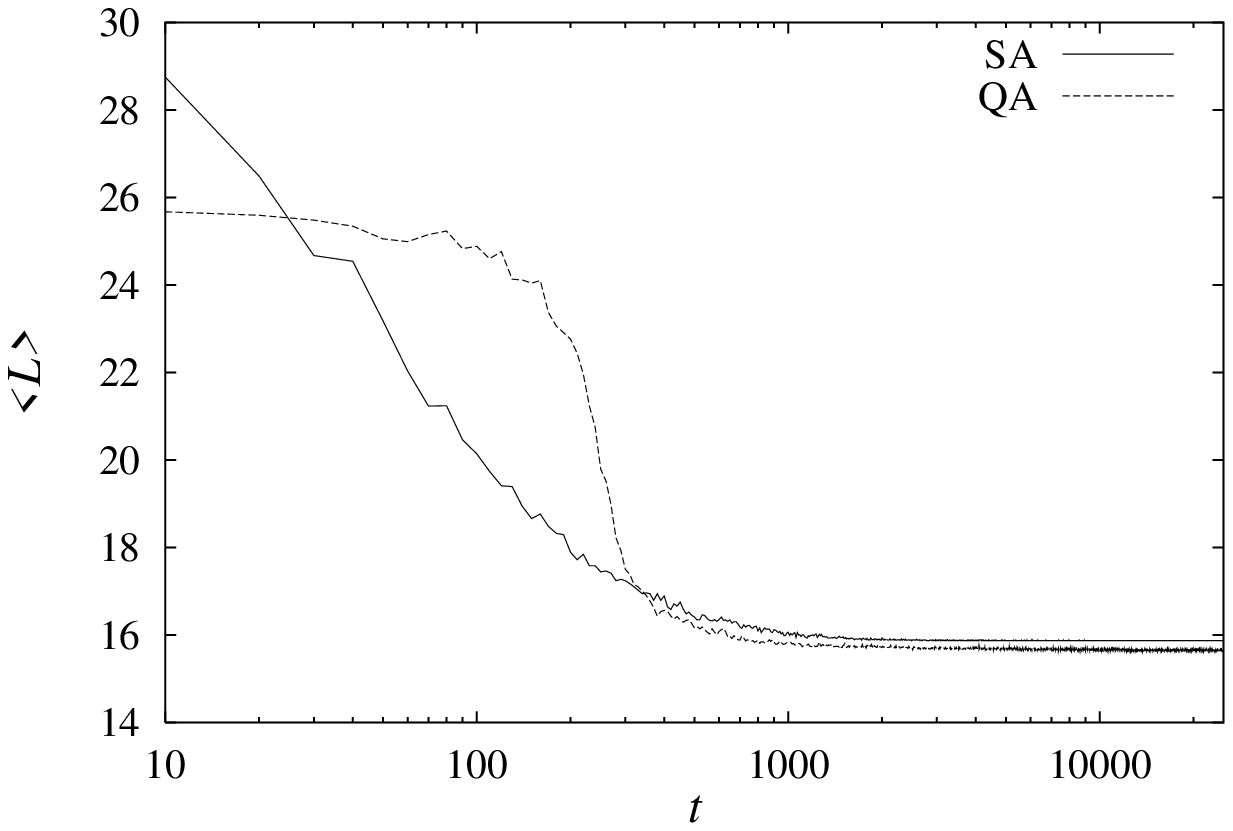}}
\scalebox{0.5}{\includegraphics{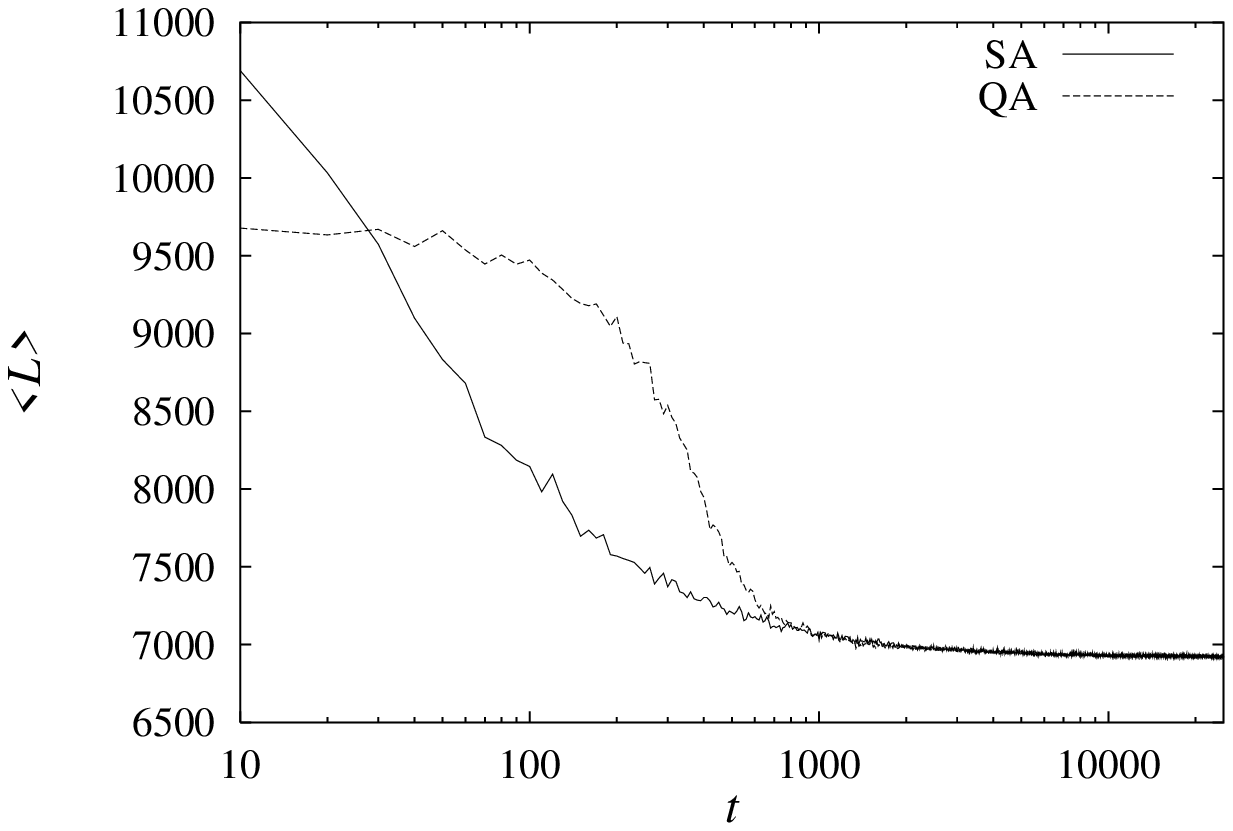}}
\caption{
Average of the length.
The scheduling is $5/\sqrt{t}$.
} 
\label{fig_tsp_l_2}
\end{figure}

\begin{figure}[p]
\scalebox{0.5}{\includegraphics{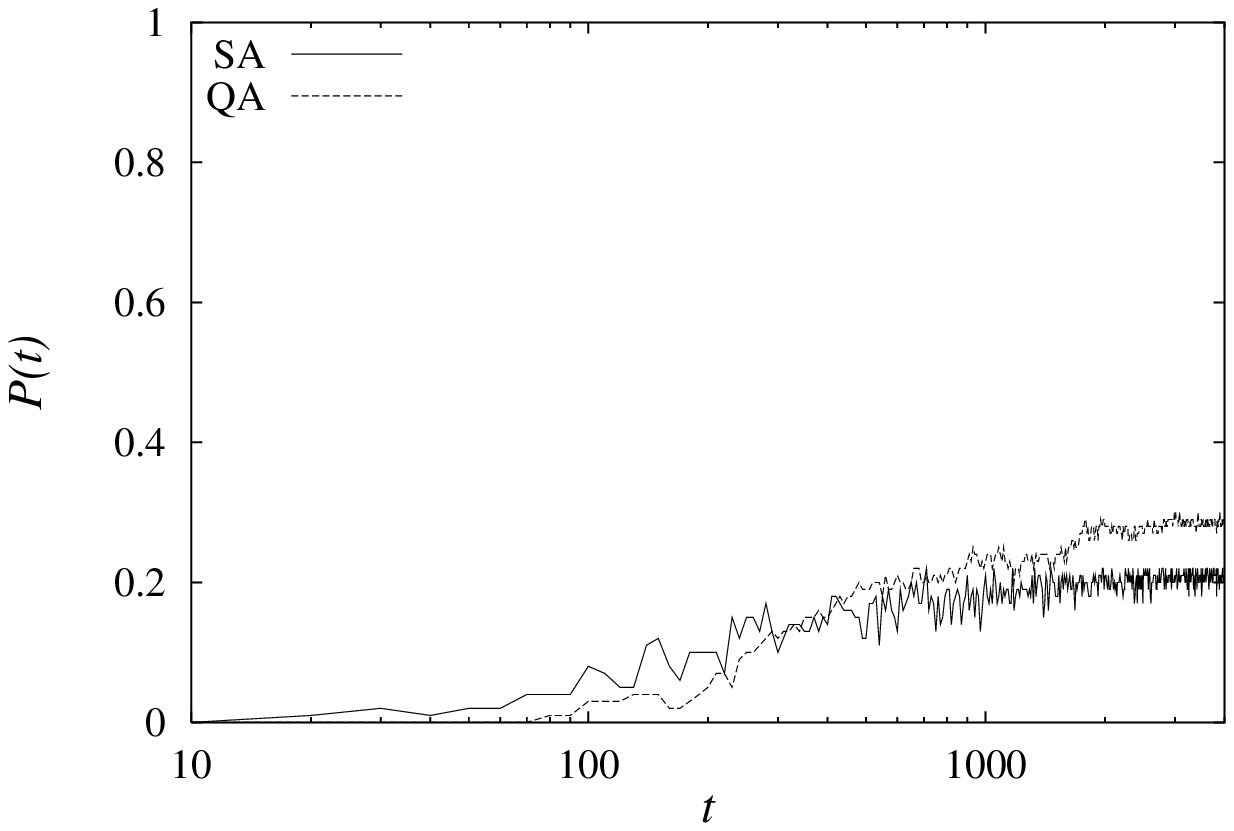}}
\scalebox{0.5}{\includegraphics{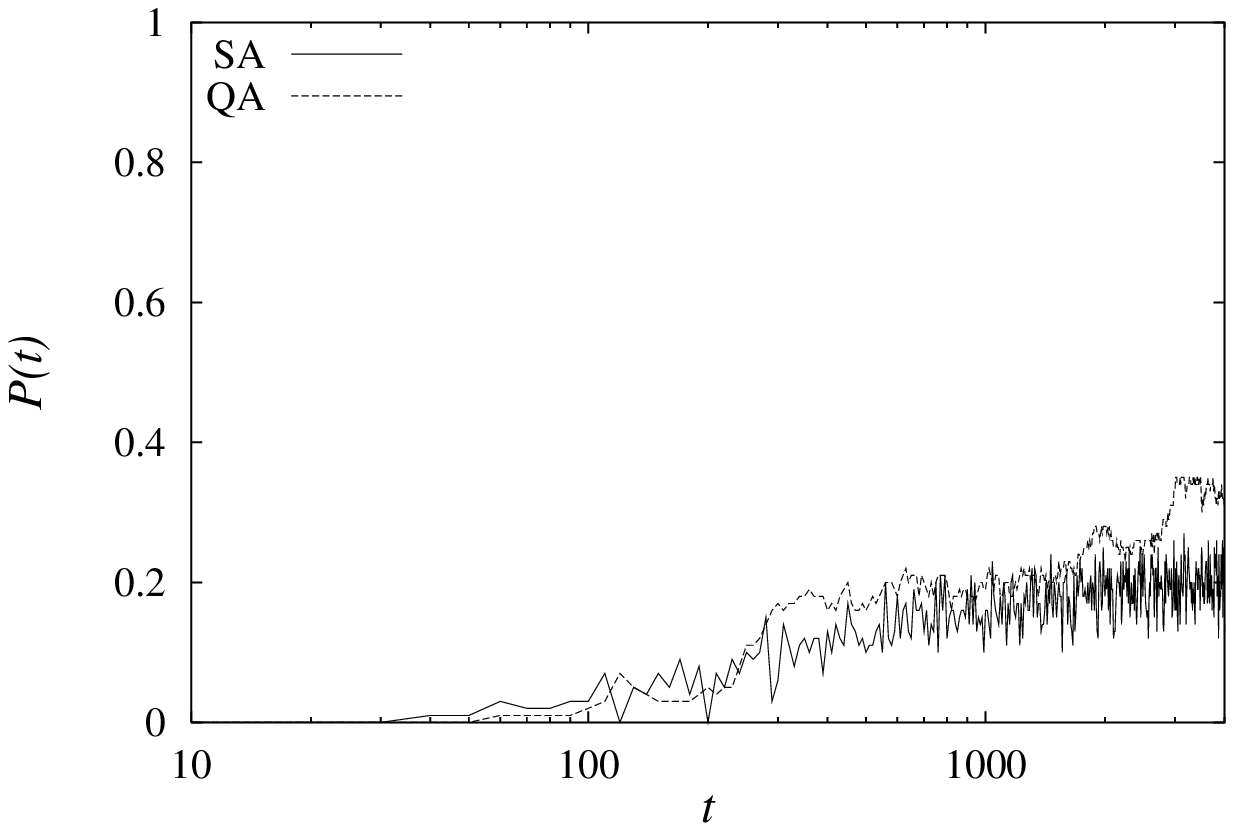}}
\\
\scalebox{0.5}{\includegraphics{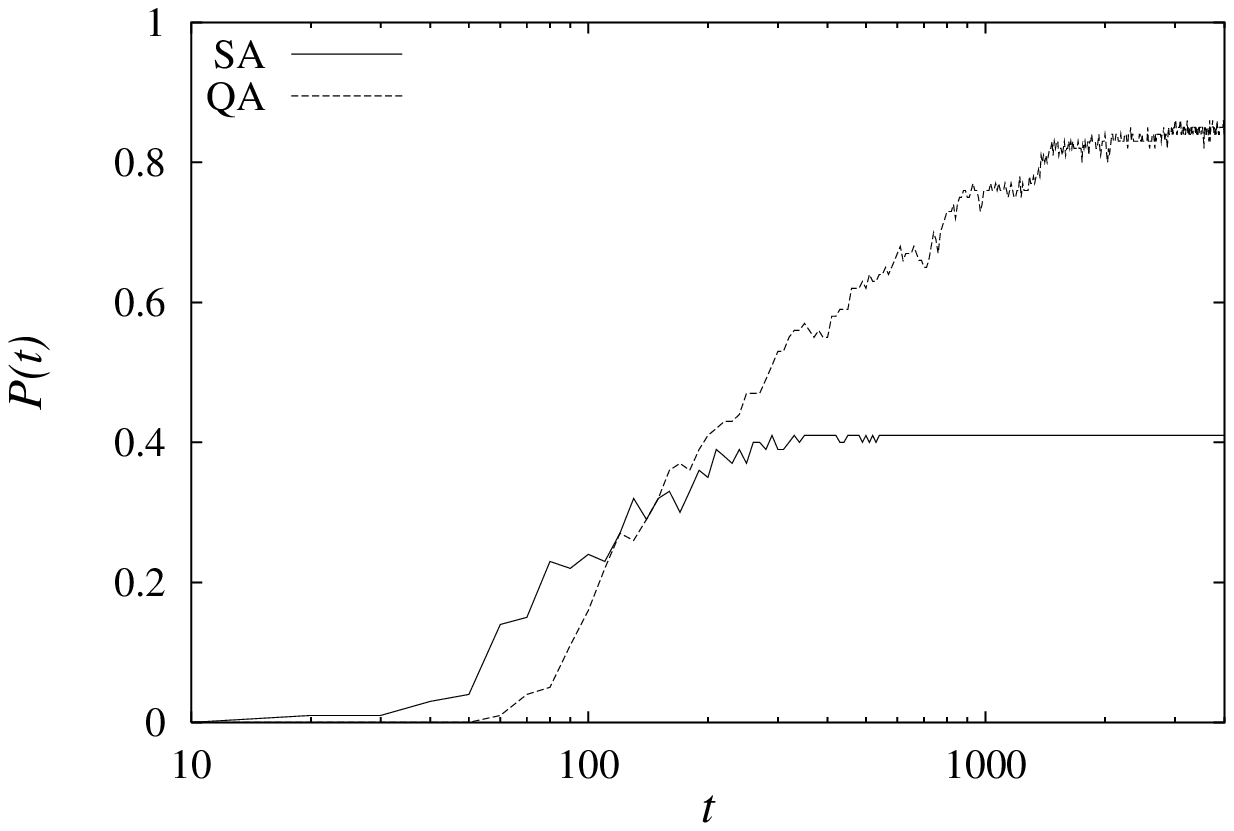}}
\scalebox{0.5}{\includegraphics{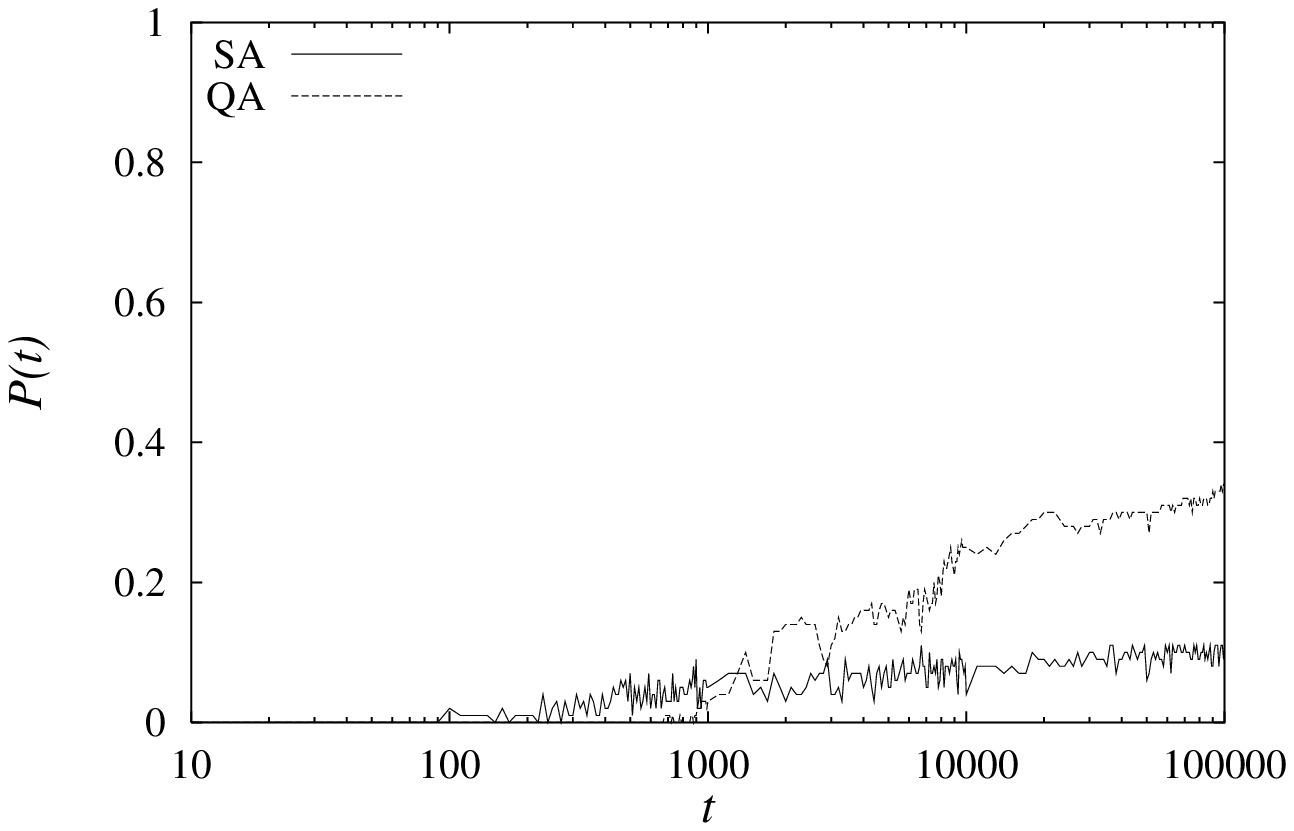}}
\caption{
Probability to find the minimum-length.
The scheduling is $2/\sqrt{t}$.
} 
\label{fig_tsp_p_3}
\end{figure}

\begin{figure}[p]
\scalebox{0.5}{\includegraphics{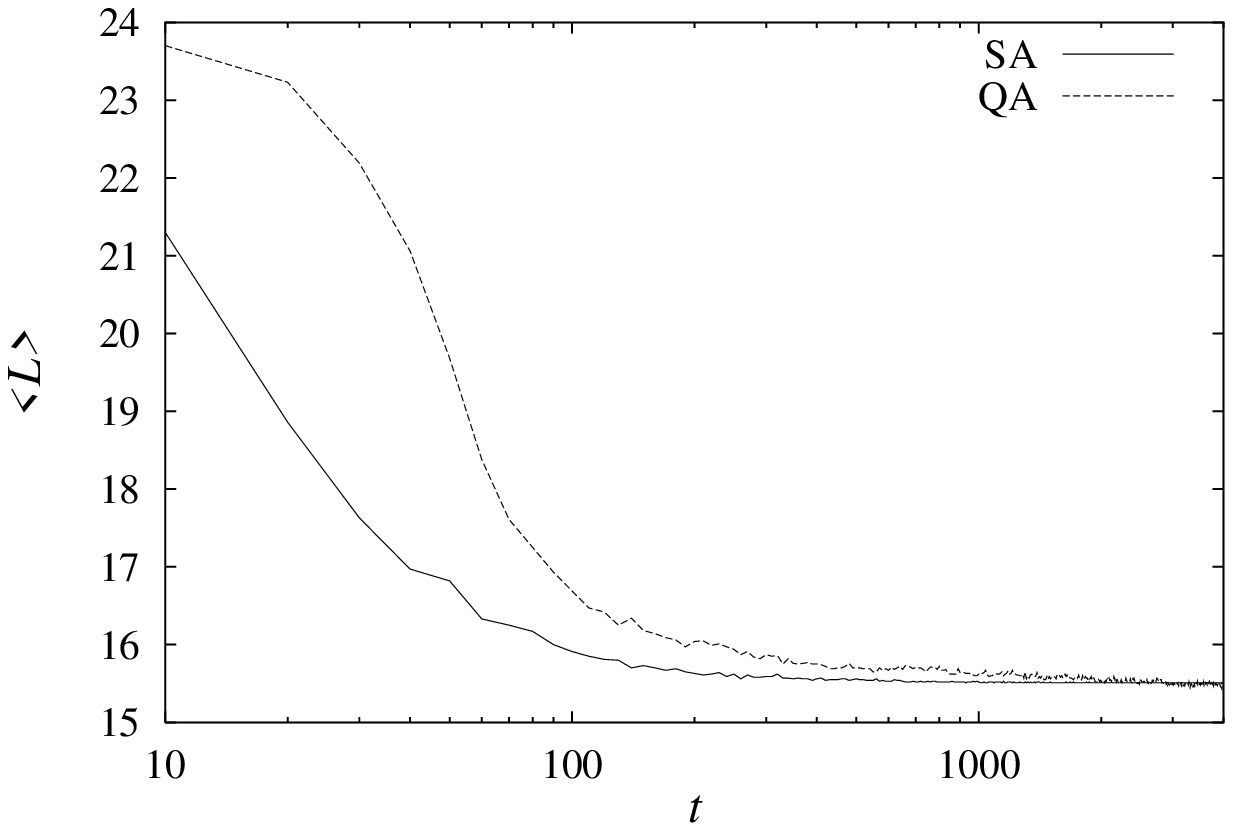}}
\scalebox{0.5}{\includegraphics{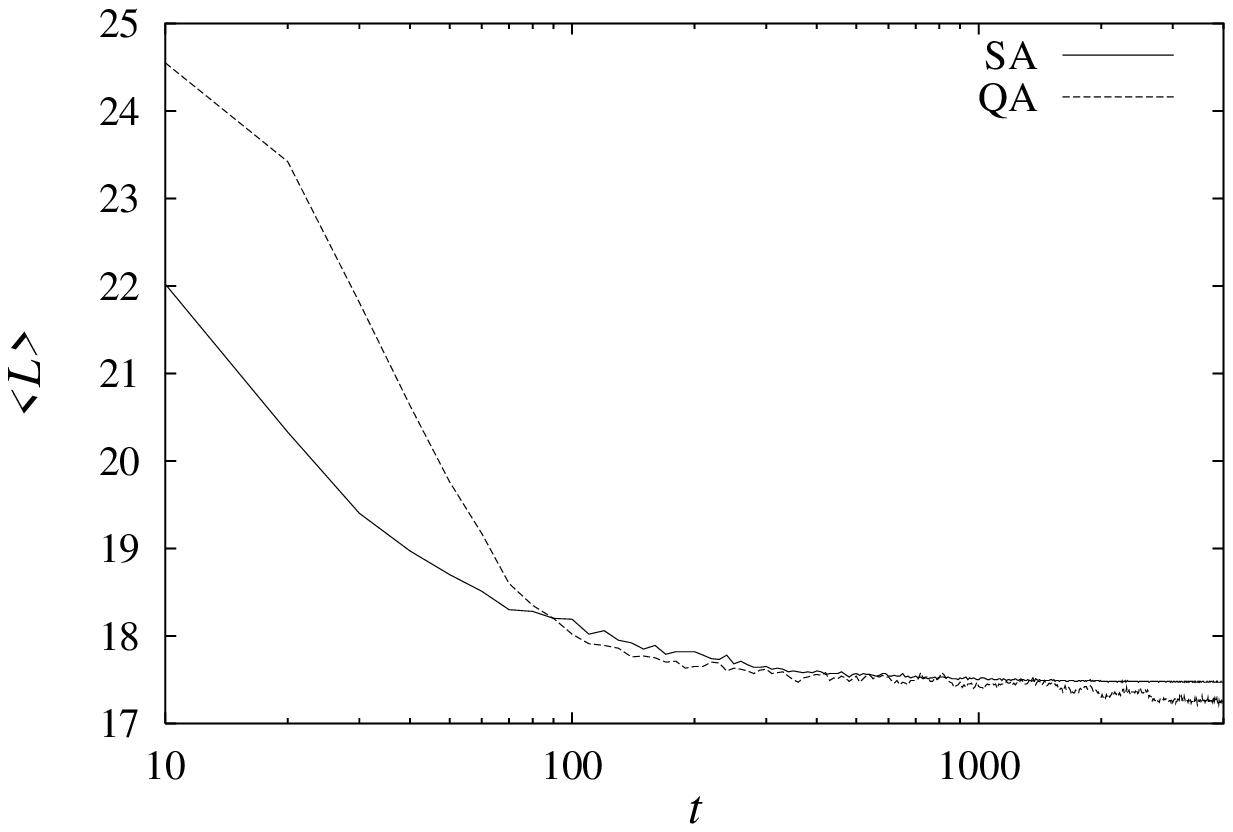}}
\\
\scalebox{0.5}{\includegraphics{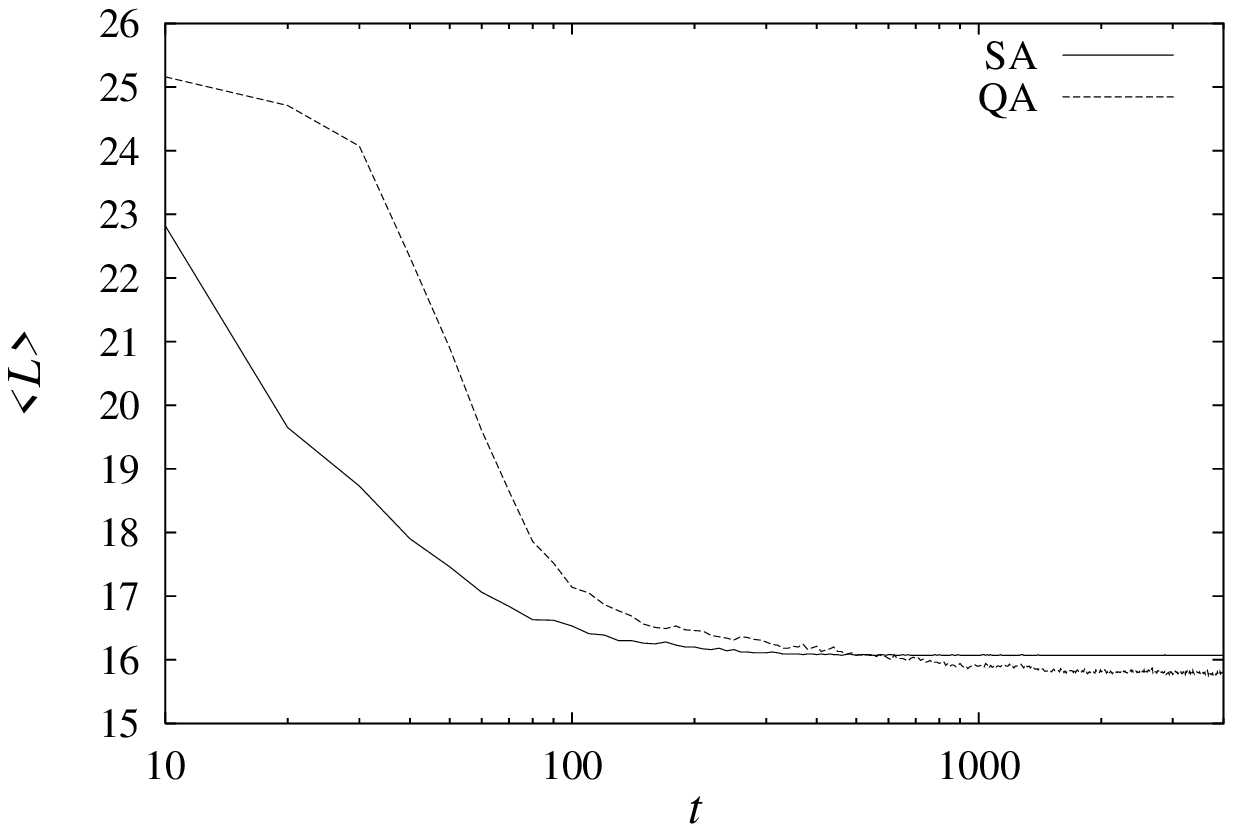}}
\scalebox{0.5}{\includegraphics{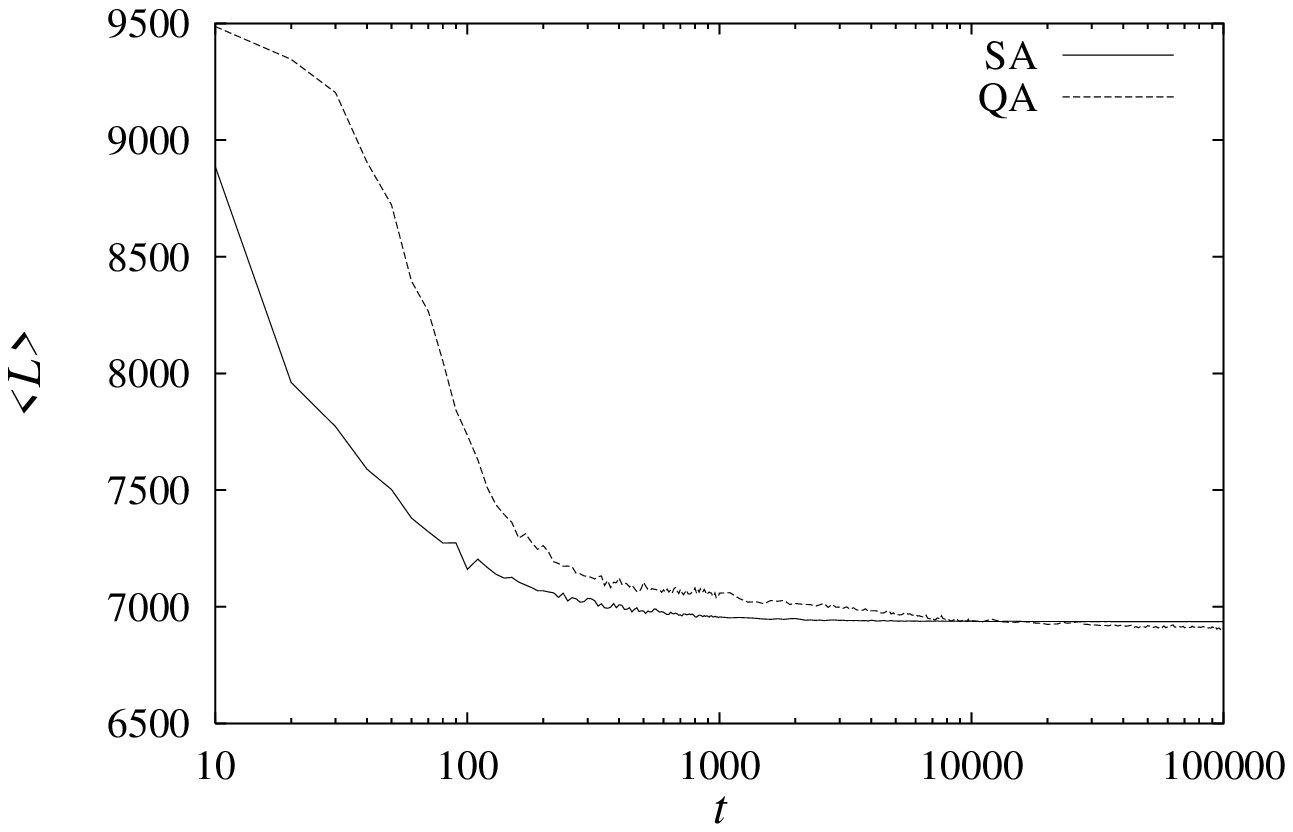}}
\caption{
Average of the length.
The scheduling is $2/\sqrt{t}$.
} 
\label{fig_tsp_l_3}
\end{figure}

Simulations are performed with $500$ Trotter slices for QA and $100$
independent runs for SA.
We observe two quantities, the probability to find the minimum-length
$P(t)$ and the average of length $\langle L \rangle$.
The probability $P(t)$ is obtained from the ratio of the number of the
ground state configurations for each time in $100$ independent runs for
SA and $500$ Trotter slice for QA.
The probabilities of SA and QA for the four problems with the $10/\sqrt{t}$
scheduling are plotted in Fig.~\ref{fig_tsp_p_1} and the energy values are
in Fig.~\ref{fig_tsp_l_1}.
The plots are located in the same order as Fig.~\ref{fig_tsp_opt}.
In this schedule, QA has a better performance than SA for all the four
problems.
Moreover, we also calculated for the case of $5/\sqrt{t}$ and $2/\sqrt{t}$
schedules (see Fig.~\ref{fig_tsp_p_2}--\ref{fig_tsp_l_3}).
Results show that QA is better than SA in both of the average length
and the probability for all problems.
The probabilities of SA for the ``H-character'' problem are obviously
saturated and fall short of probability one, and the probability of QA
almost reaches one.
This is similar to the situation in the differential equations for
small-size systems for SA and QA.

\begin{figure}[htb]
\scalebox{1}{\includegraphics{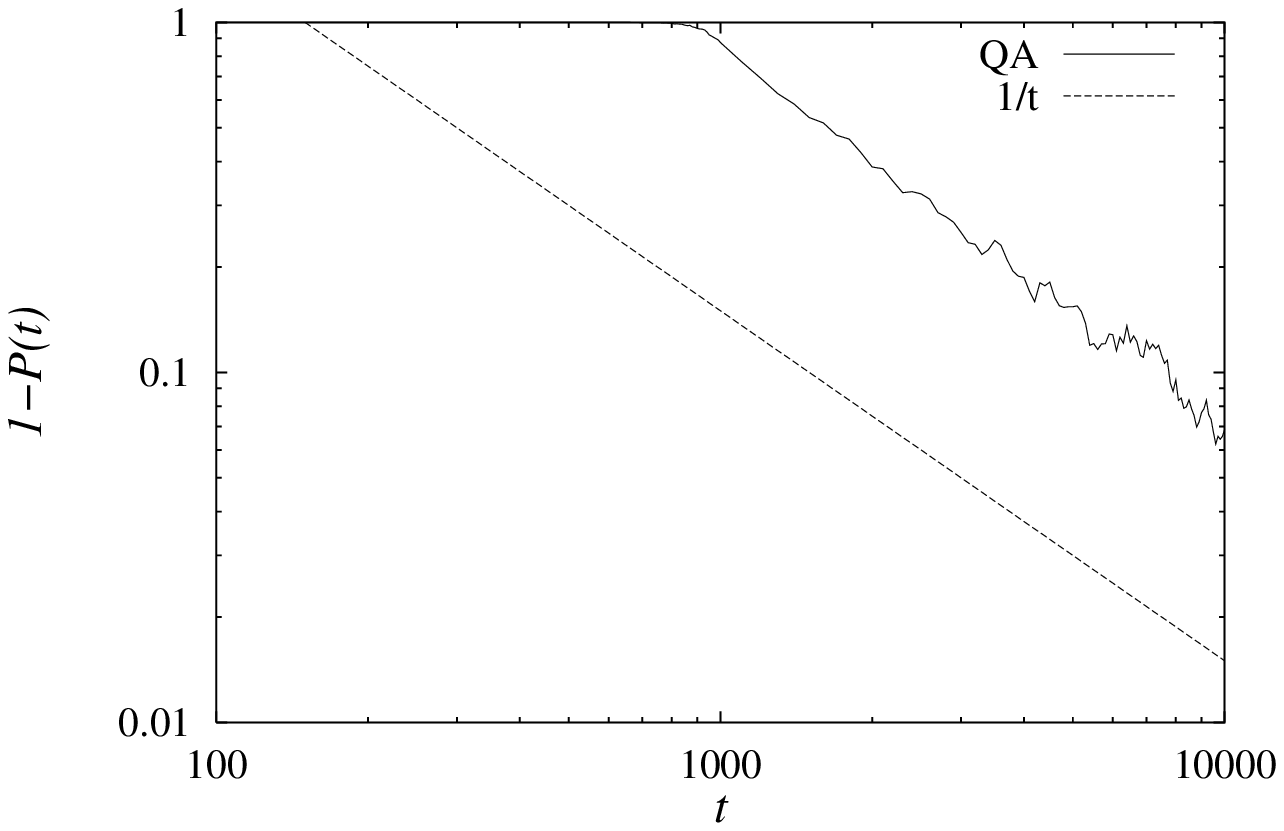}}
\caption{
The quantity $1-P(t)$ is plotted.
The transverse field is scheduled as $\Gamma=10/\sqrt{t}$.
} 
\label{fig_rand_1-p}
\end{figure}

Next, we investigate the asymptotic property of finding the optimal
state.
The quantity $1-P(t)$ is plotted in Fig.~\ref{fig_rand_1-p}
under the log-log scale.
The curve is proportional to $1/t$ in the asymptotic region.
This implies that the system almost follows the stationary state and the
annealing schedule is sufficiently slow.
The reason is discussed in Sec.~\ref{sec:2.4}.

\begin{figure}[htb]
\scalebox{1}{\includegraphics{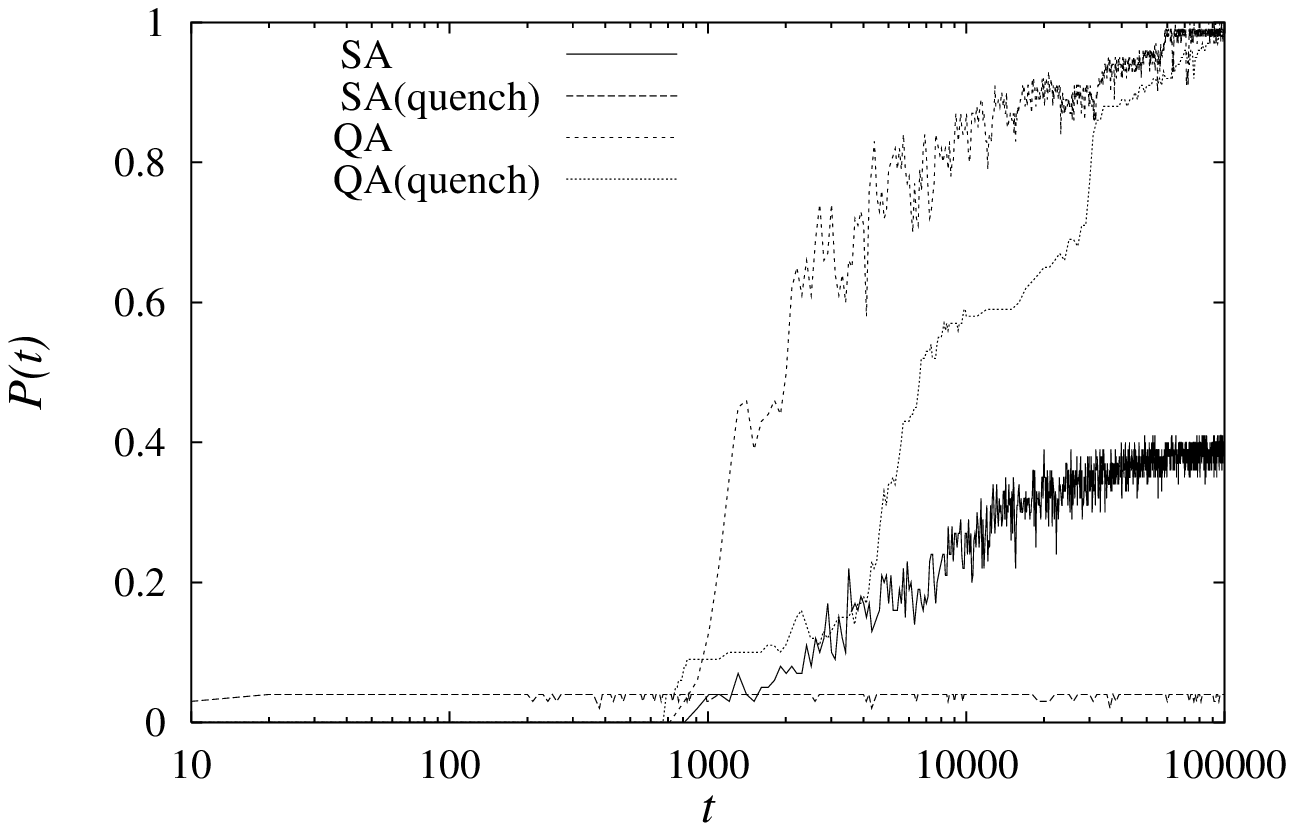}}
\\
\scalebox{1}{\includegraphics{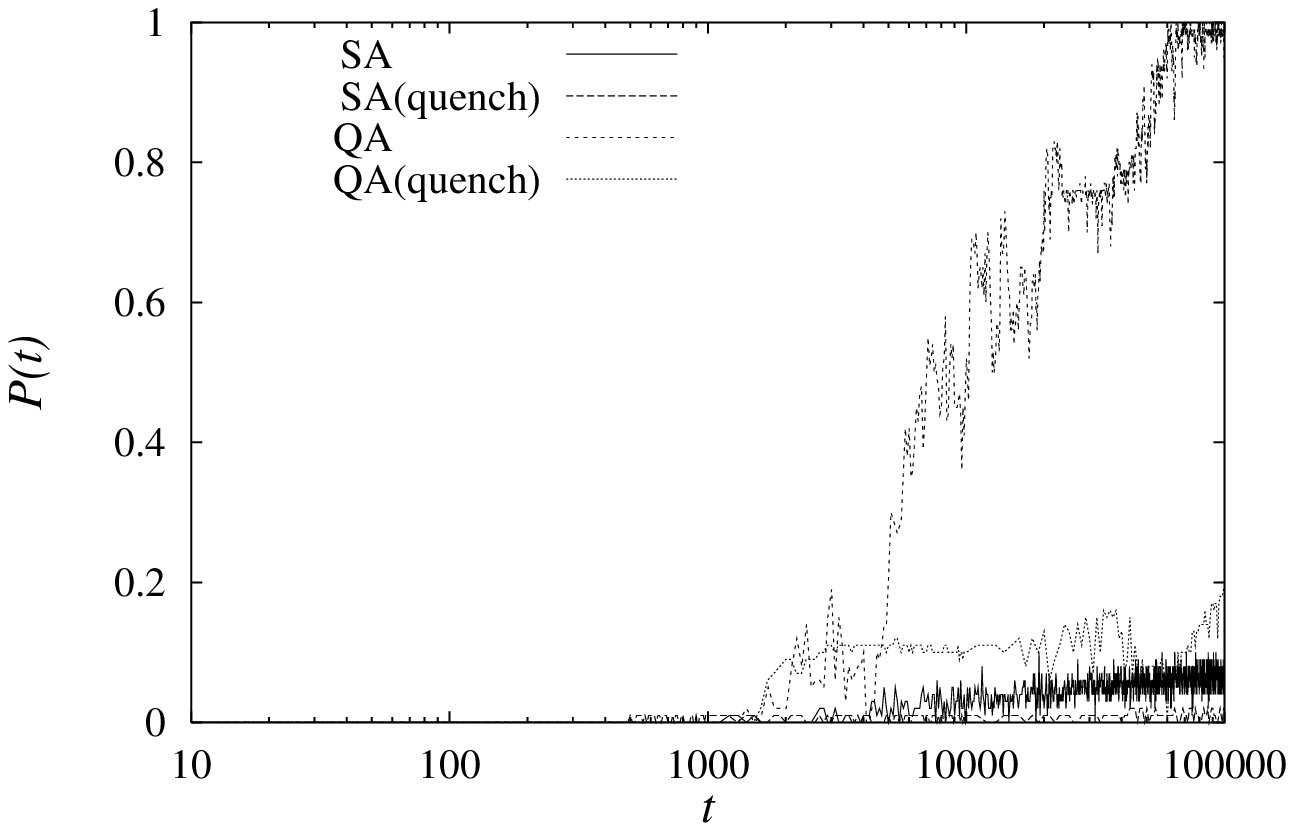}}
\caption{
Difference in the probability between annealed and quenched systems.
Top and bottom figures are the results of ``random'' and ``ulysses16''
respectively.
} 
\label{fig_tsp_relax_p}
\end{figure}

\begin{figure}[htb]
\scalebox{1}{\includegraphics{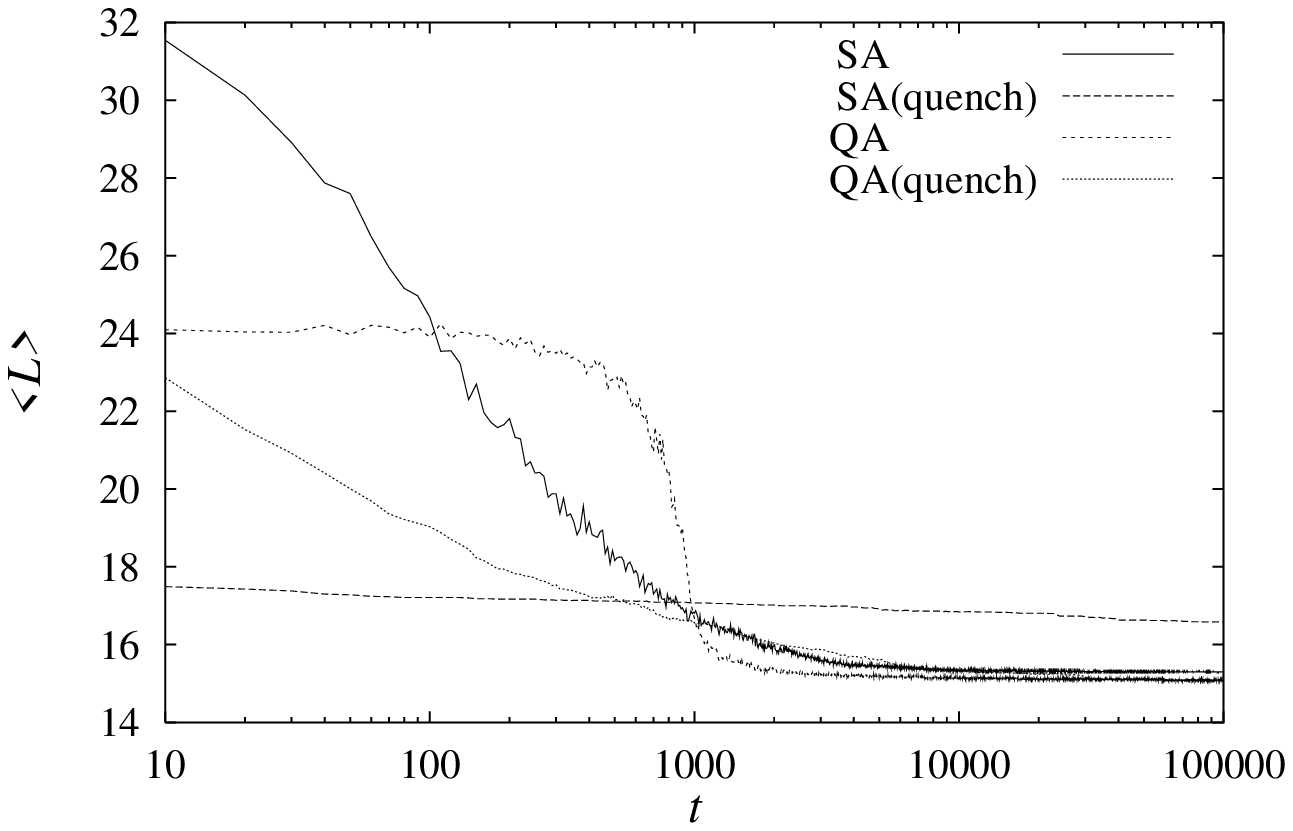}}
\\
\scalebox{1}{\includegraphics{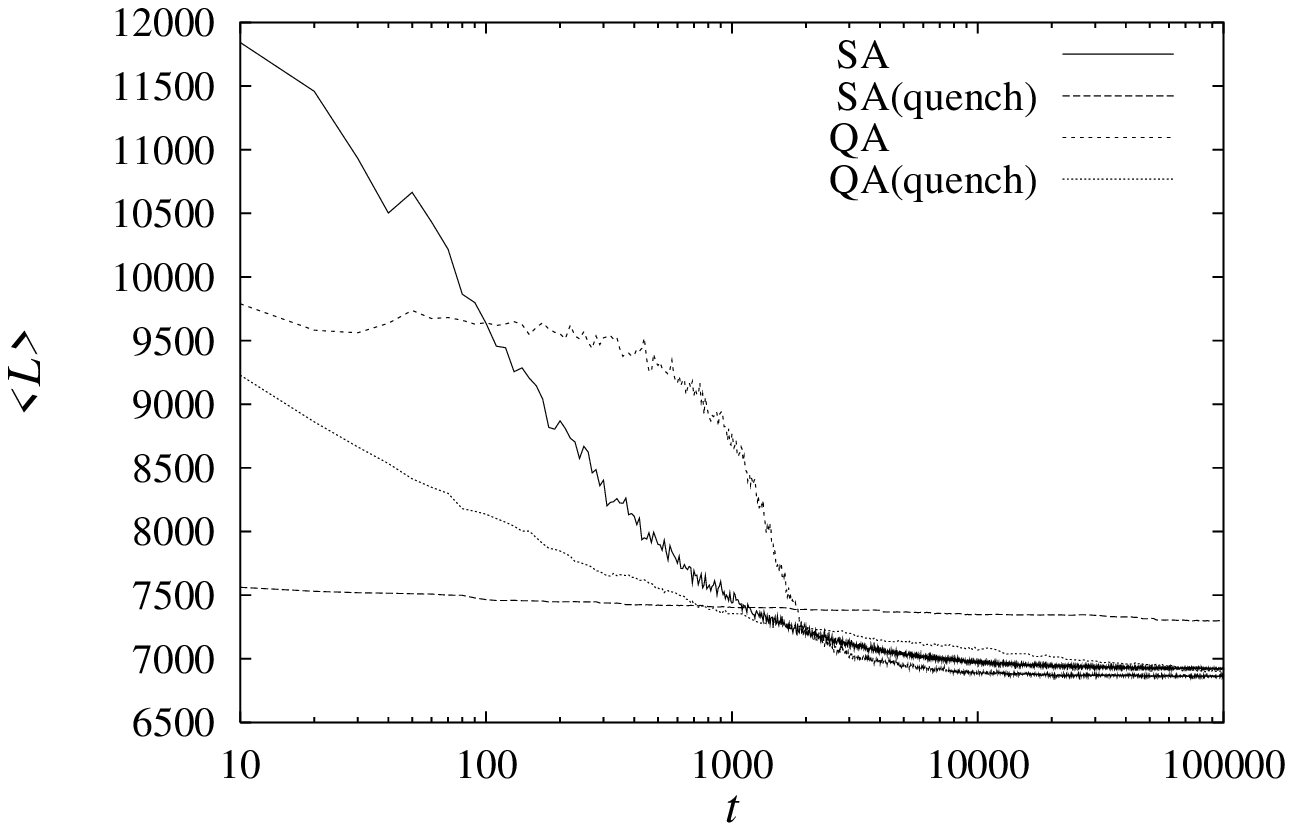}}
\caption{
Difference in the average energy between annealed and quenched systems.
Top and bottom figures are the results of ``random'' and ``ulysses16''
respectively.
} 
\label{fig_tsp_relax}
\end{figure}

Lastly, we compare the annealed and quenched systems of SA and QA
to make the effect of the annealing clear.
Annealing is introduced to avoid slow relaxation in low or zero
temperature in the system whose landscape is complicated.
If the annealing works, the length for the annealed system is lower than the
quenched system.
The quenching parameter is the temperature $T$ for SA and the transverse
field $\Gamma$ for QA.
These values are quenched to $0.1/\sqrt{10}=0.0316 \cdots$.
In the annealed system, the temperature and the transverse field decrease
from infinity to the same value as in the quenched system.
We calculate for the two problems, ``random'' and ``ulysses16''.
The results of the probability and the average energy are shown in
Fig.~\ref{fig_tsp_relax_p} and~\ref{fig_tsp_relax}.
We can list the four system in the order of the performance to find the
ground state: (1) the quantum system with annealing ``QA'', (2) the
quantum system without annealing ``QA(quench)'', (3) the classical
system with annealing ``SA'' and (4) the classical system without
annealing ``SA(quench)''.
We can find that both of the annealed systems show better performance
than the quenched systems.
It is notable that the probability for the quenched quantum system
``QA(quench)'' is larger than SA.
In spite of annealing the temperature, SA finds the ground state less
often than the quenched quantum system.
From these results, we obtain the following conclusions.
The relaxation of the quantum systems is faster than the classical
systems whatever we control the parameters, annealing or quenching.
The annealing process accelerate the relaxations for both of the quantum and
the classical systems.

\section{Summary}
\label{sec:4.5}

We have applied QA to TSP by mapping the problem to the Ising system.
SA has a four-spin flip dynamics in the spin representation of the TSP.
QA should contain the quantum effect which flips four spins at a time.
Because of the difficulty of the calculation, we have adopted the
one-spin dynamics by applying the transverse field.

Numerical results show that QA has better performance than SA in the
probability to find the minimum-length of the closed tour and the
average of the tour length and obeys the $1/t$-law asymptotically.
We only calculated the case in which the Trotter number is $500$ for QA.
The dependence on the Trotter number was not studied, and we did not
estimate how many Trotter slices the quantum Monte Carlo simulation
need.
We also did not check the actual performance of QA.
QA needs the computational power Trotter-number times as large as the
calculation of SA, but we did not take into account this disadvantage.
We only calculated the quantities as a function of the Monte Carlo
steps, not the real computational time.
In other words, we cannot conclude which methods find the
minimum-length tour of TSP in small computer power, SA or QA,
and only conclude so far that QA can be applied to TSP and QA improves
in terms of the Monte Carlo step not the real computational time.
These are the future problems.
Another future problem is to study the effect of the four-spin
interaction term as a quantum effect.
It is not clear that the multi-spin interaction makes the system
converge to the optimal state faster or not.

If QA for TSP accelerates the search for the optimal solution actually,
QA may be applied to various problems which can be mapped to the
Ising-spin systems.
We should have used a four-spin transition term as a quantum effect in QA
for TSP to have correspondence with the four-spin flip dynamics of SA, but
we adopt a one-spin transition term $-\Gamma \sum_i \sigma_i^x$.
The results show that QA finds the optimal solution and the one-spin
transition term works.
This term forces neighboring Trotter slices to be in the same
configuration in the representation of Suzuki-Trotter decomposition.
Assuming that this effect is the essence of QA, we can import such an
effect to non-Ising systems.
For instance, it is possible to extend our method to the problems
represented in terms of Potts spins.

\chapter{Summary}

We have proposed a new idea on the general method to solve
combinatorial optimization problems.
The idea is quantum annealing (QA) in which quantum tunneling effects cause
transitions between states, in contrast to the usual thermal transitions in
simulated annealing (SA).
First, we studied QA by the time-dependent Schr\"odinger equation for
small-size systems.
Next, we extended QA to the stochastic process, the quantum Monte Carlo
method.
The two dynamics, the Schr\"odinger equation and the quantum Monte
Carlo dynamics, are not equivalent to each other.
However, for the purpose of finding the ground state, both of them work
and the performance is improved in comparison with SA.
Finally, we presented the possibility of the implementation of QA for
the general combinatorial optimization problems.
The test-bed was the traveling salesman problem (TSP).
The chance of finding the optimal solution by QA was larger than by SA.

The idea of QA is based on SA which was proposed by Kirkpatrick
{\it et al.} to find the optimal state of the optimization problems by
introducing the annealing process~\cite{Kirk}.
Annealing is performed by decreasing the temperature of the system
from a high temperature to a low temperature.
For instance, let us consider the silicon oxide $\text{SiO}_2$.
If the decreasing rate is slow enough, the system is in crystalline
state.
If the rate is fast, the glass state appears.
SA imitates this process by the Monte Carlo simulation.
We consider QA by replacing thermal fluctuations in SA with quantum
fluctuations.

First, we adopt the transverse field as the quantum effect of the Ising
spin systems.
The task is to find the ground state of the system without the
transverse field.
The transverse field term is controlled and vanishes in the infinite-lime
limit so that the system approaches the ground state.
The dynamics of this system is not described by the Monte Carlo
simulation used for SA.
The natural dynamics of the present system is provided by the
Schr\"odinger equation.
We checked the performance of QA in finding the ground state in comparison
with SA by numerical calculations.
In these calculations, we applied the same function to the time-dependent
schedules of the temperature $T$ and the amplitude of the transverse field
$\Gamma$.
For the $c/\sqrt{t}$ schedule, the behaviors of QA and SA were quite
different.
By QA we find the ground state with probability one, while we do not
find the ground state by SA.
We also found the remarkable property that the probability to find the
ground state converges to one asymptotically as $1/t$.
This property suggests that the system follows the stationary
(or almost stationary) state, because the probability to find the system
in the ground state is expressed using the perturbation in terms of
$\Gamma (\ll 1)$ as
\begin{equation}
\nonumber
 P_{\text{QA}}(\Gamma)
  \sim  1-\Gamma^2\sum_{i\neq 0}\frac{1}{(E_0^{(0)}-E_i^{(0)})^2},
\end{equation}
where $E_i^{(0)}$ is the energy of the $i$th state of
the non-perturbed (classical) system
and $E_0^{(0)}$ is the ground-state energy.
When we put $\Gamma=c/\sqrt{t}$, the $1/t$-law appears.

Secondly, we have considered the calculation for the large-size systems.
The calculation of the Schr\"odinger equation faces difficulties for
larger $N$ because the number of states increases exponentially as $2^N$
and the size of the storage for the Hamiltonian reaches the
storage limit of the computer.
The Monte Carlo method can be used to avoid this difficulty.
There are two ideas to replace the dynamics of the Schr\"odinger equation
with the Monte Carlo method.
One is the path-integral Monte Carlo~\cite{Creutz,Bonca} which regards
the Trotter direction as the time (more preciously, the imaginary time).
The other is the quantum Monte Carlo with ``the Monte Carlo
step''-dependent interaction between Trotter slices.
We regard the Monte Carlo step as the time in the Schr\"odinger equation.
From the similarity to SA, we adopted the quantum Monte Carlo for
large-size systems of QA.

We found that QA improves a performance in finding the ground state even
if we take into account that QA needs $M$ times more calculations than SA.
The Trotter number we adopt is $M=100$, which is large enough for our
calculations (for the two-dimensional EA model with $N=625$).
The $1/t$ convergence of the probability to find the ground state
still appears in the Monte Carlo calculation of QA.
We also found that the relaxation of the quantum system is faster than the
classical system, if the amplitude of the transverse field $\Gamma$ in
the quantum system and the temperature in the classical system have the
same value.
This implies that the quantum effect accelerates relaxations.
We consider that this fact is the reason for the better performance of QA
than SA.

Lastly, we have applied QA to TSP by expressing the problem
in terms of the Ising system.
Numerical results show that QA has a better performance than SA in the probability
to find the minimum-length of the closed tour and the average of the tour
length at the same Monte Carlo step.
Moreover, the $1/t$ convergence of the probability to find the optimal
tour still appears in TSP by QA.
These results imply that the QA works also in TSP.
However, we did not check the actual performance of QA.
QA needs the computational power Trotter-number times in comparison with SA,
but we did not take into account this disadvantage.
We only calculated the quantities as a function of the Monte Carlo
step, not the real computational time.
In other words, we can not conclude which method finds the
minimum-length tour of TSP with smaller computer power, SA or QA,
and only conclude so far that QA can be applied to TSP and QA improves
in terms of Monte Carlo steps, not in the real computational time.

The findings in our study are summarized briefly as follows:
QA provides a general and faster algorithm to find the ground state of
the Ising system whose energy landscape is complicated.
If we adopt the schedule of the transverse field as $\Gamma=c/\sqrt{t}$,
the asymptotic form of the probability to miss the ground state is
proportional to $1/t$.
The relaxation of the quantum system is faster than the classical
system under the condition of the control parameter $\Gamma = T$.
From the application side, the optimal schedule is not known in general.
If the relaxation is fast, we do not have to take into account the
possibility to be stuck in local minima more seriously in QA than SA.
QA may also work for the combinatorial optimization problems which can
be mapped to the Ising spin system.

The future problems are the followings:
The analytical foundations of the better performance of QA than SA and
the critical rate of the decreasing the transverse field are needed.
(The same method of the analysis of SA by Geman and Geman~\cite{Geman}
may be applied to the analysis of the Trotter decomposed system of QA.)
The application to the study of the ground state property of the Ising
spin systems, especially the three-dimensional EA model, is attractive.
Whether the same picture as the SK model, the Parisi
picture~\cite{Parisi3,Dominicis}, can be applied to the
three-dimensional EA model or not is an open problem.
For some problems in SA the multi-spin flip is performed and the
corresponding interaction term representing quantum effect in QA is the
multi-spin interaction term.
For instance, SA for TSP needs four-spin flip, but we adopt only a
single-spin interaction term (the transverse field).
To study the effect of the multi-spin interaction term is needed.
It is not clear whether the multi-spin interaction makes the system
converge to the optimal state faster or not.
On the other hand, the transverse field is mapped to the interaction
between the Trotter slices and the strength of the interaction is
controlled by the strength of the field.
This interaction forces neighbor Trotter slices to be in the same
configuration in the representation of Suzuki-Trotter decomposition.
Assuming that this effect is the essence of QA, we can import such an
effect to a wide class of non-Ising systems.
For example, it is possible to extend our method to the problems
represented in terms of Potts spins.
This is one possible way to apply QA to general problems which can
not be expressed in terms of the Ising spin.

\appendix
\chapter{Single-spin problem}
\addcontentsline{toc}{chapter}{Appendix}

In this Appendix we explain some technical aspects to derive
the exact solution of the time-dependent Schr\"odinger equation
for the transverse Ising model with a single spin.
The three cases of
$\Gamma(t) = - c t, c / t$ and  $c / \sqrt{t}$
will  be discussed.

%- - - - - - - - - - - - - - - - - - - - - - - - - - - - - - - - - - - - - - -
\section[Case of $\Gamma(t) = - c t$ (Landau-Zener model)]
{Case of $\Gamma(t) = - c t$ (Landau-Zener model)
\protect\cite{Miya1,Zener,Miya2,Miya3,Raedt}}

Let us express the solution of the Schr\"odinger equation at time $t$
by the parameters $a=\langle +|\psi (t)\rangle$ and
$b=\langle -|\psi (t)\rangle$.
The Schr\"odinger equation (\ref{Schroedinger}) with
${\cal H}=-h\sigma^z-\Gamma \sigma^x$ is expressed as
a set of first order differential equations for $a$ and $b$.
It is convenient to change the variables as
\begin{equation}
 \tilde{a}=\frac{1}{\sqrt{2}}(a+b), \hspace{1em}
 \tilde{b}=\frac{1}{\sqrt{2}}(a-b),
\end{equation}
by which the Schr\"odinger equation is now
\begin{equation}
 \frac{d^2\tilde{b}(t)}{d t^2}+(-ic+h^2+c^2t^2)\tilde{b}(t)=0.
\end{equation}
By using the notation
\begin{eqnarray}
 z & = & \sqrt{2c}e^{-\pi i/4}t, \\
 \lambda & = & \frac{ih^2}{2c},
\end{eqnarray}
we find
\begin{equation}
 \frac{d^2\tilde{b}(t)}{d z^2}
  +(\lambda+\frac{1}{2}-\frac{1}{4}z^2)\tilde{b}(t)=0.
 \label{diff1}
\end{equation}

The initial state is specified as $a=b=1/\sqrt{2}$ or
$\tilde{b}=0$ as $t\to -\infty$.
The solution of (\ref{diff1}) satisfying this condition is the
parabolic cylinder function $D_{-\lambda-1}(-iz)$ \cite{specialf}.
Thus, we obtain the solution as
\begin{eqnarray}
 \tilde{a}(t) & = & \frac{1}{h}\left(-ct\tilde{b}(t)
                    -i\frac{d\tilde{b}(t)}{dt}\right), \\
 \tilde{b}(t) & = & C_1D_{-\lambda-1}(iz),
\end{eqnarray}
where $C_1$ is a constant.
To fix $C_1$, we use the condition
\begin{equation}
|\tilde{a}(-\infty)| = \frac{2C_1ce^{\pi h^2/8c}}{h\sqrt{2c}} = 1.
\end{equation}
Then we have
\begin{equation}
 C_1=\frac{h}{\sqrt{2c}}e^{-\pi h^2/8c}.
\end{equation}
The wave function of this system is given in Eq. (\ref{eq:lin}).

%- - - - - - - - - - - - - - - - - - - - - - - - - - - - - - - - - - - - - - -
\section{Case of $\Gamma(t) = c/t$}

We next consider the case of $\Gamma(t) = c/t$.
By eliminating $a$ from the Schr\"odinger equation, we obtain
\begin{eqnarray}
\nonumber
\lefteqn{\frac{d^2 b(t)}{d t^2}
 - \frac{1}{\Gamma(t)}\frac{d\Gamma(t)}{dt}\frac{db(t)}{dt}} \\
\label{eq:sch_e}
 & & + \left(h^2+\Gamma^2(t)-\frac{ih}{\Gamma(t)}\frac{d\Gamma(t)}{dt}\right)
 b(t) = 0.
\end{eqnarray}
Substituting $\Gamma(t) = c/t$, we have
\begin{equation}
 \frac{d^2 b(t)}{d t^2}
 - \frac{1}{t}\frac{d b(t)}{d t}
 + \left( h^2 + \frac{i h}{t} + \frac{c^2}{t^2}\right) b(t) = 0.
\end{equation}
The solutions of this equation are expressed by the confluent $P$ function
\cite{specialf}
\begin{eqnarray}
\nonumber
\lefteqn{\tilde{P} \left\{
 \begin{array}{cccc}
  \infty & 0 & \\
  \overbrace{ \makebox[5ex]{$ih$} \makebox[3ex]{$1$} } & ic & t \\
  \makebox[5ex]{$-ih$} \makebox[3ex]{$0$} & -ic &
 \end{array}
 \right\}} \\
 & & = e^{iht} t^{ic} \tilde{P} \left\{
 \begin{array}{cccc}
  \infty & 0 & \\
  \overbrace{ \makebox[5ex]{$0$} \makebox[6ex]{$1+ic$} } & 0 & -2iht \\
  \makebox[5ex]{$1$} \makebox[6ex]{$ic$} & -2ic &
 \end{array}
 \right\},
\end{eqnarray}
the right-hand side of which has two independent expressions
in terms of the confluent hypergeometric function
\begin{eqnarray}
 f(t) & = & e^{iht} t^{ic} F(1+ic,1+2ic;-2iht), \\
\nonumber
 g(t) & = & e^{iht} t^{ic} (-2iht)^{-2ic} \\
      &   & \times F(1-ic,1-2ic;-2iht).
\end{eqnarray}
The general solution is $b(t) = C_1 f(t) + C_2 g(t)$.
Using the initial condition
\begin{eqnarray}
b(0) & = & C_1+C_2 = \frac{1}{\sqrt{2}}, \\
a(0) & = & C_1-C_2 = \frac{1}{\sqrt{2}},
\end{eqnarray}
we find
\begin{equation}
b(t) = \frac{1}{\sqrt{2}} e^{iht} t^{ic} F(1+ic,1+2ic;-2iht).
\end{equation}
The asymptotic forms of $b(t)$ and $|b(t)|^2$ are then given as
\begin{eqnarray}
\nonumber
b(t) & \sim & \frac{\sqrt{2} (2h)^{-ic} \Gamma(2ic)}{\Gamma(ic)} \\
 & &  \times \left\{e^{-iht-\pi c/2} + c e^{iht+\pi c/2} (2ht)^{-1} \right\},
\end{eqnarray}
\begin{equation}
|b(t)|^2 \sim \frac{\sinh(\pi c)}{\sinh(2\pi c)} \left\{
e^{-\pi c} + \frac{c \cos(2ht)}{ht} + \frac{c^2 e^{\pi c}}{4h^2 t^2} \right\}.
\end{equation}

%- - - - - - - - - - - - - - - - - - - - - - - - - - - - - - - - - - - - - - -
\section{Case of $\Gamma(t) = c/\sqrt{t}$}

The final solvable model has $\Gamma(t) = c/\sqrt{t}$.
The Schr\"odinger equation (\ref{eq:sch_e}) is then expressed as
\begin{equation}
\frac{d^2 b(t)}{d t^2}
 - \frac{1}{2t}\frac{d b(t)}{d t}
 + \left( h^2 + \frac{2c^2+ih}{2t} \right) b(t) = 0.
 \label{diff2}
\end{equation}
The solution is the confluent $P$ function \cite{specialf}
\begin{eqnarray}
\nonumber
\lefteqn{\tilde{P} \left\{
 \begin{array}{cccc}
  \infty & 0 & \\
  \overbrace{\makebox[5ex]{$ih$}\makebox[9ex]{$\frac{1}{2}-i\gamma$}} & 0 & t\\
  \makebox[5ex]{$-ih$} \makebox[9ex]{$i\gamma$} & \frac{1}{2} &
 \end{array}
 \right\}} \\
 & & = e^{iht} \tilde{P} \left\{
 \begin{array}{cccc}
  \infty & 0 & \\
  \overbrace{\makebox[5ex]{$0$}\makebox[9ex]{$\frac{1}{2}-i\gamma$}} & 0 &
   -2iht \\
  \makebox[5ex]{$1$} \makebox[9ex]{$i\gamma$} & \frac{1}{2} &
 \end{array}
 \right\},
\end{eqnarray}
where $\gamma = c^2/2h$.
The two independent solutions are thus \cite{specialf}
\begin{eqnarray}
 f(t) & = & e^{iht} F(\frac{1}{2}-i\gamma,\frac{1}{2};-2iht), \\
 g(t) & = & e^{iht} (-2iht)^{1/2} F(1-i\gamma,\frac{3}{2};-2iht).
\end{eqnarray}
The general solution of (\ref{diff2}) is therefore the linear combination
of the above two functions
\begin{equation}
\label{eq:sol_lin}
 b(t) = C_1 f(t) + C_2 g(t).
\end{equation}
The constants $C_1$ and $C_2$ are fixed by the requirement
\begin{eqnarray}
 b(0) & = & C_1 = \frac{1}{\sqrt{2}}, \\
 a(0) & = & \frac{\sqrt{h}}{\sqrt{2}c}e^{(5/4)\pi i}C_2  = \frac{1}{\sqrt{2}}.
\end{eqnarray}
Substituting $C_1$ and $C_2$ into Eq. (\ref{eq:sol_lin}),
we find
\begin{eqnarray}
\nonumber
b(t) & = & \frac{1}{\sqrt{2}} e^{iht} 
    F\left(\frac{1}{2}-i\gamma,\frac{1}{2};-2iht \right) \\
& &     + \frac{c}{\sqrt{h}} e^{(3/4)\pi i} e^{iht} (-2iht)^{1/2}
                   F\left(1-i\gamma,\frac{3}{2};-2iht \right).
\end{eqnarray}
The asymptotic form is
%
%\begin{equation}
%\label{asym2}
%b(t) \sim \sqrt{\pi} e^{-\pi c^2/4h} \Biggl[
%           e^{-iht} (2ht)^{-i\gamma} \left\{
%           \frac{1}{\sqrt{2}\Gamma(\frac{1}{2}-i\gamma)}
%           + \frac{\sqrt{h} e^{(5/4)\pi i}}{c\Gamma(-i\gamma)} \right\}
%         + e^{iht} (2ht)^{-1/2+i\gamma} \left\{
%           \frac{e^{-(1/4)\pi i}}{\sqrt{2}\Gamma(i\gamma)}
%           + \frac{c}{2\sqrt{h}\Gamma(\frac{1}{2}+i\gamma)} \right\} \Biggr].
%\end{equation}
\begin{eqnarray}
\nonumber
b(t) & \sim & \sqrt{\pi} e^{-\pi c^2/4h} \Biggl[
           e^{-iht} (2ht)^{-i\gamma} \left\{
           \frac{1}{\sqrt{2}\Gamma(\frac{1}{2}-i\gamma)}
           + \frac{\sqrt{h} e^{(5/4)\pi i}}{c\Gamma(-i\gamma)} \right\} \\
\label{asym2}
 & &     + e^{iht} (2ht)^{-1/2+i\gamma} \left\{
           \frac{e^{-(1/4)\pi i}}{\sqrt{2}\Gamma(i\gamma)}
           + \frac{c}{2\sqrt{h}\Gamma(\frac{1}{2}+i\gamma)} \right\} \Biggr].
\end{eqnarray}
The probability $|b(\infty)|^2$ that the system remains in the excited state
can be calculated as the asymptotic
form of (\ref{asym2}) with the condition $c^2/h\gg 1$
\begin{eqnarray}
|b(\infty)|^2 & = & \frac{\pi e^{-\gamma \pi}}{2}
                \left|\frac{1}{\Gamma(\frac{1}{2}-i\gamma)}
             +\frac{\gamma^{-1/2} e^{(5/4)\pi i}}{\Gamma(-i\gamma)}\right|^2 \\
 & \sim & \frac{e^{-\gamma \pi}}{4}
                \left|e^{1/2}\left(\frac{1}{2}-i\gamma\right)^{i\gamma}
               +\gamma^{-1/2}e^{(5/4)\pi i}(-i\gamma)^{i\gamma+1/2}\right|^2 \\
 & \sim & \frac{e^{-\gamma\pi}}{4}\left|(-i\gamma)^{i\gamma}\frac{i}{8\gamma}
               \right|^2 = \frac{1}{256\gamma} = \frac{h^2}{64c^4}.
\end{eqnarray}

%- - - - - - - - - - - - - - - - - - - - - - - - - - - - - - - - - - - - - - -
\section[Final value dependence on initial condition]
{Dependence of the final value on the initial condition}

We show that we can choose the initial condition so that
the final state is the ground state when $\Gamma=c/\sqrt{t}$.
This is not possible for $\Gamma=-ct$ or $c/t$.
From Eq. (\ref{eq:sol_lin}), the asymptotic form of
the solution as $t\rightarrow\infty$ is
\begin{eqnarray}
\nonumber
b(t) & \sim & C_1 \frac{\sqrt{\pi}e^{-\pi c^2/4h-iht}(2ht)^{-ic^2/2h}}
           {\Gamma(1/2-ic^2/2h)} \\
     &   &+C_2 \frac{i\sqrt{\pi}he^{-\pi c^2/4h-iht}(2ht)^{-ic^2/2h}}
           {c^2\Gamma(-ic^2/2h)}.
\end{eqnarray}
The coefficients $C_1$ and $C_2$ are fixed under the conditions
$b(\infty)=0$ and $|a(0)|^2+|b(0)|^2=1$ as
\begin{eqnarray}
C_1 & = & \left\{1
         +\frac{\sinh(\pi c^2/h)}{2\sinh^2(\pi c^2/2h)}\right\}^{-1/2},\\
C_2 & = & \frac{ic^2\Gamma(-ic^2/2h)}{h\Gamma(1/2-ic^2/2h)}C_1.
\end{eqnarray}
This solution is not the ground state of the Hamiltonian ${\cal H}(0)$.

The reason why one cannot obtain such a solution for the other schedules
($\Gamma=c/t,-ct$) is the following:
The general solution for $\Gamma=c/t$ also has two coefficients,
and the initial state is represented as the linear combination
of two terms whose phases are indefinite:
\begin{eqnarray}
\label{eq:indet1}
a(0) & = & \left. C_1 t^{ic} \right|_{t\rightarrow 0} -
           \left. C_2 t^{-ic} \right|_{t\rightarrow 0},\\
\label{eq:indet2}
b(0) & = & \left. C_1 t^{ic} \right|_{t\rightarrow 0} +
           \left. C_2 t^{-ic} \right|_{t\rightarrow 0}.
\end{eqnarray}
The lowest-energy state at $t=0$ corresponds to $a(0)=b(0)=1/\sqrt{2}$
(times an arbitrary phase factor), which is realized by choosing $C_2=0$
in Eqs. (\ref{eq:indet1}) and (\ref{eq:indet2}).
The indefiniteness of $t^{ic}$ as $t\rightarrow 0$ is irrelevant
because this is only the overall phase factor.
Such a situation does not happen for other values of $a(0)$ and $b(0)$,
leading to a serious difficulty to determine the wave function at $t=0$.
Thus we cannot choose an initial condition other than $a(0)=b(0)=1/\sqrt{2}$.
A similar fact exists in the case of $\Gamma=-ct$.

\end{document}